\documentclass[useAMS,usenatbib,usegraphicx,usepsfig]{mn2e}
\begin{document}
 \topmargin=-1.3cm

\title[An ultraviolet view of NGC 5846 group]{
  Galaxy evolution in nearby galaxy groups. III. \\ A {\it GALEX} view of NGC 5846, the largest group in the local universe}
\author[A. Marino, P. Mazzei, R. Rampazzo, L. Bianchi]{Antonietta Marino$^{1}$\thanks{E-mail:
antonietta.marino@oapd.inaf.it}, Paola Mazzei$^{1}$, Roberto Rampazzo$^{1}$, Luciana Bianchi$^{2}$ \\  
 $^{1}$INAF Osservatorio Astronomico di Padova, vicolo dell'Osservatorio~5, I-35122  Padova, Italy\\
$^{2}$Dept. of Physics and Astronomy, Johns Hopkins University, 3400 North Charles
           Street, Baltimore, MD 21218  USA }

\date{Accepted. Received}

\pagerange{\pageref{firstpage}--\pageref{lastpage}} \pubyear{2016}

\maketitle

\label{firstpage}

\begin{abstract}
We  explore the co-evolution of galaxies in nearby  groups  ($V_{hel}\leq 3000$
km~s$^{-1}$) with  a multi-wavelength approach.  We analyze GALEX far-UV (FUV) and near-UV (NUV) imaging, and SDSS u,g,r,i,z  data   of groups  spanning a large range of dynamical phases. \\
We characterize the photometric properties of spectroscopically-confirmed galaxy members and investigate the global properties of the groups through a dynamical analysis. \\
Here we focus on NGC 5846, the third most massive association of Early-Type Galaxies (ETG) after the Virgo and Fornax clusters. The group, composed of 90 members, is dominated by ETGs (about 80 per cent), and among ETGs about 40 per cent are dwarfs.  
Results are compared with those obtained for three groups in the LeoII cloud, which are radically different both in member-galaxy population  and  dynamical properties. The FUV-NUV cumulative colour distribution and the normalized UV luminosity function (LF) significantly differ due to the  different fraction of late-type galaxy population.  \\
 The UV LF of NGC 5846 resembles that of the Virgo cluster, however our analysis suggests that star-formation episodes are still occurring in most of the group galaxies, including ETGs. The NUV-i colour distribution, the optical-UV colour-colour diagram, and NUV-r vs. M$_r$ colour-magnitude relation suggest that the gas contribution cannot be neglected in the evolution of ETG-type group members. Our analysis highlights that NGC 5846  is still in an active phase of its evolution, notwithstanding the dominance of dwarf and bright ETGs and its virialized configuration.

\end{abstract}


\section{Introduction}

The study of  the co-evolution of galaxies in groups is crucial to 
address the problem of the star formation quenching and galaxy morphological 
transformations, as groups contain most ($\sim$ 60 per cent)  of the galaxies in the universe  
at the present day \citep[e.g.][]{Tully88, Ramella02, Eke04, Tago08}, and most of the
stellar mass is formed in groups. The transition between galaxy properties 
typical of field and clusters happens just   at the characteristic 
densities of  groups, suggesting the existence of evolutionary mechanisms 
acting before galaxies in groups fall into clusters \citep{Lewis02, Gomez03}.
Such re-processing mechanisms act by transforming field, i.e. spiral galaxies, 
to cluster-like galaxies, i.e. ETGs, and basically drive groups 
from an ``active" (star forming) phase, typical of field, to a more ``passive"
phase, typical of clusters \citep[e.g.][]{Goto03, Gomez03}.  Evidence in this sense comes from
a number of ``classical'' studies on the impact of environment on 
galaxy properties showing that ETGs are more strongly clustered than 
late-type galaxies \citep[e.g][]{Davis76}. \citet{Dressler80} showed that the 
fraction of elliptical and S0 galaxies is higher in denser environments. 
By now, it is widely accepted that galaxies in clusters tend to have depressed star formation rates (SFRs) in comparison with the field population \citep[e.g.][and references therein]{Balogh04, Poggianti06}.\\
Several  physical processes are believed to play a role in galaxy evolution and star 
formation variations. They have been inferred both from observations and simulations
and there is a wide consensus that they are different in rich and poor galaxy environments.  
Mergers can transform spirals into ellipticals \citep{Toomre72, Barnes02}, 
and  quench star formation by ejecting the interstellar medium  via starburst, 
AGN or shock-driven winds \citep[e.g.][]{Dimatteo05}.    
Since velocity dispersions of groups are comparable to the velocity dispersion 
of individual galaxies, both interactions and merging  are more favored  in groups 
than in clusters  \citep{Mamon92} as well as phenomena like ``galaxy strangulation" 
\citep{Kawata2008}. Transforming mechanisms in rich environments should 
act through galaxy-galaxy ``harassment'' and ram-pressure \citep[see e.g.][]{Moore1996}. 
\\
 \indent
 The dramatic evidence of  galaxy transformation has been obtained with 
  GALEX.   UV - optical  colour  magnitude diagrams (CMD)
 evidenced  not only a sequence of  red  galaxies,  mostly ETGs, and a 
 ``blue cloud'' mainly composed of late-type galaxies, but also a intermediate region, the ``green valley",  populated  of transforming galaxies \citep{Schawinski2007}.
 
  We are exploring the co-evolution of galaxies in groups in the local Universe by adopting a multi-wavelength approach.  We use UV  and  optical imaging to analyze a set of nearby groups spanning a large range of evolutionary phases.
 In particular,   GALEX  UV wide-field imaging, made it possible  to directly map  present-day 
 star formation, ranking groups according to their blue and red galaxy population. 
 Moreover, we  analyze group compactness,  a signature of the evolutionary phase, through a 
 kinematic and dynamical analysis. 
To further investigate the transition from active groups to more evolved passive systems,
we  select  groups  with different properties, 
 from late-type galaxy dominated  groups, analogs of our Local Group 
  \citep[][hereafter Paper I]{Marino10}, to groups with an increasing 
 fraction of  ETGs, increasing signatures of interaction, and advanced stage of 
 virialization \citep[][hereafter  Paper II]{Marino13}.\\
 \indent
 In order to investigate the transition between active groups and more evolved passive systems
we analyze in this paper  the NGC 5846 group, that   in the \citet{Ramella02}  catalogue 
 is labeled as USGC U677.\\
 \indent
 \citet{Ferguson1991} showed that the dwarf-to-giant ratio increases with the richness of
 the group.  \citet{Eige10} estimated the early-type-dwarf-to-giant-ratio (EDGR)
  (i.e. dE+dE,N+dS0 to E+S0) for NGC~5486 group obtaining EDGR=2.69. In the Local Supercluster,
  this value is exceeded only by Virgo (EDGR=5.77) and Fornax (EDGR=3.83) clusters 
  indicating that  NGC~5846  is the third more massive galaxy aggregate.
The group is dominated by two bright ETGs, NGC 5846 and NGC 5813, that do not show 
clear signatures of interaction in  UV and optical images. X-ray studies revealed the 
presence of an extended (50 -- 100 kpc radius) halo of NGC 5846,  and numerous peculiar features,  cavities and bubbles  in both galaxies \citep{Trinchieri02, Mulchaey03, Werner2009, Machacek2011, Randall2011, Werner2014}. 
These two X-ray halos suggested the presence of sub-structures in the NGC 5846 group and
motivated a wide  literature about the determination of group members  
  \citep[e.g.][]{Tully87, Haynes1991, Nolthenius93, Giuricin00, Ramella02, Mahdavi05, Eige10}.
  The low optical luminosity galaxy population has been studied  by \cite{Mahdavi05} and \cite{Eige10}: nucleated dwarfs reside near  the bright members,
  and only  four dwarfs show fine structures or interaction signatures. At odd, signatures of
  activity are found in the inner part of the two bright members. 
  \citet{Werner2014} detected H$\alpha$+[NII] at kpc scale 
  and [CII] $\lambda=157~\mu$m emission in  both galaxies. \citet{Rampazzo2013},
  using Spitzer, detected mid infrared lines like  [NeII] $\lambda=12.81~\mu$m,
  [NeIII]$\lambda=15.55~\mu$m and some H$_2$0-0 lines   in both galaxies.\\
   \indent
  Although  NGC 5846  is the third most  massive association of ETGs after  
  Virgo and Fornax, its UV properties 
  are still unknown and this motivates the present study.
 The paper is arranged as follows. Section~2  discusses
the criteria  adopted to select the galaxy members, and  the group 
 kinematical and dynamical properties. 
  Section~3 presents the UV and optical observations 
and  data  reduction. The photometric   
 results  are given in Section~4.  Section~5  and 6 focus on the discussion and
 our  conclusions, respectively. 
H$_0$=75 km~s$^{-1}$Mpc$^{-1}$ is used throughout the paper but for luminosity function, where 
in agreement with \citet{Bos14}  a value of 70 km~s$^{-1}$Mpc$^{-1}$ is adopted.

\section{Membership, and dynamical analysis of the group}
\label{Selection}
\begin{table*}
	\caption{Journal of the galaxy members$^a$.}	 
	 \scriptsize
	\begin{tabular}{lllllllllllll}
	\hline\hline
	Id. &   Galaxy  & RA & Dec.&  Morph.& Mean Hel. & B$_T$   & D$_{25}$ &logr$_{25}$ & P.A. &  Incl. & E(B-V)\\ 
	Nr.    &            & (J2000) & (J2000) & type & Vel &  &  &   & &  &\\
	         &	       & [deg] & [deg]  &  [$^a$ | $^b$] & [km/s] & [mag] & [\arcsec]  &   &[deg] &[deg]&  [mag] \\
				  
	\hline\hline
1 &   PGC053384                & 224.00505  & 2.46358 & S0-a | \dots & 2109 $\pm$ 22 & 15.20 $\pm$ 0.44 & 58.6 & 0.45 & 171.3 & 90 &0.038 \\
2 &   PGC1186917               & 224.14275  & 1.17921 & S0 | \dots & 1918 $\pm$ 6 & 17.24 $\pm$ 0.29 & 27.4 & 0.27 & 41.1 & 72.8   &0.034 \\
3 &   PGC1179522               & 224.47140  & 0.93426 & S? | E/S0 & 1887 $\pm$ 39 & 16.90 $\pm$ 0.39 & 48.8 & 0.54 & 58.7 & 90     &0.034 \\
4 &   PGC184851                & 224.58825  & 1.84533 & E-S0 | E & 1870 $\pm$ 5 & 15.99 $\pm$ 0.29 & 0.87 & 0.26 & 85.6 & 80       &0.041  \\
5 &   SDSSJ145824.22+020511.0     & 224.59995  & 2.08630 & E | dE & 2369 $\pm$ 60 &  -    &  -  &  -  &  -  &  -                   &0.037   \\
6 &   SDSSJ145828.64+013234.6  & 224.61930  & 1.54303 & E | dE,N & 1494 $\pm$ 20 & 17.64 $\pm$ 0.5 & 28.1 & 0.07 & 154.9 & 46      &0.039   \\
7 &   PGC1223766               & 224.67030  & 2.33986 & E | dE,N & 1559 $\pm$ 24 & 18.36 $\pm$ 0.41 & 15.1 & 0.07 & 69.1 & 45.4     &0.036   \\
8 &   PGC1242097               & 224.69205  & 2.96899 & E  | E & 1791 $\pm$ 40 & 16.39 $\pm$ 0.45 & 23.9 & 0.06 & 129.5 & 40.3      &0.036   \\
9 &   PGC053521                & 224.70300  & 2.02350  & E  | E& 1805 $\pm$ 2 & 14.87 $\pm$ 0.32 & 56.0 & 0.2 & 2.6 & 77.5         &0.037    \\
10 &   SDSSJ145944.77+020752.1 & 224.93640  & 2.13106 & E | dE,N & 1458 $\pm$ 59 & 18.39 $\pm$ 0.5 & 18.1 & 0.01 &  -  & 16.5      &0.035    \\
11 &   NGC5806                 & 225.00165  & 1.89128 & Sb | Scd  & 1348 $\pm$ 3 & 12.35 $\pm$ 0.06 & 181.2 & 0.27 & 172.5 & 60.4   &0.039    \\
12 &   PGC053587               & 225.06915  & 2.30069 & S0 | S0 & 1819 $\pm$ 3 & 15.50 $\pm$ 0.26 & 58.6 & 0.4 & 10.1 & 90         &0.034     \\
13 &   SDSSJ150019.17+005700.3 & 225.08010  & 0.95001 & E | dE, N& 1961 $\pm$ 60 & 17.56 $\pm$ 0.5 & 28.1 & 0.07 & 41.2 & 44.1     &0.042      \\
14 &   NGC5846:[MTT2005]046    & 225.11205  & 1.47526 &  \dots | pec & 1501 $\pm$ 60 &  -  &  -  &  -  &  -  &  -                  &0.039      \\
15 &   NGC5811                 & 225.11235  & 1.62362 & SBm | dE& 1535 $\pm$ 6 & 14.76 $\pm$ 0.32 & 61.4 & 0.06 & 96.8 & 31.3      &0.039      \\
16 &   SDSSJ150033.02+021349.1 & 225.13755  & 2.23036 & E | dE & 1278 $\pm$ 37 & 17.24 $\pm$ 0.5 & 33.0 & 0.1 & 31.3 & 58.2        &0.034      \\
17 &   PGC1193898              & 225.21915  & 1.40493 & E | dE,N & 1885 $\pm$ 10 & 16.87 $\pm$ 0.39 & 36.3 & 0.34 & 9.6 & 90       &0.038      \\
18 &   SDSSJ150059.35+015236.1 & 225.24705  & 1.87670 & E | dE,N & 2196 $\pm$ 48 & 18.76 $\pm$ 0.5 & 16.1 & 0.05 & 96.2 & 38.1     &0.035	\\
19 &   SDSSJ150059.35+013857.0 & 225.24735  & 1.64913 & E | E & 2363 $\pm$ 18 & 18.18 $\pm$ 0.35 & 17.3 & 0.14 & 14.7 & 68.7       &0.038	\\
20 &   SDSSJ150100.85+010049.8 & 225.25350  & 1.01382 & E | dE/I & 1737 $\pm$ 5 & 18.03 $\pm$ 0.35 & 23.9 & 0.21 & 113.6 & 90       &0.039	\\
21 &   PGC053636               & 225.26295  & 0.70764 & Sb | S0/a & 1724 $\pm$ 3 & 15.88 $\pm$ 0.29 & 32.3 & 0.11 & 172.6 & 40.5   &0.043	\\
22 &   SDSSJ150106.96+020525.1 & 225.27900  & 2.09031 & E | dE,N & 1943 $\pm$ 25 & 18.33 $\pm$ 0.35 & 20.8 & 0.12 & 154.9 & 62.3   &0.034	\\
23 &   NGC5813                 & 225.29700  & 1.70201 & E | E & 1956 $\pm$ 7 & 11.52 $\pm$ 0.19 & 250.1 & 0.18 & 142.5 & 90         &0.037	\\
24 &   PGC1196740              & 225.31395  & 1.49826 & E | dE & 2139 $\pm$ 5 & 17.75 $\pm$ 0.39 & 25.1 & 0.13 & 0.9 & 68.3        &0.039	\\
25 &   PGC1205406              & 225.31635  & 1.77348 & E | dE/I & 1343 $\pm$ 15 & 18.06 $\pm$ 0.45 & 25.6 & 0.33 & 111.5 & 90     &0.036	\\
26 &   SDSSJ150138.39+014319.8 & 225.40995  & 1.72204 & E | dE,N & 2290 $\pm$ 15 & 17.83 $\pm$ 0.35 & 23.9 & 0.03 &  -  & 28.7      &0.037	 \\
27 &   PGC1208589              & 225.41085  & 1.87018 & E | dE,N & 2152 $\pm$ 2 & 17.90 $\pm$ 0.62 & 14.7 & 0.09 & 106.5 & 52.5    &0.034	 \\
28 &   UGC09661                & 225.51465  & 1.84102 & SBd | Sdm & 1243 $\pm$ 2 & 14.81 $\pm$ 0.3 & 62.8 & 0.04 & 139 & 28.5      &0.034	 \\
29 &   PGC1192611              & 225.61740  & 1.36422 & E | dE & 1516 $\pm$ 55 & 18.49 $\pm$ 0.47 & 15.8 & 0.07 &  -  & 45.6       &0.036	 \\
30 &   SDSSJ150233.03+015608.3 & 225.63750  & 1.93582 & E | dE/I & 1647 $\pm$ 60 & 18.18 $\pm$ 0.5 & 21.8 & 0.14 & 17.2 & 69.2     &0.031	 \\
31 &   SDSSJ150236.05+020139.6 & 225.65010  & 2.02759 & E | dE & 1992 $\pm$ 20 & 18.11 $\pm$ 0.35 & 30.1 & 0.42 & 108.5 & 90        &0.031	 \\
32 &   PGC1230503              & 225.93435  & 2.55236 & E | dE & 1782 $\pm$ 17 & 17.56 $\pm$ 0.35 & 20.8 & 0.14 & 122.2 & 69.1     &0.032	 \\
33 &   SDSSJ150349.93+005831.7 & 225.95790  & 0.97651 & I | dI & 2002 $\pm$ 48 & 16.99 $\pm$ 0.5 & 43.5 & 0.22 & 70.8 & 61.7       &0.037	 \\
34 &   PGC1185375              & 225.95955  & 1.12684 & S0 | E & 1575 $\pm$ 7 & 16.48 $\pm$ 0.35 & 27.4 & 0.17 & 102.8 & 59.3      &0.034	 \\
35 &   PGC087108               & 225.98430  & 0.42954 & I | \dots & 1581 $\pm$ 2 & 17.32 $\pm$ 0.58 & 26.2& 0.15 & 179.1 & 50.8   &0.032	  \\
36 &   NGC5831                 & 226.02900  & 1.21994 & E | E & 1631 $\pm$ 2 & 12.44 $\pm$ 0.11 & 134.3 & 0.05 & 128.7 & 38.5       &0.034	  \\
37 &   PGC1197513              & 226.03515  & 1.52454 & S0-a | S0/a & 1837 $\pm$ 2 & 16.43 $\pm$ 0.38 & 35.7 & 0.25 & 11.1 & 63    &0.035	 \\
38 &   PGC1230189              & 226.05450  & 2.54297 & E | E & 1909 $\pm$ 7 & 15.89 $\pm$ 0.31 & 42.5 & 0.15 & 179.3 & 73.5       &0.029	 \\
39 &   PGC1179083              & 226.09935  & 0.91839 & E | dE & 1657 $\pm$ 60 & 18.12 $\pm$ 0.5 & 18.1 & 0.05 & 127 & 38.4        &0.034	 \\
40 &   PGC1216386              & 226.10295  & 2.11462 & E | E & 1704 $\pm$ 13 & 17.44 $\pm$ 0.34 & 28.1 & 0.27 & 97.9 & 90  	   &0.034  \\
41 &   NGC5846:[MTT2005]139    & 226.14300  & 1.03243 & E | dE,N & 2184 $\pm$ 60 &  -       &  -  &  -  &  -  &  -                 &0.034	      \\
42 &   PGC1190315              & 226.17870  & 1.29088 & E | dE & 1967 $\pm$ 9 & 16.96 $\pm$ 0.42 & 23.9 & 0.04 &  -  & 33.7         &0.032	       \\
43 &   SDSSJ150448.49+015851.3 & 226.20225  & 1.98084 & E | dE & 1960 $\pm$ 23 & 18.07 $\pm$ 0.5 & 21.3 & 0.09 & 18.3 & 53.9       &0.034	     \\
44 &   PGC1211621              & 226.26825  & 1.96426 & E | E & 2381 $\pm$ 2 & 17.60 $\pm$ 0.42 & 16.9 & 0.07 & 160.7 & 45.6       &0.034	      \\
45 &   NGC5838                 & 226.35960  & 2.09949 & E-S0 | S0 & 1252 $\pm$ 4 & 11.79 $\pm$ 0.12 & 233.4 & 0.47 & 38.8 & 90      &0.029	       \\
46 &   NGC5839                 & 226.36455  & 1.63474 & S0 | S0 & 1227 $\pm$ 32 & 13.69 $\pm$ 0.06 & 86.7 & 0.05 & 103.1 & 30      &0.035		\\
47 &   PGC1190358              & 226.36890  & 1.29233 & I | dE & 2304 $\pm$ 2 & 17.79 $\pm$ 0.41 & 28.1 & 0.1 & 157.1 & 41.1       &0.037		 \\
48 &   PGC1199471              & 226.38255  & 1.58772 & E | dE,N & 919 $\pm$ 41 & 18.15 $\pm$ 0.46 & 17.3 & 0.1 & 126.1 & 56.6     &NA	 		 \\
49 &   PGC1190714              & 226.40715  & 1.30309 &  E? | E/dE & 2074 $\pm$ 17 & 17.43 $\pm$ 0.37 & 22.3 & 0.06 & 114.1 & 30   &0.037		  \\
50 &   PGC1209872              & 226.46055  & 1.90834 & E | dE  & 1721 $\pm$ 9 & 16.93 $\pm$ 0.31 & 29.4 & 0.11 & 177.2 & 61.1     &0.032		 \\
51 &   PGC1213020              & 226.47165  & 2.00775 & I | dI & 1300 $\pm$ 31 & 18.35 $\pm$ 0.46 & 21.3 & 0.27 & 143.9 & 67.4     &0.032		  \\
52 &   NGC5845                 & 226.50330  & 1.63397 & E | E & 1450 $\pm$ 9 & 13.44 $\pm$ 0.15 & 60 & 0.15 & 152.9 & 72            &0.034		  \\
53 &   PGC1218738              & 226.51410  & 2.18486 & Sm | Sm & 1659 $\pm$ 4 & 16.34 $\pm$ 0.31 & 41.5 & 0.06 & 148.4 & 32.6     &0.030		  \\
54 &   PGC1191322              & 226.52805  & 1.32242 &  E? | E/dE & 2300 $\pm$ 14 & 18.01 $\pm$ 0.34 & 19.4 & 0.21 & 66 & 53.5    &0.032		  \\
55 &   PGC1215798              & 226.54695  & 2.09585 & Scd | Scd & 1824 $\pm$ 1 & 17.64 $\pm$ 1.92 & 51.1 & 0.68 & 5.2 & 82       &0.030		  \\
56 &   NGC5846                 & 226.62180  & 1.60629 & E | E & 1750 $\pm$ 32 & 11.09 $\pm$ 0.16 & 255.9 & 0.02 &  -  & 25          &0.035		  \\
57 &   NGC5846A                & 226.62150  & 1.59494 & E | E & 2251 $\pm$ 18 & 12.72 $\pm$ 0.35 & 189.7 & 0.15 & 111.7 & 66.7       &0.035		  \\
58 &   SDSSJ150634.25+001255.6 & 226.64265  & 0.21556 &  E? | \dots & 2006 $\pm$ 75 & 17.87 $\pm$ 0.5 & 28.7 & 0.21 & 22.5 & 53.3  &0.035		  \\
59 &   PGC3119319              & 226.64265  & 1.55883 & E | E & 1509 $\pm$ 2 & 16.13 $\pm$ 0.35 &  -  &  -  & 140 &  -             &0.035		  \\
60 &   NGC5841                 & 226.64580  & 2.00488 & S0-a | S0 & 1257 $\pm$ 2 & 14.55 $\pm$ 0.34 & 70.5 & 0.39 & 152.9 & 90     &0.030		  \\
 \hline
\end{tabular}
 										   
$^a$data from \tt HYPERLEDA http://leda.univ-lyon1.fr \citep{Makarov14}.  $^b$data from \citet{Mahdavi05}.					   
 
 \label{tab1}
\end{table*}

\begin{table*}
\addtocounter{table}{-1}
	\caption{continued.}	 
	 \scriptsize
	\begin{tabular}{llllllllllllll}
	\hline\hline
	 Id. &  Galaxy  & RA & Dec.&  Morph.& Mean Hel. & B$_T$   & D$_{25}$ &logr$_{25}$ & P.A. &  Incl. &E(B-V)\\ 
	  Nr. &              & (J2000) & (J2000) & type & Vel &  &  &   & & & \\
	      &	       & [deg] & [deg]  &    [$^a$ | $^b$] & [km/s] & [mag] & [\arcsec]  &   &[deg] &[deg]& [mag] \\
				  
	\hline\hline
61 &   PGC1156476              & 226.67070 & 0.07675 &  E? | \dots & 1663 $\pm$ 11 & 18.07 $\pm$ 0.32 & 20.8 & 0.23 & 8.9 & 54.7  & 0.034   \\
62 &   PGC1171244              & 226.67520 & 0.63445 & E | dE  & 2260 $\pm$ 9 & 17.99 $\pm$ 0.28 & 18.1 & 0.18 & 169.7 & 90       & 0.033   \\
63 &   NGC5846:[MTT2005]226    & 226.74300 & 1.99454 & E | dE & 1307 $\pm$ 60 &   &  -  &  -  &  -  &  -                          & 0.030    \\
64 &   NGC5850                 & 226.78185 & 1.54465 & Sb | Sb & 2547 $\pm$ 3 & 11.89 $\pm$ 0.24 & 198.7 & 0.15 & 114.4 & 46.9     & 0.035     \\
65 &   PGC1185172              & 226.89225 & 1.12043 &  S? | E/dE   & 1586 $\pm$ 8 & 17.64 $\pm$ 0.37 & 19.9 & 0.16 & 124 & 47.7  & 0.032     \\
66 &   PGC054004               & 226.90500 & 2.01954 & E |dE,N  & 1923 $\pm$ 10 & 15.86 $\pm$ 0.28 & 41.5 & 0.08 &  -  & 50.4     & 0.029      \\
67 &   NGC5854                 & 226.94880 & 2.5686 & S0-a | S0  & 1730 $\pm$ 26 & 12.65 $\pm$ 0.09 & 181.2 & 0.63 & 55 & 90       & 0.028      \\
68 &   PGC054016               & 226.94910 & 1.29209 & E | E & 2070 $\pm$ 7 & 15.67 $\pm$ 0.4 & 36.1 & 0.02 &  -  & 23.7          & 0.035      \\
69 &   PGC1217593              & 227.00580 & 2.15102 & E | E & 1073 $\pm$ 24 & 18.04 $\pm$ 0.4 & 19.0 & 0.15 & 36.5 & 76.9         & 0.027      \\
70 &   PGC054037               & 227.02335 & 1.65156 & S? | S0/a & 1843 $\pm$ 3 & 16.08 $\pm$ 0.73 & 34.5 & 0.29 & 115 & 63.4     & 0.032      \\
71 &   NGC5846:[MTT2005]258    & 227.03550 & 2.90502 & E | dE, N & 1652 $\pm$ 60 &    &  -  &  -  &  -  &  -                      & 0.027      \\
72 &   NGC5846:[MTT2005]259    & 227.03835 & 1.42058 & E | dE, N & 2314 $\pm$ 60 &  &  -  &  -  &  -  &  -                        & 0.032      \\
73 &   PGC054045               & 227.03850 & 1.60856 & I | dI & 2158 $\pm$ 21 & 16.09 $\pm$ 0.46 & 37.8 & 0.04 &  -  & 25.5        & 0.032    	\\
74 &   SDSSJ150812.35+012959.7 & 227.05155 & 1.49975 & E | dE,N & 1537 $\pm$ 28 & 18.02 $\pm$ 0.36 & 22.3 & 0.01 &  -  & 16.6     & 0.032    	  \\
75 &   NGC5846:[MTT2005]264    & 227.08275 & 1.68963 & E | dE,N & 2088 $\pm$ 60 &  &  -  &  -  &  -  &  -                         & 0.029    	   \\
76 &   PGC1206166              & 227.09445 & 1.79848 & E | dE, N & 1741 $\pm$ 13 & 18.08 $\pm$ 0.4 & 25.6 & 0.39 & 139.4 & 90     & 0.030    	   \\
77 &   NGC5846:[MTT2005]268    & 227.10690 & 1.70693 & E | dE & 2049 $\pm$ 60 &   &  -  &  -  &  -  &  -                          & 0.030    	   \\
78 &   PGC1209573              & 227.19660 & 1.9001 & E | dE & 1991 $\pm$ 8 & 16.67 $\pm$ 0.3 & 41.1 & 0.36 & 159.6 & 90          & 0.020    	    \\
79 &   PGC1176385              & 227.26785 & 0.82193 & Sa | S0/a & 1644 $\pm$ 2 & 16.81 $\pm$ 0.29 & 33.7 & 0.22 & 179 & 57.9     & 0.033    	    \\
80 &   SDSSJ150907.83+004329.7 & 227.28270 & 0.72479 & E | dE & 1666 $\pm$ 10 & 17.69 $\pm$ 0.35 & 43.5 & 0.49 & 133.8 & 90       & 0.032    	    \\
81 &   PGC1210284              & 227.31225 & 1.92142 & E | dE & 1728 $\pm$ 9 & 16.64 $\pm$ 0.29 & 34.5 & 0.2 & 87.2 & 90          & 0.020    	    \\
82 &   NGC5864                 & 227.38995 & 3.05272 & S0 | S0 & 1802 $\pm$ 21 & 12.70 $\pm$ 0.19 & 150.7 & 0.51 & 66.5 & 90        & 0.027    	    \\
83 &   NGC5869                 & 227.45580 & 0.47011 & S0 | \dots & 2074 $\pm$ 16 & 13.15 $\pm$ 0.25 & 131.3 & 0.19 & 110.7 & 61.5 & 0.032    	    \\
84 &   UGC09746                & 227.57010 & 1.93358 & Sbc | Scd & 1736 $\pm$ 4 & 14.84 $\pm$ 0.27 & 46.7 & 0.53 & 138.6 & 78.6   & 0.020    	    \\
85 &   UGC09751                & 227.74365 & 1.43753 & Sc | Sd & 1553 $\pm$ 7 & 15.97 $\pm$ 0.67 & 73.8 & 0.58 & 118.5 & 78.8     & 0.027    	    \\
86 &   PGC1202458              & 227.75550 & 1.6806 & E | dE & 1652 $\pm$ 18 & 17.28 $\pm$ 0.29 & 27.4 & 0.09 & 171.9 & 53.1      & 0.024    	      \\
87 &   SDSSJ151121.37+013639.5 & 227.83965 & 1.61079 & E | dE & 2029 $\pm$ 69 &  -  &  -  &  -   &  -  &  -                       & 0.024    	      \\
88 &   UGC09760                & 228.01050 & 1.69849 & Scd | Sd & 2021 $\pm$ 3 & 15.20 $\pm$ 0.65 & 106.7 & 0.73 & 61.2 & 85.1     & 0.025    	      \\
89 &   PGC1199418              & 228.03420 & 1.58584 & E | E & 1941 $\pm$ 3 & 17.00 $\pm$ 0.33 & 20.3 & 0.07 & 137.7 & 45.4       & 0.025    	       \\
90 &   PGC1215336              & 228.10005 & 2.07999 &  S? | \dots & 1684 $\pm$ 10 & 16.94 $\pm$ 0.29 & 30.1 & 0.19 & 96.9 & 50.9  & 0.015    	       \\
 \hline
\end{tabular}
   										   				    
 \end{table*}

 \subsection{Selection of the NGC 5846 group members}
\begin{figure}
  \centering
   \vspace{-0.5cm}  
 \includegraphics[width=8cm,angle=-90]{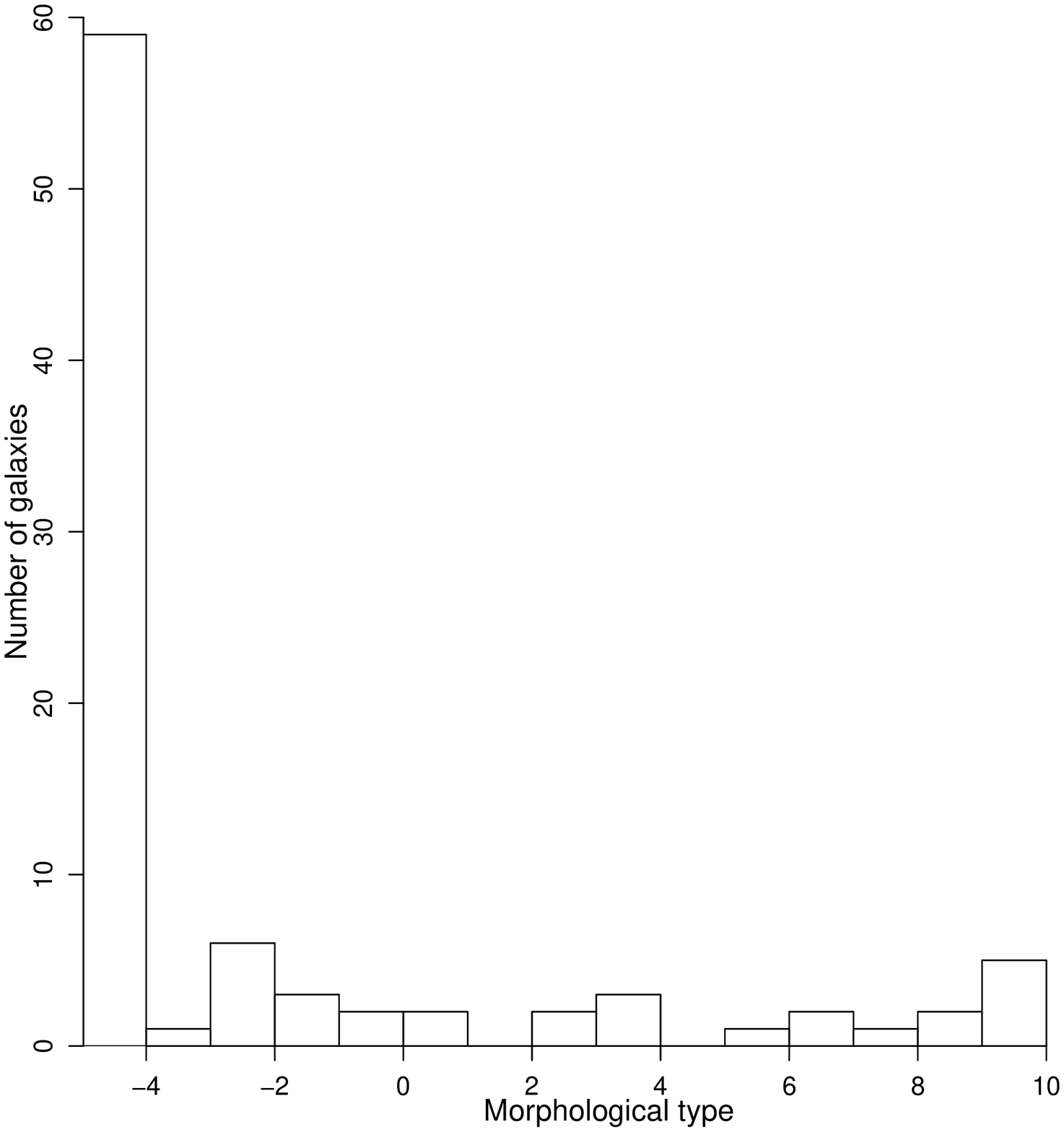}
 \includegraphics[width=8cm,angle=-90]{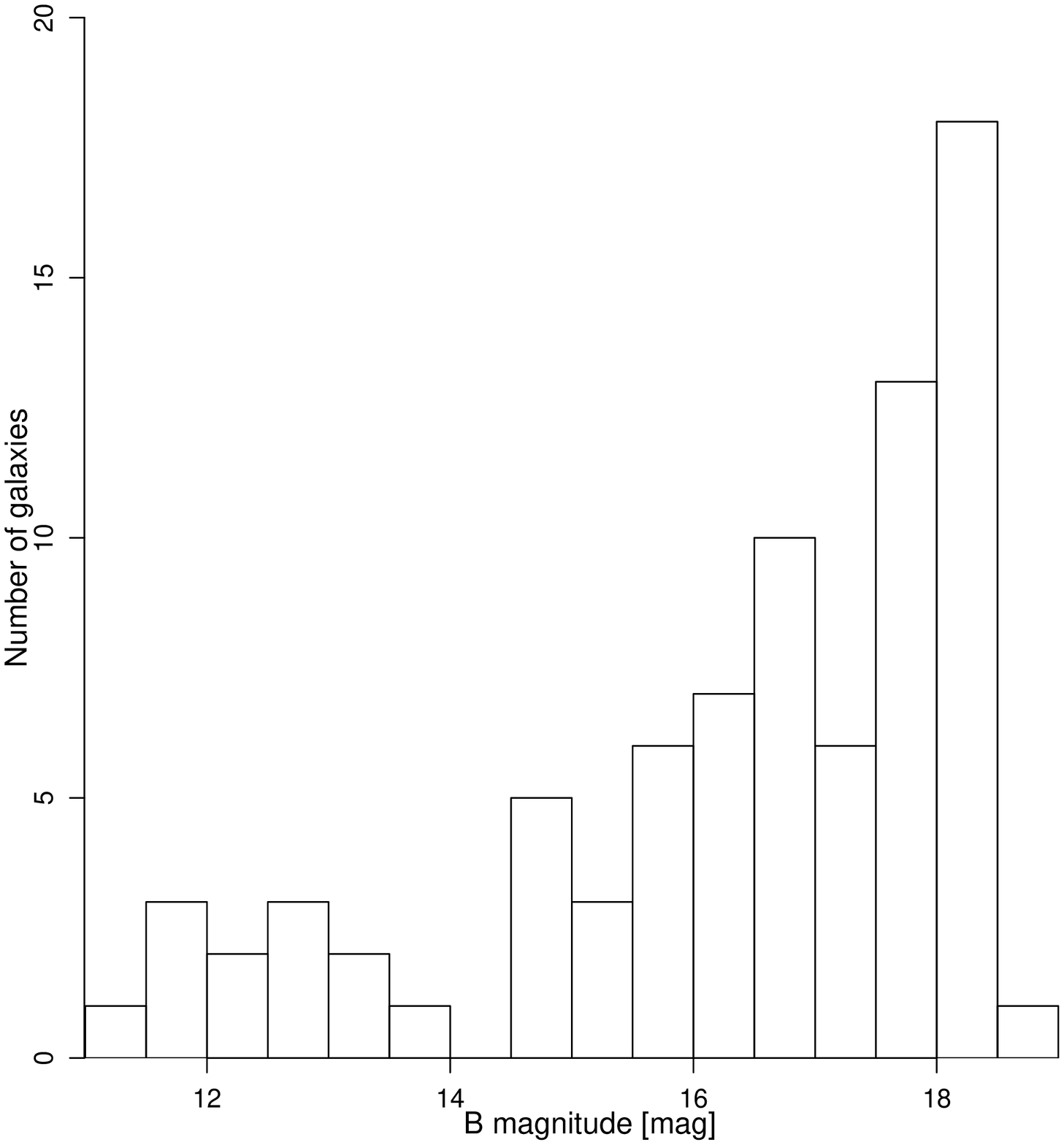}\\
    \caption{Morphological type (top)  
    and  B magnitude (bottom)  distributions of NGC 5846    
    members from {\tt HYPERLEDA}.  All members are spectroscopically 
   confirmed  (see Section~\ref{Selection}). }
  \label{f2}
  \end{figure}

We follow the approach developed in our Paper~I and II.  Briefly, once 
 characterized the group through a density analysis of a region 
of 1.5\,Mpc of diameter around the B brightest member, we revise
 the group membership using recent red-shift surveys. 
For each member galaxy, we investigate morphology and measure surface 
photometry in  FUV, NUV, and optical $u, g, r, i,  z$ bands.\\
 As  described in more detail in Paper II, the group sample has been selected 
 starting from the catalog of \citet{Ramella02} which lists 1168 groups of galaxies  
 covering  4.69 steradians to a limiting magnitude of  m$_{B}\approx$15.5.
The  member galaxies of the catalog were cross-matched  with the 
 GALEX and SDSS archives in order to select
groups covered by both surveys.
We chose only groups within 40 Mpc,  i.e. with an heliocentric radial velocity 
V$_{hel} <$ 3000 km\,s$^{-1}$,   and composed of at least 8 galaxies  to single out
intermediate  and rich structures. 
The above criteria led to a sample of 13 nearby groups having between 
8 and 47 members listed in the catalog of  \citet{Ramella02}. Their  fraction of ETGs, according to the   
{\tt Hyper-Lyon-Meudon Extragalactic Database} \citep[][{\tt HYPERLEDA} hereafter]{Makarov14},
ranges from the same fraction as in the  field, i.e. $\approx$15-20 per cent, to a value  typical 
of dense environments, $\approx$80-85 per cent \citep[e.g.][]{Dressler80}.   
  UV and optical data are available for most of their galaxy members and we
further obtained new  NUV imaging of most of the remaining 
galaxies in the GALEX GI6 program 017 (PI A. Marino).  \\
We focus here on  NGC 5846 group. 
In the catalogue of \citet{Ramella02} the group, named USGC U677, is composed of 17 members with  $\langle V_{hel} \rangle \sim$ 
1634 $\pm$ 117 kms$^{-1}$ and has an  average apparent B magnitude of $\langle B_T \rangle \sim$
11.64 $\pm$ 2.07 mag.
According to the {\tt HYPERLEDA} classification, the group is dominated by ETGs,  
whose percentage achieves $\sim$70 per cent, similar  to that in clusters. 
We further include as group members all the galaxies in the  {\tt HYPERLEDA} catalogue with 
a heliocentric radial velocity within $\pm$3$\sigma$ the group average velocity, 
$\langle V_{hel} \rangle$,  and within
a diameter of $\sim$ 1.5\,Mpc around the centre of the group given by  \citet{Ramella02}.\\
 Table \ref{tab1} lists the properties of the 90 galaxies selected using the method described above.
Columns from 1 to  12 provide  our ID member number,  the galaxy name,  J2000  coordinates,  morphological type, heliocentric  velocity, B total apparent magnitude,  
major axis diameter, D$_{25}$,  axial ratio, position angle, inclination and the foreground galactic extinction,  respectively.
 The morphological type  is taken from {\tt HYPERLEDA}. This catalogue associates a type, T,
    to the morphological classification. Galaxies with T$\leq$0
   are considered early-type. Ellipticals are in the range -5 
    $ \leq$ T $\leq$ -3, Spirals have T $\geq$ 0. Dwarf galaxies
    are not fully considered in this classification.  However,
   Sm, magellanic Irr, (T=9) and generic Irr (T=10) are dwarfs. 
   Dwarf  early-type are  not classified in the {\tt HYPERLEDA}
   scheme. We adopt for them the same classification scheme and values of T as 
   for bright ones. 
We also added in column 5 the morphological type provided by \citet{Mahdavi05}.
These authors identified dwarfs 
among Ellipticals (Es) distinguishing normal dEs and nucleated dEs,N.  
Apart from luminosity classification, in very few cases the two morphological classifications 
differ. 
Figure \ref{f2}  shows the morphological type distribution (top panel) and the apparent B-band 
 magnitudes (bottom panel) of the 90 members of the group. Members with morphological type T
 $\leq$ -4 and apparent B magnitudes $\geq$ 16 dominate.

We compared the members of the group  by {\tt HYPERLEDA}  and  \citet{Mahdavi05}  selection criteria with those
identified in the literature. \citet{Mahdavi05}  argue, on statistical grounds, that
a total of 251$\pm$10 galaxies, listed in their Table~1 composed of 324
candidates, are members of the group. In their Table 2 \citet{Mahdavi05} provide the number of spectroscopically confirmed
members, which amount to 84, belonging to different classes statistically established.
These classes range from 0 to 5, i.e from members confirmed by spectroscopy, priority rating 0, to galaxies excluded by their statistical surface brightness criteria, priority rating 5. 
Our selection of 90 members includes all 84 spectroscopically confirmed members of \citet{Mahdavi05},  four dwarfs with spectroscopic redshift included in the \cite{Eige10} list, plus other two ETGs in {\tt HYPERLEDA}. 
We report in col. 3 of Table \ref{exp}  the identification number provided by \citet{Mahdavi05}
and by \citet{Eige10}. All dwarfs in the \citet{Eige10} list,  but two, NGC~5486\_56 and NGC~5846\_51,
are included by our selection criteria. PGC087108 according to \citet{Eige10} are two HII regions classified as
individual galaxies in SDSS e in PGC. UV imaging indicated that it is a galaxy (see Section 4.1).\\
Summarizing, our selection procedure includes all spectroscopically confirmed members
present in the \citet{Mahdavi05} and \citet{Eige10} lists within a diameter of $\sim$1.5\,Mpc about the group centre. As explained in the following section we extended the  search of spectroscopic possible members in a wider area of 4\,Mpc.  Additional galaxies, not included in \citet{Mahdavi05} and \citet{Eige10}, with a redshift measure are listed  in  Table \ref{a2}.
 Although compatible with membership in the  redshift space, they are  distant 
from the centre of mass of the group and outside the box of 1.8\,Mpc 
considered by  \citet{Mahdavi05}.\\

\subsection{Substructures}
The presence of substructures in a galaxy group is  a signature of recent accretion
and therefore  probes  the evolution of its members \citep[e.g][]{Firth06, Hou12}.
Substructures manifest  as a deviation in the spatial and/or velocity arrangement of the system.
 
If a group is  a dynamically relaxed system, the spatial distribution  of its galaxies should be approximately spherical and their velocity distribution  Gaussian.  The presence of substructures 
indicates a departure from this quasi-equilibrium state. As already discussed in Paper II,
at least one the following characteristics shows the presence of substructures:
(i) significant multiple peaks in the galaxy position distribution;
(ii) significant departures from a single-Gaussian velocity distribution;
(iii) correlated deviations from the global velocity and position
distribution. 

\begin{figure}
   \vspace{-0.3cm}
  \includegraphics[width=8cm]{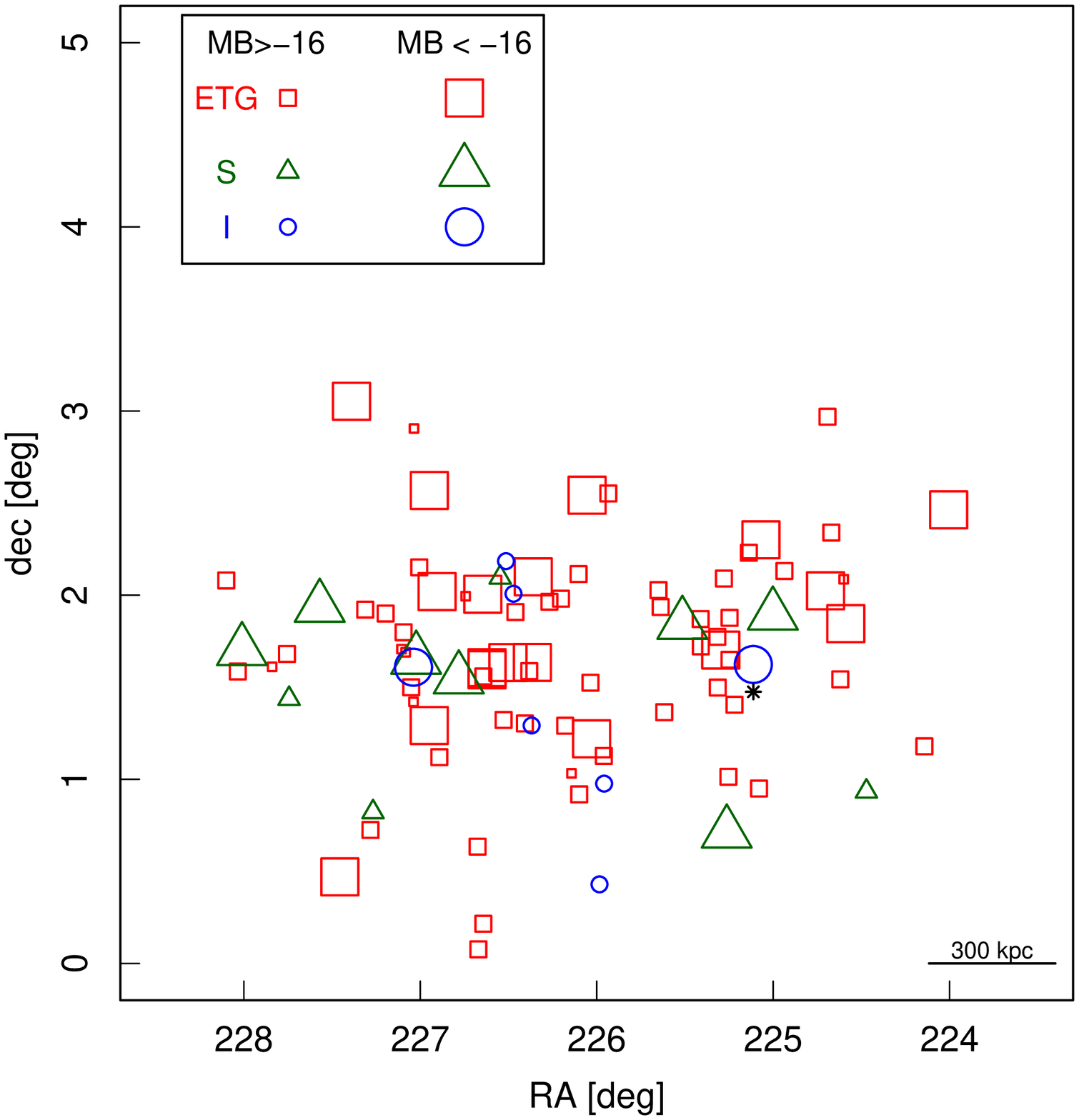}\\
  \vspace{-0.3cm}  
 \includegraphics[width=8cm,angle=-90]{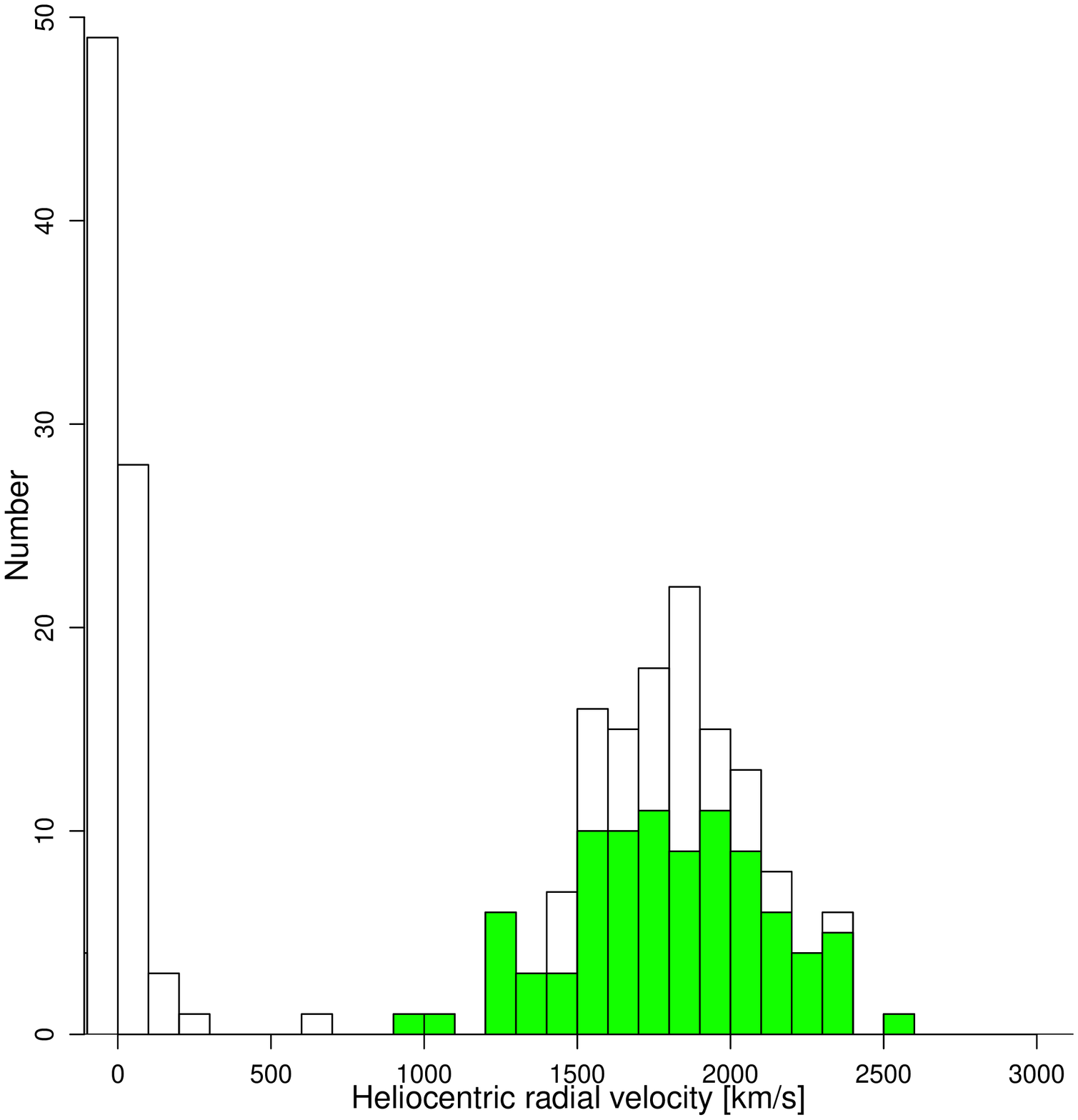}\\
    \caption{{\it Top:} Spatial distribution of galaxy members 
separated in B magnitude bins and  morphological type.  
The smallest symbols are galaxies with no B magnitudes. Black asterisk  refer to
galaxies with no B magnitudes and no morphological types.   
 {\it Bottom:}
 Histogram of the heliocentric radial velocity (10-3000 kms$^{-1}$ )
of galaxies within a box of 4\,Mpc$^2$ centred on NGC~5846. The width of the velocity bins is 100 km\,s$^{-1}$.
Green filled bins show the velocity distribution of the 90 members listed in
Table~1. }
 \label{f1}
  \end{figure}

Fig.~\ref{f1} shows the projected spatial distribution of the group members (top panel). 
Galaxies are separated in B magnitude bins and  morphological types. 
ETGs,  Spirals and Irregulars, with absolute magnitudes M$_B >$ -16 and, M$_B <$ -16 are 
indicated with  squares, triangles and circles of increasing size, respectively. 
The group is dominated by ETGs (72 per cent),  60 per cent  Ellipticals\footnote{Morphological type $\leq$ -3}  approximately  homogeneously distributed. Two peaks may be present in the projected spatial distribution.  \\
\indent
 The velocity distribution of group members is shown in Figure \ref{f1} (bottom panel, filled bins). To discern if the heliocentric radial velocity  has a Gaussian  distribution, we applied the Anderson-Darling normality test and found it   
 does not significantly depart (p-value = 0.92) from normality.\\

We also performed  the  \citet{Dressler88} test (DS test hereafter), which uses both spatial information and velocity, to find substructures in our group.
The DS test identifies a fixed number, NN, of nearest neighbours on the sky around each galaxy, 
computes the local mean velocity and velocity dispersion of this subsample, and compares 
these values with the average velocity and velocity dispersion of the entire group,   $\bar{v}$ and  $\sigma_{gr}$ respectively. The deviations  of the local average velocity and the dispersions from the global values are summed.
 In particular, for  the galaxy $i$, the deviation of its projected neighbors is defined as 
$\delta_i$ = (NN +1)/$\sigma_{gr}$ [ (v$_{loc}$ - $\bar{v})^2$ + ($\sigma_{loc} - \sigma)^2$],
where v$_{loc}$ and $\sigma_{loc}$ are the local average velocity and velocity dispersion.
The total deviation, $\Delta$, is  the sum of the local deviations, $\delta_i$ :
$\Delta$ = $\sum_i^N\delta_i$,
where N is the number of the group members. 
If the group velocity distribution is close to  Gaussian and the local variations are only random 
fluctuations,   $\Delta$ will be of the order N.  If  $\Delta$ varies significantly from N there is probable substructure. 

To compute $\delta$, we set the number of neighbors, NN,  at  10 $\approx$  N$^{1/2}$  \citep[see e.g.][]{Silverman86}.  
Since $\delta_i$ are not statistically independent 
 it is necessary to calibrate the $\Delta$ statistic by performing a Monte Carlo analysis.
The velocities are randomly shuffled among the positions and  $\Delta_{sim}$  is
recomputed 10000 times  to provide  the probability that the measured  $\Delta$ is a random result. 
The significance of having substructure, given by the p-value, is quantified by the ratio of the  number of the simulations in  which the value of $\Delta_{sim}$ is larger than the observed value, and the total number of simulations: p = (N($\Delta_{sim} > \Delta)/N_{sim}$). 

It should be noted that the $\Delta$ statistic is insensitive to subgroups that are well superimposed. It relies on some displacement of the centroids of the subgroups.

The p-value measures the probability that a value of $\Delta_{sim}\ge \Delta$ occurs by chance;  a  p-value $>$0.10  gives a  high significance level to the presence of substructures, values ranging from 0.01 and 0.10 give substructures from marginal to probable.

\begin{figure}
 \vspace{-0.3cm}
 \includegraphics[width=8cm,angle=-90]{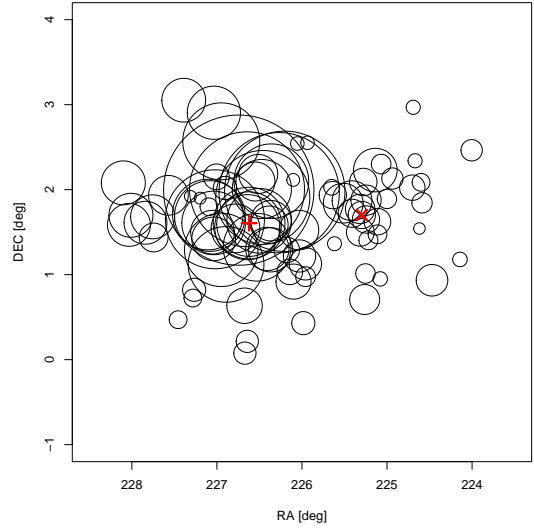}
  \caption{  \citet{Dressler88} `bubble-plot'  based on the ten nearest galaxies. The bubble size is proportional to the squared deviation of the local velocity distribution from the group velocity distribution.  Red plus and  cross show the position of NGC 5846 and NGC 5813 respectively.
  }
 \label{f11}
  \end{figure}

In  Figure \ref{f11}, we show  the Dressler-Shectman `bubble-plot'; 
each galaxy in the group is marked by a circle whose diameter scales with  $e^{\delta}$. 
Larger circles indicate larger deviations in the local kinematics compared to the global one. 
 Many large circles in an area indicate a correlated spatial and kinematical variation, i.e. a substructure.  We also show the position of NGC 5846 and NGC 5813.
The group does not present significant substructures (p= 0.07).

\begin{figure}
  \centering
 \includegraphics[width=8cm,angle=-90]{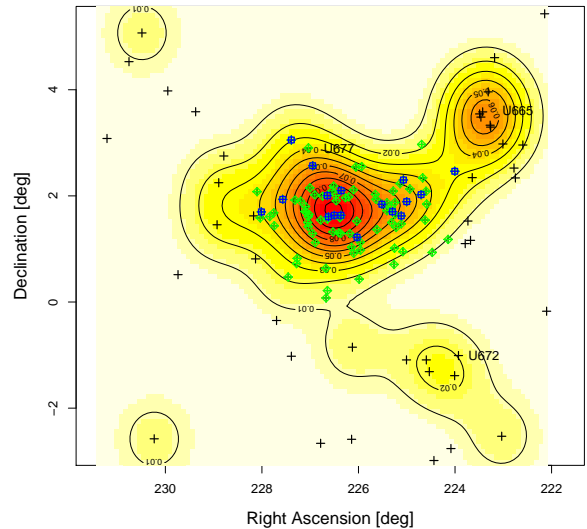}
  \caption{Spatial  distribution of galaxies within a box of 4 $\times$ 4 Mpc$^2$ centred on NGC 5846 (shaded yellow/red area).  Blue squares show the members of the  group listed in the catalogue of 
  \citet{Ramella02} and green diamonds indicate the added members (Section 2). 
   The map is normalized to the total density.
  The 2D binned kernel-smoothed number density contours for the galaxies with m$_B \le$ 15.5 mag (circle + cross) are shown.  The value of 
  m$_B$ = 15.5 mag is the magnitude limit of the \citet{Ramella02} catalogue. }
  \label{f3}
 \end{figure}

 \subsection{Group's environment density analysis}

In order to characterize the environment of the group, we have considered the galaxy distribution within a box of 4 $\times$ 4 Mpc$^2$ (about two times the size of a typical group) centred on  NGC 5846. 
From the {\tt HYPERLEDA} database we have selected all the galaxies within such box  with a heliocentric radial velocity within  $\pm$ 3$\sigma$ of  the group mean velocity, as given in the catalog of \citet{Ramella02}. We have found  136  galaxies (Table \ref{tab1}  and  Table \ref{a2} in Appendix B). In the bottom panel of Figure \ref{f1}, we highlight the velocity distribution of group members (green filled bins) given in Table 1 superposed on  that  of galaxies in the 4 $\times$ 4 Mpc$^2$ box.  
Among these galaxies, we have selected only those more luminous than 15.5 mag in the B band 
(i.e. the magnitude limit of the galaxies in \citet{Ramella02}, and 
on this sample we performed a density analysis. 
The 2D binned kernel-smoothed number density contour map is shown 
in  Figure \ref{f3}. Density in the 4 $\times$ 4 Mpc$^2$  box (shaded yellow)  is
colour coded. The highest densities correspond to the red regions;  in the
normalized map, density levels above 0.05 are in yellow. 
Blue squares show the members of the group  from
 the catalogue of  \citet{Ramella02} and green diamonds indicate the new members that we have added (as explained in Section 2.1).

A  high density region  approximately centred on NGC 5846, and  elongated towards NGC 5813, the second B-band brighter member, appears. There are also two poor groups, USGC U665 and USGC U672, likely falling toward NGC 5846.

\subsection{Group dynamical properties}     
\label{secdyn}
The virial theorem provides the standard method to estimate the mass of a self-gravitating 
system from dynamical parameters, positions and velocities of the group members. 
It applies if  the system analyzed is in dynamical equilibrium and  its luminosity is a  tracer of the mass.  

We  derived the kinematic and dynamical properties of NGC 5846, following
the approach described in \citet [their Table 6]{Firth06} and already used in  \citet{Marino10} for
LGG 225 and  \citet{Marino13a} for USGC U268 and USGC U376.   
 
\begin{figure}
 \vspace{-0.3cm}
  \includegraphics[width=8cm, angle=-90]{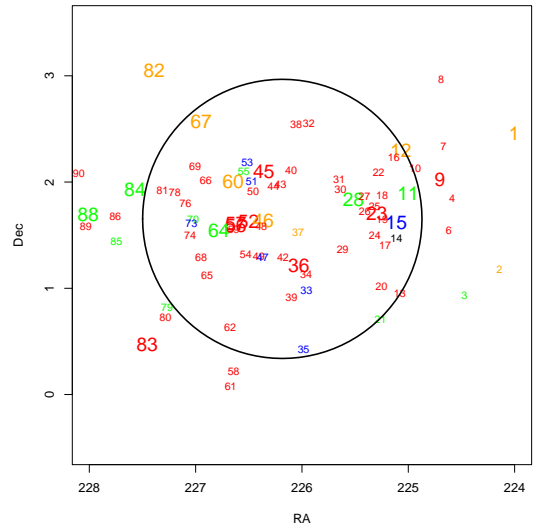}
 \caption{Projected distribution of the galaxy members of the NGC 5846 group. Numbers refer to their identification in Table \ref{tab1}. 
 Different morphological types correspond to different colours  as in Fig.  \ref{f1}. Galaxies 
 with B magnitude $\leq$ 15.5 are labeled with bigger numbers than those with B magnitude $>$ 15.5 or unknown (in black).  The circle (solid line)  centred on the  centre of mass of the group encloses galaxies within the virial radius of 550 kpc. }
 
	 \label{col}
\end{figure}

Results are summarized in Table~\ref{Dyn}. The errors have been computed via jackknife simulations \citep[e.g.][]{Efron82}.
 In order to obtain an estimate of the compactness of the group, 
we  computed the harmonic mean radius  using
the projected separations r$_{ij}$ between the i-th and j-th group member.
Figure \ref{col} shows the relative positions of the group members with superposed 
 a circle centred on the centre of mass of the group  and  radius corresponding to the virial radius.

The   projected  mass is about two times the virial mass.  The contribution of the 30 galaxies surrounding NGC 5813 (Table \ref{tab1})  is about   20 per cent of the virial mass and  34\%  of the projected one (see the second line in Table~\ref{Dyn}).\\
For comparison,  the  projected mass of the groups USGC 268 and USGC U376  \citep{Marino13a},  is  a factor  3-4 times  the virial one.
 Using N-body simulations, \citet{Perea90a} showed that the virial mass estimate is  better than the
projected mass estimate since it is less sensitive to anisotropies  or sub-clustering. However, it may be affected  
by the presence of interlopers, i.e. unbound galaxies, and by a mass spectrum.
Both factors would cause an overestimation of the group mass. Therefore the estimated masses are
upper limits.  Another caveat concerning the virial mass is that the groups may not be virialized \citep[e.g.][]{Ferguson90}.\\
 The larger difference in the virial and projected mass estimates of USGC U268 and  USGC U376 may suggest a larger probability of interlopers,
although  these groups are, on average, closer than NGC5846 group \citep[15 and 19 Mpc respectively][]{Marino13a}, or  that these are not yet virialized.\\
 The crossing time  is usually compared to the Hubble time to 
determine whether the groups are gravitationally bound systems. The derived crossing time
of NGC 5846 group (see Table 2) suggests that it  could be virialized  \citep[e.g.][]{Firth06}. %(ref!!).\\

\begin{table*}
\centering
\caption{Kinematical and dynamical properties of NGC 5846}
\scriptsize
\begin{tabular}{lclllllllllll}
\hline\hline\noalign{\smallskip}
Group & Center  &  V$_{group}$  & Velocity  & D  & Harmonic &Virial  & Projected
 & Crossing\\
 name & of mass &               & dispersion & & radius & mass &  mass &time$\times$H$_0$ \\
   & RA [deg] ~ Dec &[km/s] &[km/s]& [Mpc] & [Mpc] & [10$^{13}$ M$_{\odot}$] &  
[10$^{13}$ M$_{\odot}$] &   \\
\hline
NGC 5846    &  226.18501    1.65321 &1798$^{+8}_{-10}$   & 327$^{+12}_{-2}$ & 24.0$^{+0.1}_{-0.1}$ & 0.35$^{+0.01}_{-0.01}$ &  4.15$^{+0.24}_{-0.16}$  &  8.24$^{+0.56}_{-0.13}$   & 0.12$^{+0.00}_{-0.01}$ &  \\
NGC5846$^1$   &  226.76139    1.65059 &1800$^{+13}_{-15}$   & 332$^{+18}_{-3}$ & 24.0$^{+0.2}_{-0.2}$ & 0.27$^{+0.01}_{-0.01}$ &  3.24$^{+0.29}_{-0.20}$  &  5.42$^{+0.43}_{-0.13}$   & 0.09$^{+0.01}_{-0.00}$ &  \\
\hline
\end{tabular}

$^1$ The same quantities computed excluding the first thirty galaxies in Table \ref{tab1}, i.e., NGC 5813 and its surroundings.
\label{Dyn}
 \end{table*}

\begin{table*}
	\caption{The NGC 5846 group members and the journal of the UV observations with {\it GALEX}. }	 
	 \tiny
	\begin{tabular}{lllcccll }
	\hline\hline
Id     & Galaxy & MTT05 &  P & NUV & FUV & Survey  \\  
Nr.    &	   	     &               &       & Exp. Time & Exp. Time & \\  
	   &	      &              &      & [sec]) & [sec] & \\  
 \hline
1 & PGC053384 & \dots & \dots & 1602.1 & 1602.1 & MIS \\
2 &PGC1186917 & \dots & \dots & 1640.7 & 1640.7 & MIS \\
3 & PGC1179522 & 009 & 0  & 1640.7 & 1640.7 & MIS \\
4 & PGC184851 & 013 & 2 &1648.0 & 162.0 &GI6 ~ AIS \\
5& SDSS J145824.22+020511.0 &  014 &1  & 1648.0 & 162.0 &  GI6 ~ AIS& \\
6& SDSSJ145828.64+013234.6 & 017  & 2 & 1640.7 & 1640.7 & MIS \\
7 & PGC1223766 & 018 &  3 &1648.0 & 162.0 &GI6 ~  AIS \\
8& PGC1242097 & 020 & 5 & 2107.5 & 2107.5 & MIS \\
9& PGC053521 & 021 & 0  &1648.0 & 162.0 &GI6 ~  AIS \\
10 & SDSSJ145944.77+020752.1 & 030 & 2 & 1648.0 & 162.0 &GI6 ~  AIS \\
11 & NGC5806 & 037 & 0  & 2521.2 & 2521.2 & GI3 \\
12 & PGC053587 & 042 & 0  & 1648.0 & 162.0 &GI6 ~ AIS \\
13 & SDSSJ150019.17+005700.3 & 045 & 1 & 335.0 & 159.0 & AIS \\
14 &  NGC 5846:[MTT2005] 046  &  046 & 0  & 2521.2 & 2521.2 & GI3 \\
15 & NGC5811 & 047& 3 & 2521.2 & 2521.2 & GI3 \\
16 & SDSSJ150033.02+021349.1 & 048 & 1 & 1648.0 & 162.0 &GI6 ~  AIS \\
17 & PGC1193898 & 055 & 2 & 2521.2 & 2521.2 & GI3 \\
18 & SDSSJ150059.35+015236.1 & 058 & 2 & 2521.2 & 2521.2 & GI3 \\
19 & SDSSJ150059.35+013857.0 & 059 & 5 & 2521.2 & 2521.2 & GI3 \\
20 & SDSSJ150100.85+010049.8 & 060 & 2 & 335.0 & 159.0 & AIS \\
21 & PGC053636$^a$ & 061 & 0  & & & \\
22 & SDSSJ150106.96+020525.1 & 063 & 2 & 2521.2 & 2521.2 & GI3 \\
23 & NGC5813 & 064 & 0  & 2521.2 & 2521.2 & GI3 \\
24 & PGC1196740 & 068 & 2 & 2521.2 & 2521.2 & GI3 \\
25 & PGC1205406 & 069 & 3 & 2521.2 & 2521.2 & GI3 \\
26 & SDSSJ150138.39+014319.8 & 073 & 2  & 2521.2 & 2521.2 & GI3 \\
27 & PGC1208589 & 075 & 3 & 2521.2 & 2521.2 & GI3 \\
28 & UGC09661 & 083 & 0  & 2521.2 & 2521.2 & GI3 \\
29 & PGC1192611 & 088 & 3 & 5299.3 & 2359.1 & GI1 \\
30 & SDSSJ150233.03+015608.3 & 090 & 1 & 2521.2 & 2521.2 & GI3 \\
31 & SDSSJ150236.05+020139.6 & 091 & 3 & 2521.2 & 2521.2 & GI3 \\
32 & PGC1230503 & 113 & 3 & 2399.4 & 2399.4 & MIS \\
33 & SDSSJ150349.93+005831.7 & 114 & 0  & 5299.3 & 2359.1 & GI1 \\
34 & PGC1185375 & 115 & 0  & 5299.3 & 2359.1 & GI1 \\
35 & PGC087108 & NGC5846$_{41/42}$& Eigenthaler & 1692.0 & 1692.0 & MIS \\
36 & NGC5831 & 122 & 0  & 5299.3 & 2359.1 & GI1 \\
37 & PGC1197513 & 124 & 0  & 5299.3 & 2359.1 & GI1 \\
38 & PGC1230189 & 125 & 3 & 2399.4 & 2399.4 & MIS \\
39 & PGC1179083 & 131 & 2 & 5299.3 & 2359.1 & GI1 \\
40 & PGC1216386 & 132 & 3 & 2399.4 & 2399.4 & MIS \\
41 & SDSSJ150434.31+010156.9& 139 & 1 & 5299.3 & 2359.1 & GI1 \\
42 & PGC1190315 & 142 & 0  & 5299.3 & 2359.1 & GI1 \\
43 & SDSSJ150448.49+015851.3 & 144 & 2 & 2399.4 & 2399.4 & MIS \\
44 & PGC1211621 & 148 & 0  & 2399.4 & 2399.4 & MIS \\
45 & NGC5838 & 159 & 0  & 2399.4 & 2399.4 & MIS \\
46 & NGC5839 & 160 & 0  & 2484.2 & 2484.2 & MIS \\
47 & PGC1190358 & 162 & 0  & 5299.3 & 2359.1 & GI1 \\
48 & PGC1199471 & 165 & 3 & 2484.2 & 2484.2 & MIS \\
49 & PGC1190714 & 167 & 0  & 2484.2 & 2484.2 & MIS \\
50 & PGC1209872 & 177 & 0  & 1466.0 & 1466.0 & GI3 \\
51 & PGC1213020 & 180 & 3 & 1466.0 & 1466.0 & GI3 \\
52 & NGC5845 & 184 & 0  & 2484.2 & 2484.2 & MIS \\
53 & PGC1218738 & 187 & 2 & 1466.0 & 1466.0 & GI3 \\
54 & PGC1191322 & 191 & 0  & 2484.2 & 2484.2 & MIS \\
55 & PGC1215798 & 192 & 0  & 1466.0 & 1466.0 & GI3 \\
56  & NGC5846A & 201 & 0  & 2484.2 & 2484.2 & MIS \\
57 & NGC5846 & 202 & 0  & 2484.2 & 2484.2 & MIS \\
58 & SDSSJ150634.25+001255.6 &  NGC5846$_{44}$ & Eigenthaler & 1695.1 & 1694.1 & MIS \\
59 & PGC3119319 & 205 & 5 & 2484.2 & 2484.2 & MIS \\
60 & NGC5841 & 206 & 0 & 1466.0 & 1466.0 & GI3 \\
61 & PGC1156476 & NGC5846$_{50}$ & Eigenthaler & 164.0 & 164.0 & AIS \\
62 & PGC1171244 & 212 & 3 & 1695.1 & 1694.1 & MIS \\
63 & SDSS J150658.37+015939.5   & 226 & 2 & 1466.0 & 1466.0 & GI3 \\
64 & NGC5850 & 233 & 0  & 2484.2 & 2484.2 & MIS \\
65 & PGC1185172 & 241 & 3 & 163.0 & 163.0 & AIS \\
66 & PGC054004 & 244 & 0  & 2376.0 & 2375.0 & MIS \\
67 & NGC5854 & 246 & 0   & 2376.0 & 2375.0 & MIS \\
68 & PGC054016$^a$ & 247 & 0  & & &\\
69 & PGC1217593 & 252 & 5 & 2376.0 & 2375.0 & MIS \\
70 & PGC054037$^b$ & 256 & 4 & & &\\
 71 & SDSSJ150808.43+025416.5 & 258 & 3 & 2376.0 & 2375.0 & MIS \\
72 &   NGC5846:[MTT2005]259$^b$ & 259 & 3 & & &\\
73 & PGC054045$^b$ & 260 & 0  & & &\\
74 & SDSSJ150812.35+012959.7$^b$ & 261 & 3 & & &\\
75 & NGC 5846:[MTT2005]$^b$  & 264 & 2 & & & &\\
76 & PGC1206166$^b$ & 266 & 2 & & &\\
77 & SDSSJ150825.57+014224.8$^b$ & 268 & 2& \dots & & &\\
78 & PGC1209573$^a$ & 276 & 3 & & &\\
79 & PGC1176385 & 283 & 0 & 1696.0 & 1696.0 & MIS \\
80 & SDSSJ150907.83+004329.7 & 287 & 2 & 1696.0 & 1696.0 & MIS \\
81 & PGC1210284$^a$ & 290 & 3 & & &\\
82 & NGC5864$^b$ & 299 & 0  & & &\\
83 & NGC5869 & NGC5869 & Eigenthaler & 1450.6 &1450.6& MIS\\
84 & UGC09746 & 305 & 0  & 1655.0 &$^a$ &GI6\\
85 & UGC09751 & 311 & 0  & 1696.0 & 1696 & MIS \\
86 & PGC1202458 & 313 & 2 & 1655.0 &$^a$ &GI6\\
87 & SDSSJ151121.37+013639.5& 317 & 1 & 1655.0 &$^b$ & GI6\\
88 & UCG09760 & 321 & 0  & 1655.0 &111.0 & GI6 ~ AIS \\
89 & PGC1199418 & 323 & 5 & 1655.0 &$^a$ & GI6\\
90 & PGC1215336 & NGC5846$_{52}$ & Eigenthaler & 2905.9 & 1732.3 & MIS \\

 \hline			 							        			   
\end{tabular}

Notes. -- Col.~1 and col.~2: galaxy identification; col.~3 and col.~4 galaxy identification and
membership probability in \citet{Mahdavi05} Table~1, respectively. P values are: 
0, no-SDSS spectroscopic redshift; 1, probable member; 2, possible member; 3, 
conceivable member; 4 and 5 likely not a member.
$^a$ The UV images have a distance from the center of the field of view $>$50\arcmin.
$^b$	 There are no FUV GALEX images for these galaxies.
\label{exp}	 											   
\end{table*}

\section[]{Observations and data reduction}

 \subsection{UV and optical data}	   	    
The UV imaging was obtained from {\it  GALEX} \citep{Martin05, Morrissey07} GI program 017 (PI A. Marino) and  archival data  in two ultraviolet bands, FUV (1344 -- 1786\AA) 
 and NUV   (1771 -- 2831\AA). The instrument has a very wide field of view (1\degr.25 
 diameter) and a spatial resolution of $\approx$ 4\farcs2  and 5\farcs3 FWHM in FUV and NUV
 respectively, sampled with 1\farcs 5$\times$1\farcs 5 pixels \citep{Morrissey07}.

We used only UV images having a distance from the center of the field of view $\leq$0.5 deg,
as generally the photometric quality is better in the central part of the field  \citep{Bianchi11, Bianchi14}.
In case of multiple observations of the same galaxy, we chose the one with  longer exposure time.

This yields  {\it  GALEX} data for 78 of the 90 member galaxies, of which all  but four (UGC09746, PGC1202458, SDSSJ151121.37+013639.5, PGC1199418) 
were observed in both FUV and NUV (see Table~2).
			   	    
The exposure times (see Table \ref{exp}) for most of our  sample are $\sim$2000 sec (limit AB magnitude in FUV/NUV of $\sim$ 22.6/22.7 \citep{Bianchi09}).   We used FUV and NUV   intensity images   to compute integrated photometry of the galaxies 
and light profiles, as described in Sect. 4. \\ 
 In addition, we used optical SDSS archival data  in
  the u [2980-4130 \AA], g [3630-5830 \AA], r [5380-7230 \AA], i
[6430-8630 \AA],  and z [7730-11230 \AA] filters \citep{Ade08} to obtain  optical photometry.\\

\subsection{Aperture photometry} 
 UV and optical magnitudes of the brighter members (B$_T \leq$15.5 mag) have been obtained as follows.
  
The  UV and optical surface photometry was carried out  using  the ELLIPSE
fitting routine in the STSDAS package of IRAF \citep{Jedrzejewski87}.
The SDSS images  (corrected frames with the soft bias 
of 1000 counts subtracted) in the five bands were registered 
to the corresponding  {\it GALEX} NUV intensity images before  evaluating brightness profiles,
using the IRAF tool {\texttt register}.
We masked the foreground stars and the background galaxies in the regions where we 
measured the surface brightness profiles. 
To secure a reliable background measure, we forced the measure of 5 
 isophotes  well beyond the galaxy emission.

From the surface brightness profiles, we derived total apparent magnitudes as follows.
For each profile, we computed the integrated 
apparent magnitude within elliptical isophotes up to the radius where 
the mean isophotal intensity is 2$\sigma$ above the background.
The background was computed around  each source, as the mean of sky value  of the outer five 
isophotes.
 Errors of the UV and optical  magnitudes where estimated by propagating the 
statistical errors on the  mean isophotal
intensity provided by ELLIPSE. In addition to the statistical error, we added 
 systematic 
uncertainties in the zero-point calibration of 0.05 and 0.03 mag in FUV and NUV 
respectively \citep{Morrissey07}.
Surface photometry was corrected for galactic extinction assuming Milky Way dust {\ properties} with Rv=3.1 \citep{Cardelli89}, A$_{FUV}$/E(B-V) = 8.376, A$_{NUV}$/E(B-V) = 8.741,  and A$_{r}$/E(B-V) = 2.751.

 Table \ref{tab3} lists the measured  AB
 magnitudes both in UV and optical\footnote{We  converted SDSS counts to magnitudes following the recipe provided in {\tt http://www.sdss.org/df7/algorithms/fluxcal.html}.} bands., uncorrected for foreround Galaxy extinction.
 UV and optical  magnitudes of fainter members were
 extracted from the {\it GALEX} and SDSS pipelines. We used the FUV and NUV calibrated magnitudes  and the   Model magnitudes\footnote{http://www.sdss.org/dr5/algorithms/photometry.html}
 from the {\it GALEX} and SDSS pipelines, respectively.

\begin{table*}
	\caption{UV and optical photometry of the galaxy group. }	 
	 \tiny
	 \begin{tabular}{llllllllll}
	\hline\hline\noalign{\smallskip}
Id. &	Galaxy  & FUV & NUV & u & g & r & i &z  \\
Nr. &	        & [AB mag] &[AB mag] & [AB mag] &[AB mag] &[AB mag] &[AB mag]& [AB mag] \\
	
 \hline
%PGC053384               &  20.543 $\pm$ 0.069&  18.632 $\pm$ 0.021   &  16.135 $\pm$ 0.009 & 14.666 $\pm$ 0.002 & 14.113 $\pm$ 0.002 & 13.831 $\pm$ 0.002 & 13.653 $\pm$ 0.004 \\
1 &PGC053384               &  20.520 $\pm$ 0.090&  18.740 $\pm$ 0.040   &  16.250 $\pm$ 0.010 & 14.670 $\pm$ 0.010 & 14.080 $\pm$ 0.010 & 13.7501 $\pm$ 0.010 & 13.700 $\pm$ 0.010 \\
2 &PGC1186917              &  23.678 $\pm$ 0.405&  21.039 $\pm$ 0.100   &  18.449 $\pm$ 0.058 & 16.911 $\pm$ 0.006 & 16.236 $\pm$ 0.005 & 15.901 $\pm$ 0.005 & 15.684 $\pm$ 0.014 \\
3 & PGC1179522              &  21.368 $\pm$ 0.113&  20.067 $\pm$ 0.062   &  17.991 $\pm$ 0.040 & 16.470 $\pm$ 0.006 & 15.761 $\pm$ 0.005 & 15.453 $\pm$ 0.006 & 15.235 $\pm$ 0.014 \\
4 & PGC184851               &  21.851 $\pm$ 0.370&  19.926 $\pm$ 0.115   &  17.431 $\pm$ 0.025 & 15.806 $\pm$ 0.003 & 15.040 $\pm$ 0.003 & 14.670 $\pm$ 0.003 & 14.362 $\pm$ 0.009 \\
5 & SDSSJ145824.22+020511.0    &                  &  21.629 $\pm$ 0.220   &  20.339 $\pm$ 0.482 & 18.913 $\pm$ 0.046 & 18.425 $\pm$ 0.049 & 18.388 $\pm$ 0.070 & 18.136 $\pm$ 0.265 \\
6 & SDSSJ145828.64+013234.6 &20.617 $\pm$ 0.090  &  19.668 $\pm$ 0.048   &  19.227 $\pm$ 0.165 & 17.397 $\pm$ 0.012 & 16.751 $\pm$ 0.010 & 16.446 $\pm$ 0.011 & 16.238 $\pm$ 0.038 \\
7 & PGC1223766              &                  &  21.924 $\pm$ 0.292   &  19.372 $\pm$ 0.131 & 17.886 $\pm$ 0.015 & 17.201 $\pm$ 0.012 & 16.885 $\pm$ 0.015 & 16.848 $\pm$ 0.064 \\
8 & PGC1242097              &  18.504 $\pm$ 0.013&  17.970 $\pm$ 0.009   &  16.796 $\pm$ 0.010 & 15.769 $\pm$ 0.003 & 15.372 $\pm$ 0.003 & 15.159 $\pm$ 0.003 & 15.018 $\pm$ 0.006 \\
%PGC053521            &  21.592 $\pm$ 0.404&  18.970 $\pm$ 0.068   &  16.091 $\pm$ 0.009 & 14.392 $\pm$ 0.002 & 13.640 $\pm$ 0.002 & 13.262 $\pm$ 0.002 & 13.003 $\pm$ 0.003 \\
9 & PGC053521               &                                          &18.900 $\pm$ 0.040   & 15.920 $\pm$ 0.010 &14.330 $\pm$ 0.010 &13.620 $\pm$ 0.010 &13.220 $\pm$ 0.010& 13.080 $\pm$ 0.010 \\
10 & SDSSJ145944.77+020752.1 &                  &  21.650 $\pm$ 0.273   &  19.271 $\pm$ 0.139 & 18.114 $\pm$ 0.018 & 17.519 $\pm$ 0.017 & 17.193 $\pm$ 0.022 & 17.113 $\pm$ 0.096 \\
%NGC5806                 &   16.182 $\pm$ 0.008  &  15.186 $\pm$ 0.003   &  14.441 $\pm$ 0.005 & 12.718 $\pm$ 0.002 & 11.873 $\pm$ 0.002 & 11.422 $\pm$ 0.002 & 11.066 $\pm$ 0.002 \\
11 & NGC5806                   &15.830 $\pm$ 0.050 &15.180 $\pm$ 0.030 &13.600 $\pm$ 0.010 &11.990 $\pm$ 0.010 &11.320 $\pm$ 0.010 &10.860 $\pm$ 0.010 &10.600 $\pm$ 0.010\\
%PGC053587               &  20.041 $\pm$ 0.175&  19.108 $\pm$ 0.063   &  16.723 $\pm$ 0.021 & 15.249 $\pm$ 0.004 & 14.548 $\pm$ 0.003 & 14.198 $\pm$ 0.004 & 13.914 $\pm$ 0.008 \\
12 & PGC053587                 &     &   19.240 $\pm$ 0.040& 16.870 $\pm$ 0.010 &15.140 $\pm$ 0.010 &14.440 $\pm$ 0.010& 14.020 $\pm$ 0.010 &13.950 $\pm$ 0.010 \\
13 & SDSSJ150019.17+005700.3 &                  &  21.101 $\pm$ 0.267   &  18.957 $\pm$ 0.113 & 17.303 $\pm$ 0.010 & 16.597 $\pm$ 0.009 & 16.306 $\pm$ 0.010 & 16.158 $\pm$ 0.031 \\
14 & NGC5846:[MTT2005]046    &  23.516 $\pm$ 0.354&  23.311 $\pm$ 0.376   &   & & & & \\
%NGC5811                 &  17.320 $\pm$ 0.011&  16.775 $\pm$ 0.007   &  15.933 $\pm$ 0.016 & 14.626 $\pm$ 0.003 & 14.139 $\pm$ 0.003 & 13.841 $\pm$ 0.005 & 13.679 $\pm$ 0.012 \\
15 & NGC5811                 &    17.310 $\pm$ 0.050 & 16.780 $\pm$ 0.030 & 15.620 $\pm$ 0.010&  14.450 $\pm$ 0.010 & 13.810 $\pm$ 0.010 & 13.350 $\pm$ 0.010 & 13.400 $\pm$ 0.010 \\
16 & SDSSJ150033.02+021349.1 &                  &  21.365 $\pm$ 0.269   &  18.510 $\pm$ 0.098 & 16.979 $\pm$ 0.009 & 16.306 $\pm$ 0.008 & 16.038 $\pm$ 0.011 & 16.588 $\pm$ 0.089 \\
17 & PGC1193898              &                  &  20.438 $\pm$ 0.093   &  18.004 $\pm$ 0.061 & 16.410 $\pm$ 0.006 & 15.691 $\pm$ 0.005 & 15.339 $\pm$ 0.006 & 15.136 $\pm$ 0.022 \\
18 & SDSSJ150059.35+015236.1 &                  &  21.946 $\pm$ 0.186   &  19.614 $\pm$ 0.152 & 18.506 $\pm$ 0.022 & 17.840 $\pm$ 0.020 & 17.568 $\pm$ 0.026 & 17.496 $\pm$ 0.104 \\
19 & SDSSJ150059.35+013857.0 &                  &  23.147 $\pm$ 0.403   &  19.598 $\pm$ 0.189 & 17.968 $\pm$ 0.017 & 17.216 $\pm$ 0.015 & 16.814 $\pm$ 0.014 & 16.528 $\pm$ 0.043 \\
20 & SDSSJ150100.85+010049.8 &  21.456 $\pm$0.333&  20.303 $\pm$ 0.125   &  18.980 $\pm$ 0.110 & 17.800 $\pm$ 0.020 & 17.394 $\pm$ 0.018 & 17.487 $\pm$ 0.072 & 17.126 $\pm$ 0.093 \\
21 & PGC053636               & 	           &  		         &  16.888 $\pm$ 0.012 & 15.640 $\pm$ 0.003 & 15.004 $\pm$ 0.003 & 14.691 $\pm$ 0.003 & 14.460 $\pm$ 0.006 \\
22 & SDSSJ150106.96+020525.1 &                  &  22.252 $\pm$ 0.198   &  19.840 $\pm$ 0.245 & 18.069 $\pm$ 0.020 & 17.350 $\pm$ 0.018 & 17.124 $\pm$ 0.020 & 16.935 $\pm$ 0.074 \\
%NGC5813                 &  17.910 $\pm$ 0.020&  15.886 $\pm$ 0.008   &  14.102 $\pm$ 0.004 & 12.092 $\pm$ 0.002 & 11.218 $\pm$ 0.002 & 10.771 $\pm$ 0.002 & 10.423 $\pm$ 0.002 \\
23 & NGC5813                      &  17.910 $\pm$0.050  &16.330 $\pm$ 0.030  &13.320 $\pm$ 0.010  &11.450 $\pm$ 0.010  &10.500 $\pm$ 0.010  &10.140 $\pm$ 0.010  &   9.880 $\pm$ 0.010 \\
24 & PGC1196740              &  20.898 $\pm$ 0.068&  19.991 $\pm$ 0.049   &  18.531 $\pm$ 0.076 & 17.315 $\pm$ 0.010 & 16.851 $\pm$ 0.011 & 16.595 $\pm$ 0.013 & 16.473 $\pm$ 0.042 \\
25 & PGC1205406              &  22.333 $\pm$ 0.153&  21.003 $\pm$ 0.110   &  18.954 $\pm$ 0.101 & 17.612 $\pm$ 0.013 & 17.006 $\pm$ 0.022 & 16.745 $\pm$ 0.026 & 16.678 $\pm$ 0.062 \\
26 & SDSSJ150138.39+014319.8 &                  &  21.898 $\pm$ 0.262   &  19.030 $\pm$ 0.110 & 17.583 $\pm$ 0.012 & 16.877 $\pm$ 0.011 & 16.482 $\pm$ 0.012 & 16.589 $\pm$ 0.056 \\
27 & PGC1208589              &  22.332 $\pm$ 0.136&  21.662 $\pm$ 0.111   &  18.549 $\pm$ 0.085 & 17.259 $\pm$ 0.011 & 16.691 $\pm$ 0.011 & 16.429 $\pm$ 0.015 & 16.168 $\pm$ 0.045 \\
%UGC09661                &  16.651 $\pm$ 0.008&  16.237 $\pm$ 0.005   &  15.502 $\pm$ 0.012 & 14.357 $\pm$ 0.003 & 13.858 $\pm$ 0.003 & 13.634 $\pm$ 0.003 & 13.774 $\pm$ 0.012 \\
28 & UGC09661                  &16.690 $\pm$ 0.050 &16.420 $\pm$ 0.030 &15.190 $\pm$ 0.010 &14.230 $\pm$ 0.010 &14.040 $\pm$ 0.010& 13.280 $\pm$ 0.010 &13.670 $\pm$ 0.010 \\
29 & PGC1192611              &                  &  21.618 $\pm$ 0.094   &  19.423 $\pm$ 0.149 & 17.909 $\pm$ 0.018 & 17.349 $\pm$ 0.021 & 17.103 $\pm$ 0.032 & 17.063 $\pm$ 0.116 \\
30 & SDSSJ150233.03+015608.3 &  22.248 $\pm$ 0.141&  21.389 $\pm$ 0.106   &  19.425 $\pm$ 0.209 & 17.918 $\pm$ 0.027 & 17.294 $\pm$ 0.021 & 17.074 $\pm$ 0.039 & 17.015 $\pm$ 0.096 \\
31 & SDSSJ150236.05+020139.6 &                 &   21.659 $\pm$ 0.146   &  19.897 $\pm$ 0.205 & 17.854 $\pm$ 0.014 & 17.196 $\pm$ 0.016 & 16.876 $\pm$ 0.016 & 16.824 $\pm$ 0.052 \\
32 & PGC1230503              &  23.826 $\pm$ 0.619&      &  18.591 $\pm$ 0.076 & 17.157 $\pm$ 0.010 & 16.491 $\pm$ 0.007 & 16.202 $\pm$ 0.010 & 15.991 $\pm$ 0.027 \\
33 & SDSSJ150349.93+005831.7 &  18.110 $\pm$ 0.018&  17.644 $\pm$ 0.007   &  17.938 $\pm$ 0.050 & 16.744 $\pm$ 0.009 & 16.430 $\pm$ 0.011 & 16.058 $\pm$ 0.012 & 15.805 $\pm$ 0.032 \\
34 & PGC1185375              &  23.627 $\pm$ 0.484&  19.915 $\pm$ 0.050   &  17.670 $\pm$ 0.040 & 16.002 $\pm$ 0.004 & 15.247 $\pm$ 0.003 & 14.865 $\pm$ 0.004 & 14.598 $\pm$ 0.008 \\
35 & PGC087108               &  17.170 $\pm$ 0.013&  17.073 $\pm$ 0.008   &  17.571 $\pm$ 0.041 & 16.622 $\pm$ 0.008 & 16.404 $\pm$ 0.010 & 16.449 $\pm$ 0.017 & 16.427 $\pm$ 0.052 \\
%NGC5831                 &  19.248 $\pm$ 0.038&  16.875 $\pm$ 0.008   &  14.438 $\pm$ 0.005 & 12.487 $\pm$ 0.002 & 11.653 $\pm$ 0.001 & 11.217 $\pm$ 0.001 & 10.922 $\pm$ 0.002 \\
36 & NGC5831                     & 18.900 $\pm$ 0.060 &17.150 $\pm$ 0.030 &14.070 $\pm$ 0.010 &12.110 $\pm$ 0.010 &11.300 $\pm$ 0.010 &10.860 $\pm$ 0.010 &10.630 $\pm$ 0.010 \\
37 & PGC1197513              &  19.675 $\pm$ 0.035&  18.845 $\pm$ 0.014   &  17.190 $\pm$ 0.030 & 15.932 $\pm$ 0.004 & 15.413 $\pm$ 0.005 & 15.169 $\pm$ 0.006 & 15.052 $\pm$ 0.018 \\
38 & PGC1230189              &                  &  19.916 $\pm$ 0.058   &  17.214 $\pm$ 0.032 & 15.565 $\pm$ 0.005 & 14.853 $\pm$ 0.004 & 14.473 $\pm$ 0.005 & 14.308 $\pm$ 0.014 \\
39 & PGC1179083              &  23.494 $\pm$ 0.506& &  19.196 $\pm$ 0.114 & 17.358 $\pm$ 0.011 & 16.680 $\pm$ 0.013 & 16.516 $\pm$ 0.017 & 16.409 $\pm$ 0.033 \\
40 & PGC1216386              &  23.282 $\pm$ 0.282& 20.920 $\pm$ 0.099    &  18.466 $\pm$ 0.057 & 17.006 $\pm$ 0.007 & 16.362 $\pm$ 0.007 & 16.044 $\pm$ 0.007 & 15.828 $\pm$ 0.025 \\
41 & NGC5846:[MTT2005]139    &                  & 22.596 $\pm$ 0.241   &  19.619 $\pm$ 0.149 & 18.534 $\pm$ 0.032 & 17.808 $\pm$ 0.022 & 17.564 $\pm$ 0.028 & 17.346 $\pm$ 0.085 \\
42 & PGC1190315              &                  & 20.907 $\pm$ 0.068   &  18.222 $\pm$ 0.095 & 16.485 $\pm$ 0.006 & 15.775 $\pm$ 0.005 & 15.386 $\pm$ 0.008 & 15.207 $\pm$ 0.020 \\
43 & SDSSJ150448.49+015851.3 &  21.178 $\pm$ 0.085& 20.263 $\pm$ 0.071   & 18.979 $\pm$ 0.126 & 17.818 $\pm$ 0.018 & 17.275 $\pm$ 0.021 & 17.098 $\pm$ 0.027 & 17.033 $\pm$ 0.089 \\
44 & PGC1211621              &  19.296 $\pm$ 0.028& 18.954 $\pm$ 0.016   & 17.987 $\pm$ 0.031 & 17.110 $\pm$ 0.006 & 16.671 $\pm$ 0.007 & 16.545 $\pm$ 0.009 & 16.502 $\pm$ 0.032 \\
%NGC5838                 &  18.403 $\pm$ 0.021&  16.640 $\pm$	0.007 &	  13.914 $\pm$ 0.003 & 12.012 $\pm$ 0.002 & 11.156 $\pm$ 0.002 & 10.674 $\pm$ 0.002 & 10.256 $\pm$ 0.002 \\
45 & NGC5838                      & 18.290 $\pm$ 0.050 &16.700 $\pm$ 0.030 &13.320 $\pm$ 0.010 &11.520 $\pm$ 0.010 &10.710 $\pm$ 0.010& 10.240 $\pm$ 0.010 &  9.980 $\pm$ 0.010 \\
%NGC5839                 &  19.518 $\pm$ 0.034& 17.971 $\pm$ 0.016   &	    15.139 $\pm$ 0.006 & 13.192 $\pm$ 0.002 & 12.342 $\pm$ 0.002 & 11.916 $\pm$ 0.002 & 11.630 $\pm$ 0.002 \\
46 & NGC5839                    &   19.420 $\pm$ 0.060 &18.010 $\pm$ 0.030 &14.620 $\pm$ 0.010 &12.930 $\pm$ 0.010 &12.100 $\pm$ 0.010 &11.690 $\pm$ 0.010 &11.400 $\pm$ 0.010 \\
46  & PGC1190358              &  19.199 $\pm$ 0.029& 18.794 $\pm$ 0.013   & 18.308 $\pm$ 0.073 & 17.933 $\pm$ 0.018 & 17.612 $\pm$ 0.022 & 17.651 $\pm$ 0.038 & 18.249 $\pm$ 0.254 \\
48 & PGC1199471              &                  & 21.734 $\pm$ 0.159   &   19.198 $\pm$ 0.124 & 17.612 $\pm$ 0.012 & 16.938 $\pm$ 0.012 & 16.641 $\pm$ 0.012 & 16.456 $\pm$ 0.074 \\
49 & PGC1190714              &  23.202 $\pm$ 0.302& 20.520 $\pm$ 0.087   & 18.466 $\pm$ 0.063 & 17.004 $\pm$ 0.008 & 16.350 $\pm$ 0.006 & 16.018 $\pm$ 0.007 & 15.908 $\pm$ 0.024 \\
50 & PGC1209872              &  23.082 $\pm$ 0.399& 21.006 $\pm$ 0.120   & 18.402 $\pm$ 0.076 & 16.593 $\pm$ 0.006 & 15.860 $\pm$ 0.005 & 15.518 $\pm$ 0.005 & 15.308 $\pm$ 0.019 \\
51 & PGC1213020              &  19.849 $\pm$ 0.053& 19.574 $\pm$ 0.037 &   18.599 $\pm$ 0.074 & 17.817 $\pm$ 0.014 & 17.448 $\pm$ 0.025 & 17.337 $\pm$ 0.049 & 17.422 $\pm$ 0.103 \\
%NGC5845                 &  19.529 $\pm$ 0.030& 18.304 $\pm$ 0.014 &   14.986 $\pm$ 0.004 & 13.087 $\pm$ 0.002 & 12.254 $\pm$ 0.002 & 11.796 $\pm$ 0.002 & 11.463 $\pm$ 0.002 \\
52 & NGC5845                      & 19.470 $\pm$ 0.060 &18.290 $\pm$ 0.030 &14.860 $\pm$ 0.010 &12.990 $\pm$ 0.010 &12.150 $\pm$ 0.010 &11.710 $\pm$ 0.010 &11.430 $\pm$ 0.010 \\
53 & PGC1218738              &  18.268 $\pm$ 0.025&                    & 17.104 $\pm$ 0.034 & 15.912 $\pm$ 0.008 & 15.464 $\pm$ 0.014 & 15.441 $\pm$ 0.028 & 15.092 $\pm$ 0.039 \\
54 & PGC1191322              &                  & 21.423 $\pm$ 0.142   &   19.157 $\pm$ 0.084 & 17.625 $\pm$ 0.009 & 16.944 $\pm$ 0.008 & 16.621 $\pm$ 0.010 & 16.417 $\pm$ 0.029 \\
55 & PGC1215798              &  17.434 $\pm$ 0.016& 17.182 $\pm$ 0.011   &   17.101 $\pm$ 0.022 & 16.283 $\pm$ 0.005 & 16.055 $\pm$ 0.006 & 16.005 $\pm$ 0.008 & 15.854 $\pm$ 0.035 \\
%NGC5846A                &  20.149 $\pm$ 0.185& 18.692 $\pm$ 0.063   &   16.827 $\pm$ 0.009 & 14.770 $\pm$ 0.002 & 13.971 $\pm$ 0.002 & 13.571 $\pm$ 0.002 & 13.264 $\pm$ 0.003 \\
56 & NGC5846$+$A            &           17.050 $\pm$ 0.050 & 16.060 $\pm$ 0.030 & 12.460 $\pm$ 0.010 & 10.570 $\pm$ 0.010&   9.770 $\pm$ 0.010&   9.300 $\pm$ 0.010 &  9.000 $\pm$ 0.010 \\
%NGC5846                 &   17.059 $\pm$ 0.011 & 15.823 $\pm$ 0.006   &   13.918 $\pm$ 0.004 & 11.793 $\pm$ 0.002 & 10.910 $\pm$ 0.002 & 10.448 $\pm$ 0.002 & 10.118 $\pm$ 0.002 \\
57 & NGC5846                      &  17.120 $\pm$ 0.050  &16.100 $\pm$ 0.030  &12.800 $\pm$ 0.010  &10.840 $\pm$ 0.010  &9.980 $\pm$ 0.010  &9.470 $\pm$ 0.010  &9.240 $\pm$ 0.010 \\
58 & SDSSJ150634.25+001255.6 &                  & 22.701 $\pm$ 0.345   &   18.881 $\pm$ 0.077 & 17.597 $\pm$ 0.010 & 17.030 $\pm$ 0.010 & 16.768 $\pm$ 0.012 & 16.701 $\pm$ 0.054 \\
59 & PGC3119319              &  22.363 $\pm$ 0.147& 21.303 $\pm$ 0.089   &   17.892 $\pm$ 0.018 & 15.857 $\pm$ 0.003 & 15.011 $\pm$ 0.003 & 14.573 $\pm$ 0.003 & 14.220 $\pm$ 0.004 \\
%NGC5841                 &  20.896 $\pm$ 0.146& 18.892 $\pm$ 0.033   &   15.888 $\pm$ 0.009 & 14.114 $\pm$ 0.002 & 13.357 $\pm$ 0.002 & 12.966 $\pm$ 0.002 & 12.634 $\pm$ 0.003 \\
60 & NGC5841                     &  21.160 $\pm$ 0.140 &19.070 $\pm$ 0.040 &15.670 $\pm$ 0.010 &14.030 $\pm$ 0.010 &13.260 $\pm$ 0.010 &12.820 $\pm$ 0.010 &12.650 $\pm$ 0.010 \\
61 & PGC1156476              &                  & 21.833 $\pm$ 0.377   &   19.121 $\pm$ 0.063 & 17.715 $\pm$ 0.009 & 17.069 $\pm$ 0.008 & 16.800 $\pm$ 0.010 & 16.607 $\pm$ 0.034 \\
62 & PGC1171244              &  20.682 $\pm$ 0.081& 20.226 $\pm$ 0.054   & 18.850 $\pm$ 0.061 & 17.661 $\pm$ 0.010 & 17.224 $\pm$ 0.010 & 16.997 $\pm$ 0.015 & 16.930 $\pm$ 0.055 \\
63 & NGC5846:[MTT2005]226    &  21.704 $\pm$ 0.145& 21.268 $\pm$ 0.104   & 19.455 $\pm$ 0.128 & 18.329 $\pm$ 0.017 & 17.842 $\pm$ 0.026 & 17.589 $\pm$ 0.022 & 17.354 $\pm$ 0.077 \\
%NGC5850                 &   16.234$\pm$0.008  & 14.681 $\pm$ 0.003   &  14.858 $\pm$ 0.006 & 12.861 $\pm$ 0.002 & 11.952 $\pm$ 0.002 & 11.471 $\pm$ 0.002 & 11.128 $\pm$ 0.002 \\
64 & NGC5850                     &   15.110 $\pm$ 0.050 & 14.730 $\pm$ 0.030 & 13.030 $\pm$ 0.010 & 11.530 $\pm$ 0.010 & 10.960 $\pm$ 0.010 & 10.560 $\pm$ 0.010&  10.520 $\pm$ 0.010 \\
65 & PGC1185172              &                  &22.125 $\pm$ 0.567    &  18.628 $\pm$ 0.075 & 17.330 $\pm$ 0.009 & 16.784 $\pm$ 0.009 & 16.526 $\pm$ 0.011 & 16.310 $\pm$ 0.028 \\
66 & PGC054004               &  23.673 $\pm$ 0.372 &                   &  17.305 $\pm$ 0.037 & 15.685 $\pm$ 0.004 & 14.985 $\pm$ 0.004 & 14.621 $\pm$ 0.004 & 14.460 $\pm$ 0.013 \\
%NGC5854                 &  19.573 $\pm$ 0.046 &17.081 $\pm$ 0.009   &  14.355 $\pm$ 0.003 & 12.707 $\pm$ 0.002 & 12.018 $\pm$ 0.002 & 11.628 $\pm$ 0.002 & 11.354 $\pm$ 0.002 \\
67 & NGC5854                     &  19.480 $\pm$ 0.060 &17.140 $\pm$ 0.030 &13.950 $\pm$ 0.010 &12.340 $\pm$ 0.010 &11.640 $\pm$ 0.010 &11.230 $\pm$ 0.010 &11.070 $\pm$ 0.010 \\
68 & PGC054016               &       	&	                &  16.938 $\pm$ 0.025 & 15.607 $\pm$ 0.003 & 14.954 $\pm$ 0.003 & 14.613 $\pm$ 0.003 & 14.441 $\pm$ 0.009 \\
69 & PGC1217593              &                   &21.631 $\pm$ 0.128   &  18.972 $\pm$ 0.069 & 17.578 $\pm$ 0.009 & 16.915 $\pm$ 0.008 & 16.578 $\pm$ 0.009 & 16.439 $\pm$ 0.045 \\
70 & PGC054037               &              &		        &  17.274 $\pm$ 0.028 & 15.586 $\pm$ 0.004 & 14.796 $\pm$ 0.004 & 14.419 $\pm$ 0.005 & 14.159 $\pm$ 0.013 \\
71 & NGC5846:[MTT2005]258    &  22.202 $\pm$ 0.150&                    &  20.798 $\pm$ 0.565 & 18.820 $\pm$ 0.032 & 18.363 $\pm$ 0.033 & 18.273 $\pm$ 0.046 & 19.024 $\pm$ 0.365 \\
72 & NGC5846:[MTT2005]259    &		&	&   20.699 $\pm$ 0.413 & 19.221 $\pm$ 0.038 & 18.579 $\pm$ 0.036 & 18.594 $\pm$ 0.059 & 19.142 $\pm$ 0.368 \\
73 & PGC054045               &		&		 &   18.066 $\pm$ 0.073 & 16.316 $\pm$ 0.007 & 15.630 $\pm$ 0.007 & 15.329 $\pm$ 0.018 & 15.121 $\pm$ 0.021 \\
74 & SDSSJ150812.35+012959.7 &		&	&   19.487 $\pm$ 0.180 & 17.743 $\pm$ 0.014 & 17.059 $\pm$ 0.012 & 16.705 $\pm$ 0.013 & 16.437 $\pm$ 0.042 \\
75 & NGC5846:[MTT2005]264    &		&	&   19.821 $\pm$ 0.126 & 18.641 $\pm$ 0.018 & 18.101 $\pm$ 0.018 & 17.789 $\pm$ 0.021 & 17.740 $\pm$ 0.087 \\
76 & PGC1206166              &		&		 &   19.034 $\pm$ 0.098 & 17.634 $\pm$ 0.011 & 16.997 $\pm$ 0.010 & 16.730 $\pm$ 0.013 & 16.765 $\pm$ 0.202 \\
77 & NGC5846:[MTT2005]268    &		&	&   20.828 $\pm$ 0.444 & 18.919 $\pm$ 0.031 & 18.200 $\pm$ 0.116 & 18.126 $\pm$ 0.040 & 17.774 $\pm$ 0.122 \\
78 & PGC1209573              &		&		 &   17.871 $\pm$ 0.038 & 16.332 $\pm$ 0.005 & 15.682 $\pm$ 0.004 & 15.372 $\pm$ 0.005 & 15.182 $\pm$ 0.014 \\
79 & PGC1176385              &  19.943 $\pm$ 0.049 &19.424 $\pm $0.026 &   17.661 $\pm$ 0.021 & 16.507 $\pm$ 0.004 & 15.905 $\pm$ 0.004 & 15.607 $\pm$ 0.006 & 15.415 $\pm$ 0.014 \\
80 & SDSSJ150907.83+004329.7 &  19.994 $\pm$ 0.060 &19.637 $\pm $0.044 &    18.563 $\pm$ 0.056 & 17.429 $\pm$ 0.012 & 16.871 $\pm$ 0.012 & 16.635 $\pm$ 0.018 & 16.688 $\pm$ 0.052 \\
81 &	PGC1210284              &		 &		 &  17.999 $\pm$ 0.050 & 16.385 $\pm$ 0.005 & 15.702 $\pm$ 0.005 & 15.407 $\pm$ 0.006 & 15.227 $\pm$ 0.019 \\
%NGC5864                 &		&		 &  14.524 $\pm$ 0.005 & 12.578 $\pm$ 0.002 & 11.825 $\pm$ 0.002 & 11.429 $\pm$ 0.002 & 11.176 $\pm$ 0.002 \\
82 & NGC5864                     & & &14.240 $\pm$ 0.010 &12.400 $\pm$ 0.010 &11.630 $\pm$ 0.010 &11.230 $\pm$ 0.010 &10.980 $\pm$ 0.010 \\
%NGC5869                 &  19.181 $\pm$ 0.028& 17.315 $\pm$ 0.015 &   14.682 $\pm$ 0.004 & 12.729 $\pm$ 0.002 & 11.873 $\pm$ 0.001 & 11.489 $\pm$ 0.002 & 11.159 $\pm$ 0.002 \\
83 & NGC5869                     & 19.180 $\pm$ 0.060 &17.500 $\pm$ 0.030 &14.420 $\pm$ 0.010 &12.410 $\pm$ 0.010 &11.660 $\pm$ 0.010 &11.150 $\pm$ 0.010 &10.900 $\pm$ 0.010 \\ 
%UGC09746                &		&		 &  15.955 $\pm$ 0.016 & 14.606 $\pm$ 0.003 & 13.970 $\pm$ 0.005 & 13.647 $\pm$ 0.005 & 13.416 $\pm$ 0.009 \\
84 & UGC09746                   &  &17.150 $\pm$ 0.030 &15.610 $\pm$ 0.010 &14.440 $\pm$ 0.010 &13.860 $\pm$ 0.010& 13.530 $\pm$ 0.010 &13.270 $\pm$ 0.010 \\
85 & UGC09751                &  19.343 $\pm$ 0.044& 18.533 $\pm$ 0.025 &   17.492 $\pm$ 0.041 & 16.108 $\pm$ 0.006 & 15.577 $\pm$ 0.006 & 15.326 $\pm$ 0.009 & 15.264 $\pm$ 0.021 \\
86 & PGC1202458              &		&		 &   18.625 $\pm$ 0.083 & 17.018 $\pm$ 0.014 & 16.360 $\pm$ 0.013 & 16.046 $\pm$ 0.013 & 16.047 $\pm$ 0.099 \\
87 & SDSSJ151121.37+013639.5 &		&	 &   18.987 $\pm$ 0.151 & 17.152 $\pm$ 0.012 & 16.488 $\pm$ 0.018 & 16.288 $\pm$ 0.019 & 16.177 $\pm$ 0.044 \\
%UCG09760                &17.599 $\pm$0.062 &	16.985 $\pm$0.10 &   17.087 $\pm$ 0.039 & 15.748 $\pm$ 0.006 & 15.237 $\pm$ 0.011 & 14.864 $\pm$ 0.018 & 15.037 $\pm$ 0.024\\
88 & UCG09760                    & 17.380 $\pm$ 0.050 &17.020 $\pm$ 0.030 &16.200 $\pm$ 0.010 &15.030 $\pm$ 0.010 &14.680 $\pm$ 0.010 &14.530 $\pm$ 0.010 &14.450 $\pm$ 0.010 \\
89 & PGC1199418              &		&		 &   17.940 $\pm$ 0.030 & 16.662 $\pm$ 0.006 & 16.134 $\pm$ 0.005 & 15.864 $\pm$ 0.006 & 15.704 $\pm$ 0.014 \\
90 & PGC1215336              &       & 20.601 $\pm$ 0.066 &  18.124 $\pm$ 0.049 & 16.646 $\pm$ 0.006 & 16.007 $\pm$ 0.005 & 15.683 $\pm$ 0.006 & 15.480 $\pm$ 0.022 \\
 
 \hline			 							        			    
\end{tabular}

\label{tab3}	 											     
\end{table*}

\section{Results}
\label{xy}

Hereafter, following the definition  of \citet{Tamm94}, we considered  as dwarfs galaxies  
fainter than  M$_B =$-16 mag (M$_V =$-17 mag),  
 and as ETGs,  galaxies  with morphological type T$\leq$0 (i.e E-S0a-dE-dE,N-dS0) as in \citet{Bos14}. 

\subsection{UV vs. optical morphological classification}

Table \ref{tab1} (column 5) compares the morphological classification of members we adopted with that in \citet{Mahdavi05}.  In Appendix A we show the UV (left panels) and SDSS (right panels) colour composite images of all 90 galaxy members.
The {\tt HYPERLEDA}  optical morphological classification is in good agreement with that 
suggested by UV images.  

Here we present the galaxies of the group for which UV images suggest a  different morphological  classification.\\
\indent
ID 15: NGC ~5811  is  a blue  UV galaxy with a bar,  visible both in optical and  UV,   representing  the main body of the galaxy. It is a late type galaxy, SBm, rather than  dE. \\
\indent
ID 20: SDSSJ150100.85+010049.8, we adopted the {\tt HYPERLEDA} classification.\\
\indent
ID 28: UGC09661 shows a blue bar in UV, the SBd classification seems more appropriate than Sdm.   \\   
\indent
ID 35: PGC087108, shows two distinct bright sources in a common blue envelope in the UV image. It is classified as Irregular; the classification seems correct.\\
\indent
ID 47: PGC1190358 morphology is quite irregular  in the UV image. It does not appear a dE as in \citet{Mahdavi05}.  \\
\indent
ID 53: PGC 1218738 seems to have a blue inner bar in the UV image although it is classified Sm.\\
\indent
ID 55: PGC1215798, classified Scd, shows peculiar tidal tails, that 
we consider as possible interaction signatures.\\
\indent
ID 58: SDSSJ150634.25+001255.6. We consider this galaxy a dE (T=-5$\pm$5) as in  {\tt HYPERLEDA}.\\
\indent
ID 60: NGC 5841 is a S0-a, incipient arms are visible in the SDSS image. The UV composite image shows that the colour is consistent with a old stellar population.\\
\indent
ID 64: NGC5850 is a barred Spiral with a ring and irregular spiral arms, likely signatures of interaction. 
The  irregular  arms are markedly extended in the UV image  much further out than the optical size. We suggest a classification of SB(r)b as in RC3. \\
\indent
ID 65: PGC1185172  is classified S? (T=10$\pm$5). We adopted the dE classification of \citet{Mahdavi05}. \\
\indent
ID 67: NGC 5854  is classified S0-a in {\tt HYPERLEDA} and S0 in \citet{Mahdavi05}. The UV image suggests the presence of incipient arms, more consistent with the {\tt HYPERLEDA} classification.\\
\indent
ID 70: PGC054037 is classified S? in {\tt HYPERLEDA} (T=1$\pm$-5) and S0a in \citet{Mahdavi05}. We adopt this
latter classification.\\
\indent
ID 80:  PGC4005496 appears quite blue in UV composite image and with an irregular shape. Rather than a E/dE classification we suggest Im.\\
\indent
ID 88: UGC09760,  seen edge-on, is classified Scd/Sd. The yellow spot on the blue galaxy disk of the UV image is a star. The galaxy appears really bulge-less so we adopted the morphological classification given in RC3, i.e. Sd.\\
\indent
ID 90: PGC1215336 is classified S? (T=10$\pm$5). We classified it as dE from the SDSS image.
\indent

\begin{figure}
 \vspace{-0.3cm}
 \includegraphics[width=8.5cm]{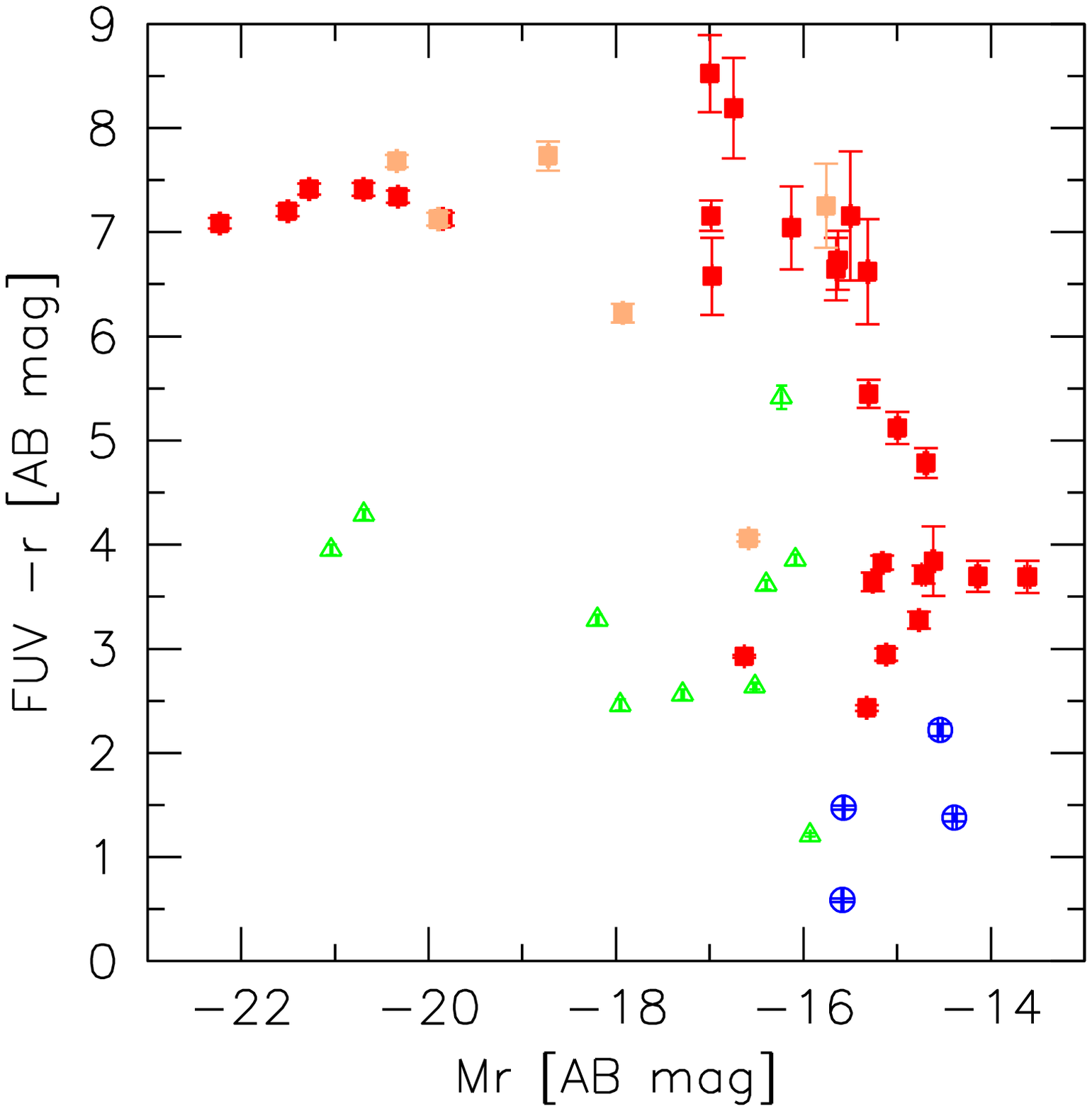}
 \includegraphics[width=8.5cm]{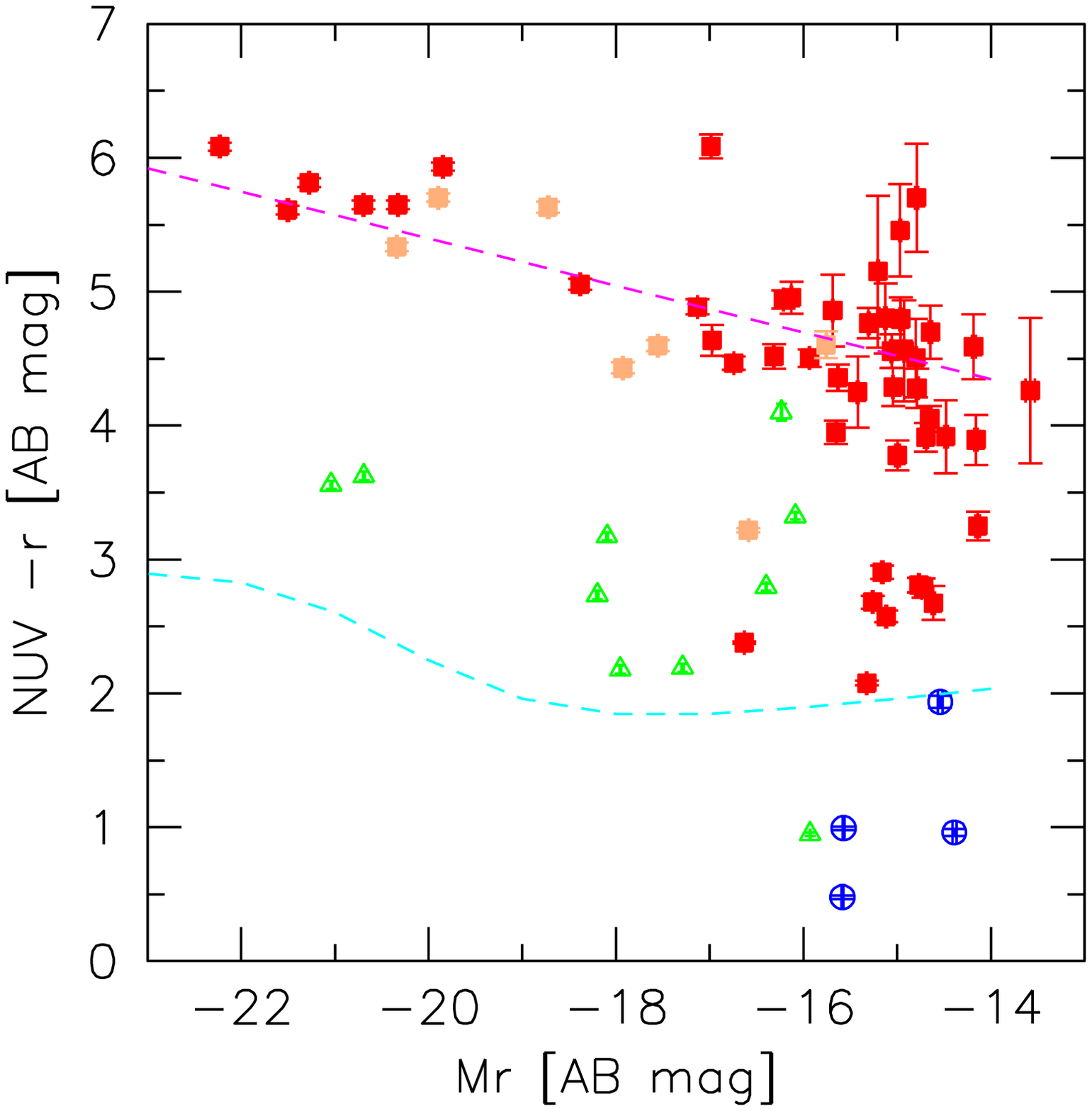}
 \caption{UV $-$ optical CMDs of NGC 5846. Top: M$_r$ versus FUV$-$r. Bottom:
M$_r$ versus NUV $-$r. In the M$_r$ versus NUV$-$r CMD, we over-plot the Wyder et al. (2007) fits to the red and blue galaxy sequences. Green triangles mark Spirals, Ellipticals and S0s are indicated with 
red  and orange squares, respectively,  and blue circles show Irregulars. The magnitudes were corrected by Galactic extinction \citep{BH82}.}
\label{CDM}
\end{figure}

  \subsection{The UV - optical  CMD}
  
Figure \ref{CDM}  shows the UV - optical CMDs of the members of NGC 5846. In the M$_r$ vs NUV - $r$ CMD (bottom panel),  there are 69
galaxies and 75 per cent are dwarfs, as previously defined. The red sequence, where passively evolving galaxies
 are located, is well defined and  populated by both Ellipticals and S0s.  ETGs represent 82 per cent
 (56/69) of the total galaxy population and 79 per cent (44/56) of them are dwarfs. The  33 per cent (14/44)   of  galaxies fainter than Mr = -18   are ETGs  lying in the ``green valley",
 i.e.  with 2$\le$NUV$-$r$\le$4,  some of them  very near to the blue sequence.   This behaviour agrees with the findings of  \citet{Maetal14}. These authors, studying the evolution of ETGs in two groups of the Leo cloud, USCG U376 and LGG 225,  found that rejuvenation episodes are more frequent in fainter ETGs (see their Fig. 5). \\

\bigskip

\begin{figure}
\vspace{-0.3cm}
\includegraphics[width=8.5cm]{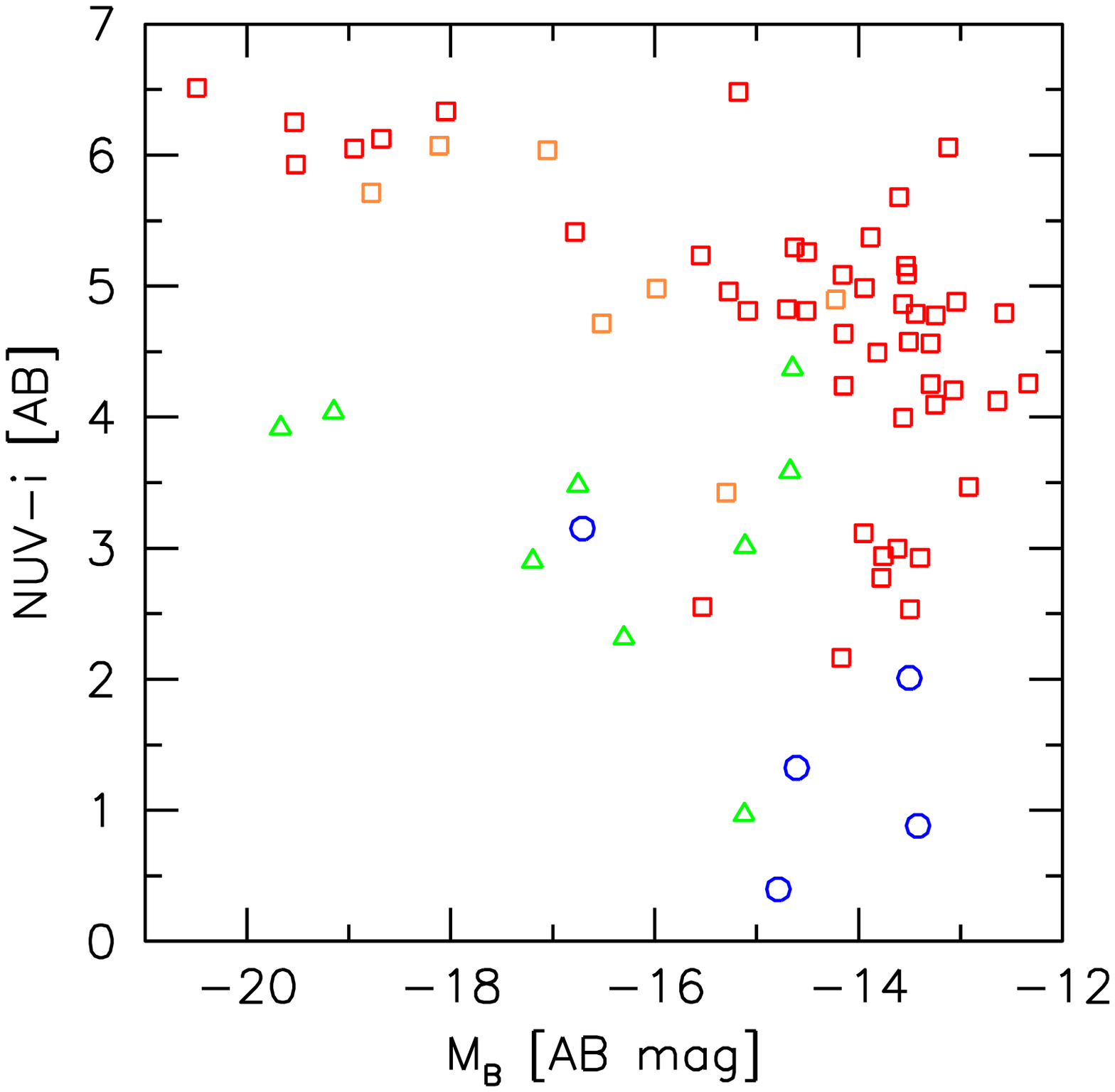}
\includegraphics[width=8.5cm]{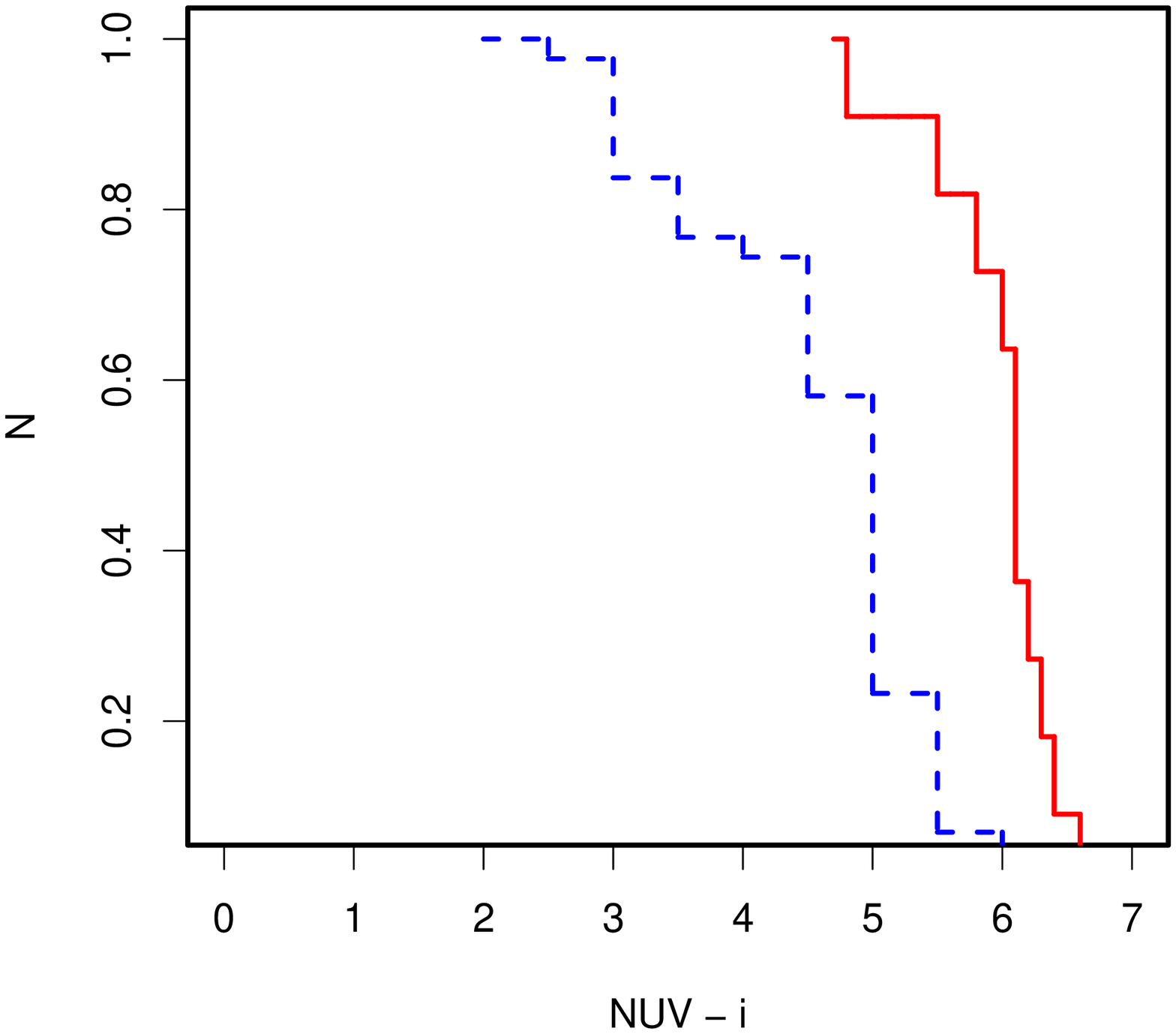}
 \caption{Top: The M$_B$ versus NUV - $i$ CMD of the group members. Symbols are as in Fig. \ref{CDM}. Bottom: Cumulative  distributions of NUV$-$i  of dwarf ( blue dashed line) and normal ( red continuous line) ETGs in our group; according to the Kolmogorov-Smirnov test the null hypothesis,   that the two distributions are drawn from a same parent distribution, can be rejected at a confidence level $>$ 99 per cent. }
\label{hist}
\end{figure}	
 Figure \ref{hist} shows the absolute B magnitude, M$_B$, versus NUV$-$i of the group members (top panel).
In the bottom panel of  Figure \ref{hist},  the cumulative
 distribution of NUV$-$i  for  normal and dwarf members of the group is shown. 
  The distribution gives the
fraction of normal and dwarf galaxies in the group that have a colour greater (redder)
than a given value of NUV$-$i. For example, $\sim$90 and 20 per cent
of normal and dwarf galaxies, respectively, have NUV$-$i $>$ 5.
 According to the Kolmogorov-Smirnov test, the null hypothesis  that two distributions are drawn from the same parent distribution can be rejected at a confidence level $>$ 99 per cent. 
  We consider the hypothesis that giant ETGs members may either
have formed through or have experienced a significant number of accretions of
dwarfs galaxies during the evolution of the group. In this hypothesis the
color distributions of the two samples should have similar characteristics.

This should be particularly true  if the accretions have been ``dry'',
i.e. the accretion has been ``sterile''  not igniting star formation episodes.

The statistically significant difference of the NUV-i  of dwarfs
and giant ETGs rules out the above formation scenario and a ``dry''
accretion scenario.
Instead, we suggest that in the star formation history of both dwarfs and normal ETGs, gas dissipation cannot be neglected. We further explore this hypothesis in the following section.

 \begin{figure}
 \vspace{-0.3cm}
 \includegraphics[width=8.5cm]{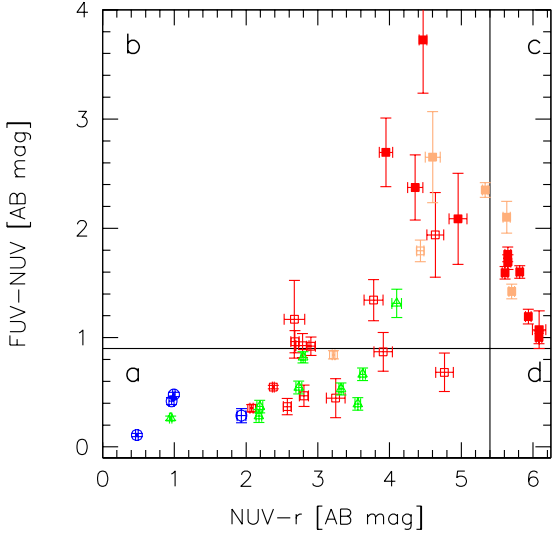}
 \caption{The FUV$-$NUV versus NUV$-$r colour$-$colour diagram of the NGC 5846 group members. Symbols are as in the previous figures.  The magnitudes were corrected by Galactic extinction following \citet{BH82}. Solid lines correspond to  FUV$-$NUV $< $0.9, i.e UV rising slope, and 
NUV$-$r $>$ 5.4, i.e. a galaxy devoid of young massive stars. These conditions,   following the UV  classification scheme by \citet{Yi2011}, separate passive evolving ETGs (region b) from star forming galaxies (region a), see text). Filled symbols are for galaxies with FUV$-$r$\ge$6.6 mag }
  \label{CC1}
\end{figure}

\subsection{The colour$-$colour FUV$-$NUV versus NUV$-$r diagram}

The slope of the UV spectrum is related to the temperature of the stars emitting in the UV and their relative contribution to the total flux. 
The value FUV$-$NUV = 0.9 indicates a flat UV spectrum in the $\lambda-$ versus F$_{\lambda}$ domain, whereas a negative FUV$-$NUV corresponds to a bluer population. \citet{Yi2011} established a colour criterion to classify ETGs according to their UV spectral morphology based on three colour. Passively evolving ETGs  would have NUV$-$r $\ge$ 5.4, and  FUV$-$r $\ge$ 6.6. 
These  values indicate the average value of the red sequence in Fig. \ref{CDM}.
ETGs showing UV upturn with no residual star formation have to obey a further  condition, FUV$-$NUV$<$0.9.
Figure \ref{CC1} shows the position of NGC 5846 group members in the colour$-$colour UV$-$optical diagram emphasising their morphological classification.
The same fraction of dwarf ETGs which stays in the ``green valley" of Fig. \ref{CDM}, i.e.
33 per cent,  lies  in the region where residual star formation is expected (region a in Fig. \ref{CC1}), in good agreement with  findings of \citet{Maetal14}.
All the  ETGs brighter than -18  in the r-band (Fig. \ref{CDM}, top panel) show red FUV$-$r colours and FUV$-$NUV$>$0.9 and  lie in the regions b) and  c) of Fig.  \ref{CC1}.
The brightest members of this group, i.e., NGC 5846 and NGC 5813,  lie in  the right upper region of this colour-colour diagram, i.e. region c), where passively evolving ETGs would stay  according the \citet{Yi2011}  criterion. 
No galaxies are found in  region d) of Fig. \ref{CC1}, where  ETGs  with UV upturn and no residual star formation would lie.

\section {Comparison with  other groups and the Virgo cluster}
	
\begin{figure}
 \vspace{-0.3cm}
 \includegraphics[width=8.5cm]{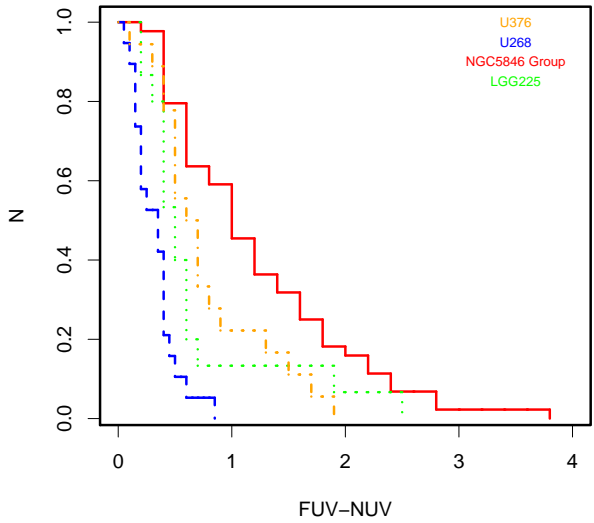}
 \caption{Cumulative  distribution of FUV$-$NUV colours of galaxies in  NGC5846 group (solid line), and  in three groups previously studied, i.e, U268 (dashed line), U376 (dot-dashed line), and LGG225 (dotted line). }
  \label{CD}
\end{figure}

\begin{figure}
\vspace{-0.3cm}
\includegraphics[width=8.5cm, angle=-90]{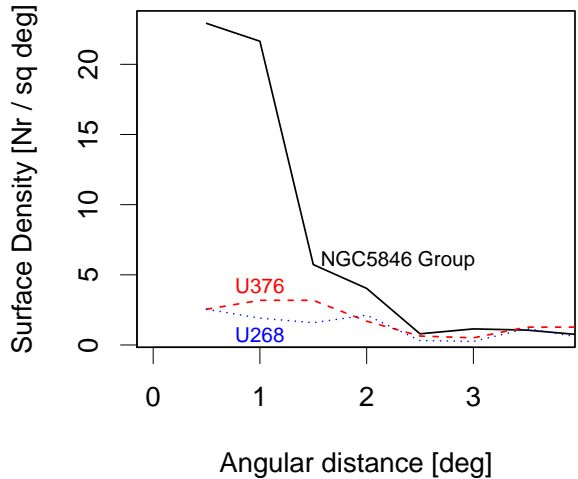}
\caption{ Surface member density  as a function of the radial distance from the
 centre of mass defined by the dynamical analysis developed in Section~\ref{secdyn}. 
 For comparison  the surface density distributions of the two groups U268
 (dotted line) and U376 (dashed line) are shown.}
\label{pho}
\end{figure}

Figure \ref{CD} shows the cumulative distribution of the FUV$-$NUV colours of NGC5846, and  of three groups already analised in the Leo cloud (Paper I and II).
This  figure points out that  the  fraction of red UV colours, i.e. FUV$-$NUV$>$0.9,    increases with increasing  number of  ETGs.
By comparing Fig. \ref{CDM} (bottom) with Figures 10 and 11 in  \citet{Marino13} we note that the  number of galaxies with red NUV$-$r colours, i.e. along  the red sequence, all ETGs,  increases with the groups  are more massive and composed by more galaxies. \\
Figure \ref{pho}  compares the density distribution versus angular distance from the dynamical centre 
of  NGC 5486  (solid line) with the density distribution of the groups previously studied, i.e,  USGC U376 (dashed line)  and USGC U268 (dotted line). These latter groups are located in the Leo cloud. The dynamical analysis by \citet{Marino13a}  suggests that USGC U268 is in a pre-virial collapse phase while U376 seems  in a more evolved phase toward virialization. 
 Notice that our distribution is different from that shown in \citet[][their Figure~7]{Mahdavi05} because 
 we use the centre of mass of the group (Table \ref{Dyn}) as centre of the density distribution. We obtain the same distribution as \citet{Mahdavi05} if we select NGC 5486  as centre of the group.

\begin{figure}
 \vspace{-0.3cm}
 \includegraphics[width=8.5cm]{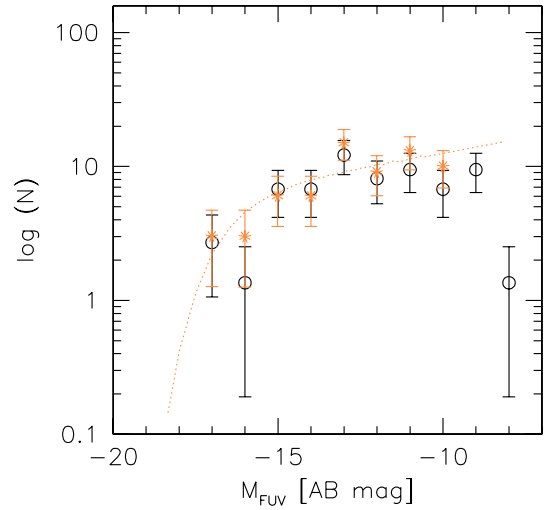}
\includegraphics[width=8.5cm]{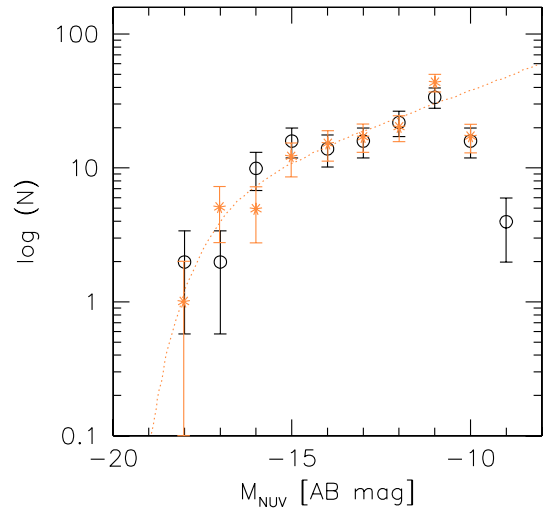}
 \caption{ The FUV and NUV LFs of NGC 5846 group (open circles) compared to those 
 of the Virgo cluster (asterisks; dotted line shows the fit limited to -13 mag by \citet{Bos14}.
  LFs have been normalized to include the same galaxy number as Virgo, i.e., 135 in NUV and 65 in FUV as in \citet{Bos14}.}
  \label{FNUV}
\end{figure}

\begin{figure}
 \vspace{-0.3cm}
\includegraphics[width=8.5cm]{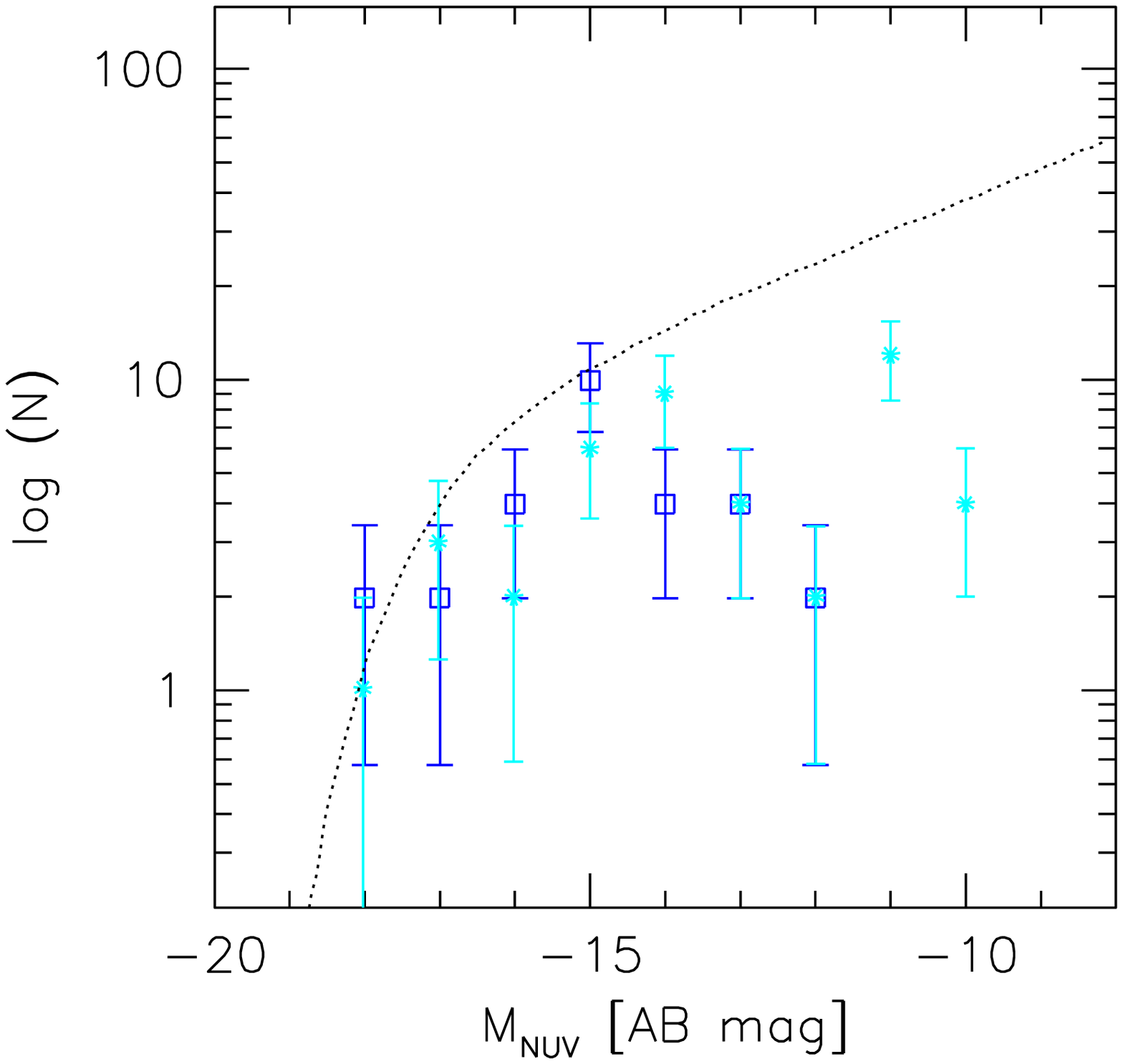}
\includegraphics[width=8.5cm]{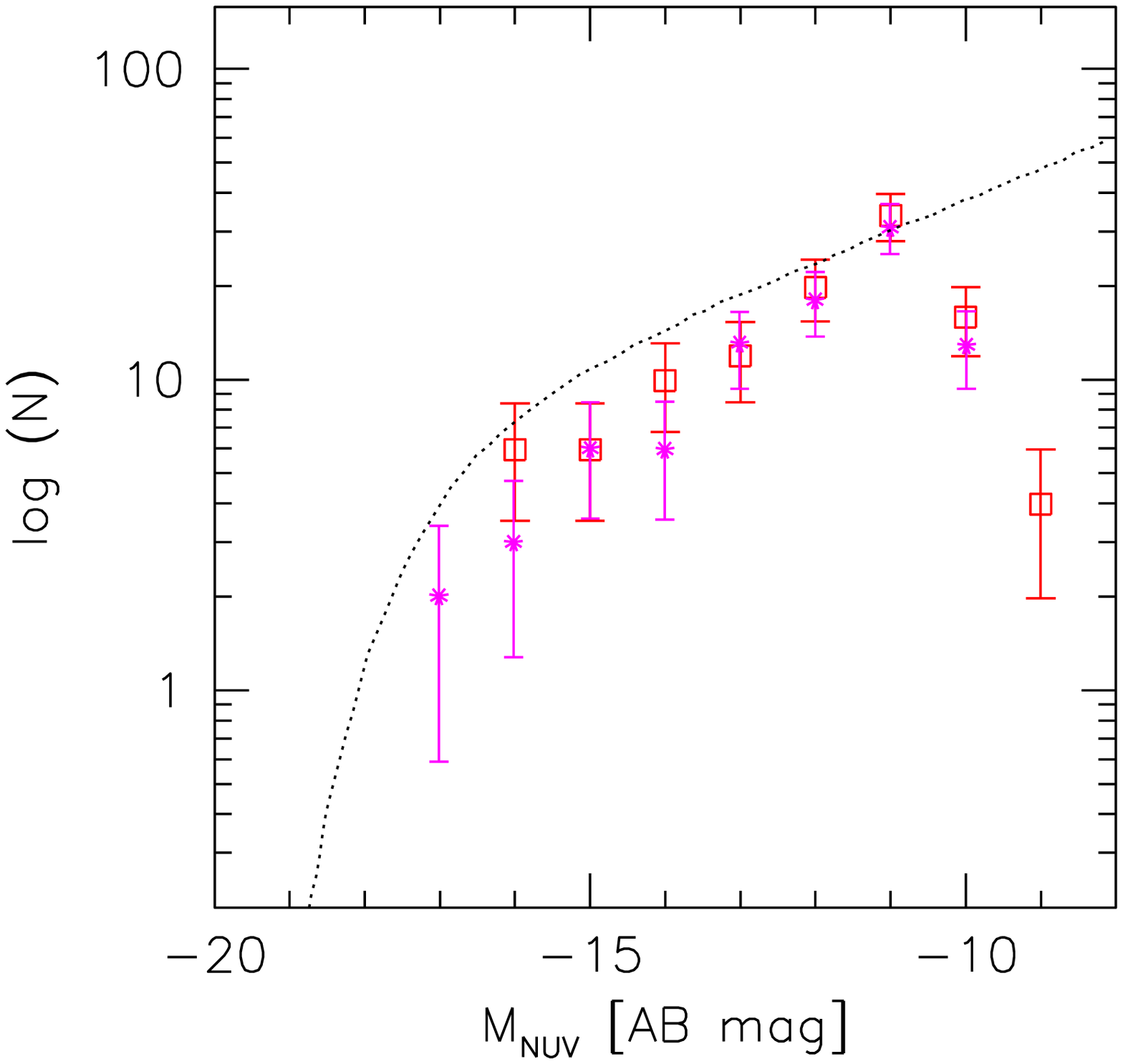}
 \caption{ The  NUV LFs of NGC 5846 group splitted in  late type galaxies (top), 
 and ETGs (bottom), compared to those  of the Virgo cluster (asterisks and dotted line as in Fig.\ref{FNUV}) from \citet[their Fig. 14]{Bos14}.}

 \label{LFtypes}
\end{figure}

Figure \ref{FNUV} shows  the UV luminosity 
function (LF hereafter; top panel FUV and  bottom panel NUV) of  NGC 5846 group. For comparison we plot, on the same scale, the UV LFs of  
the Virgo cluster (dotted line)  as in Figure~8 of \citet{Bos14}.  These authors noted that the NUV and 
FUV LFs of Virgo and the field are similar. We find that FUV and NUV LFs of NGC 5846 group are quite
similar to those of the Virgo cluster. In particular, as shown in Fig. \ref{LFtypes},  also the LFs of late type galaxies and ETGs in both groups are quite indistinguishable from those of this cluster.

\begin{figure}
 \vspace{-0.3cm}
\includegraphics[width=8.5cm]{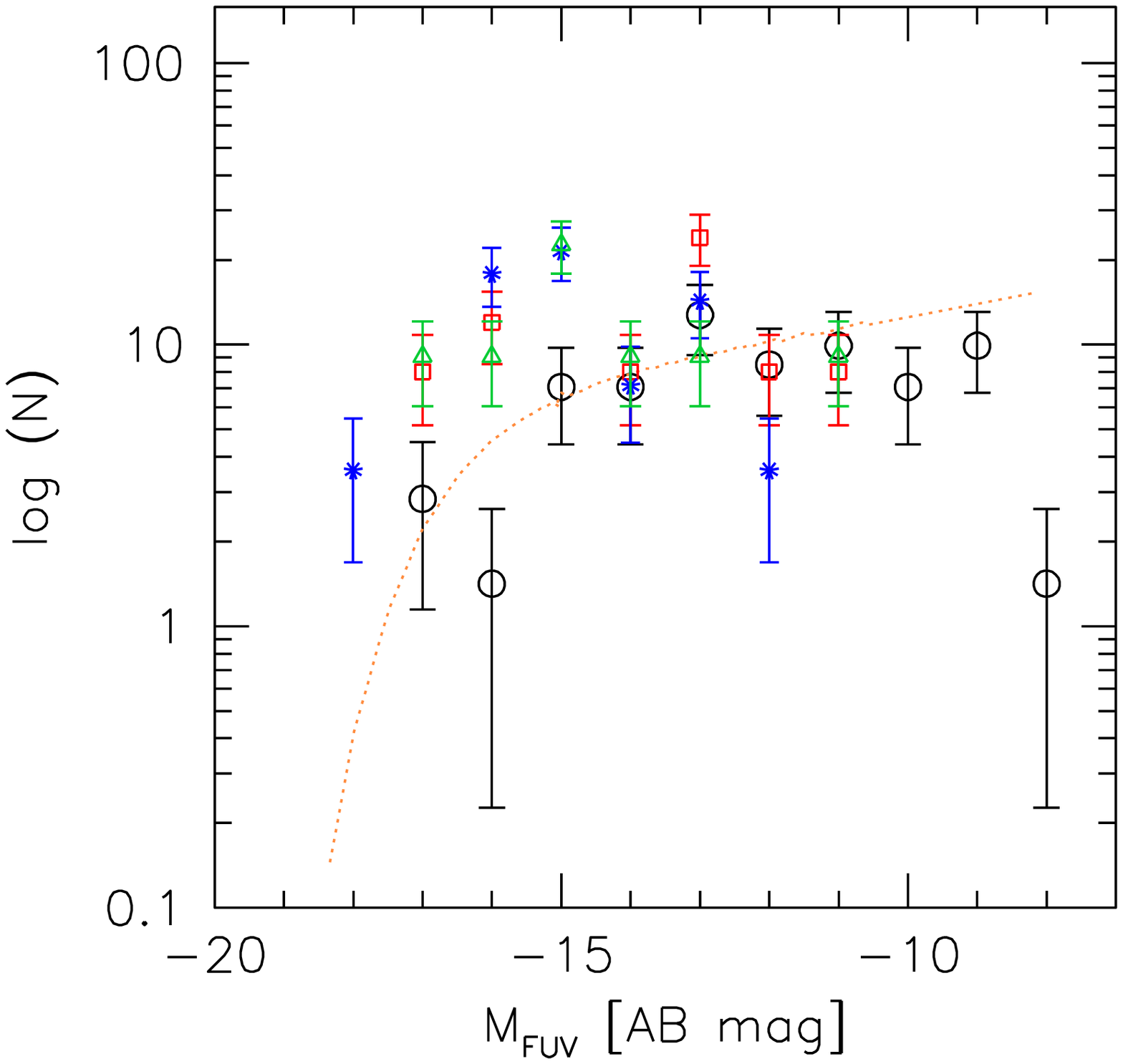}
\includegraphics[width=8.5cm]{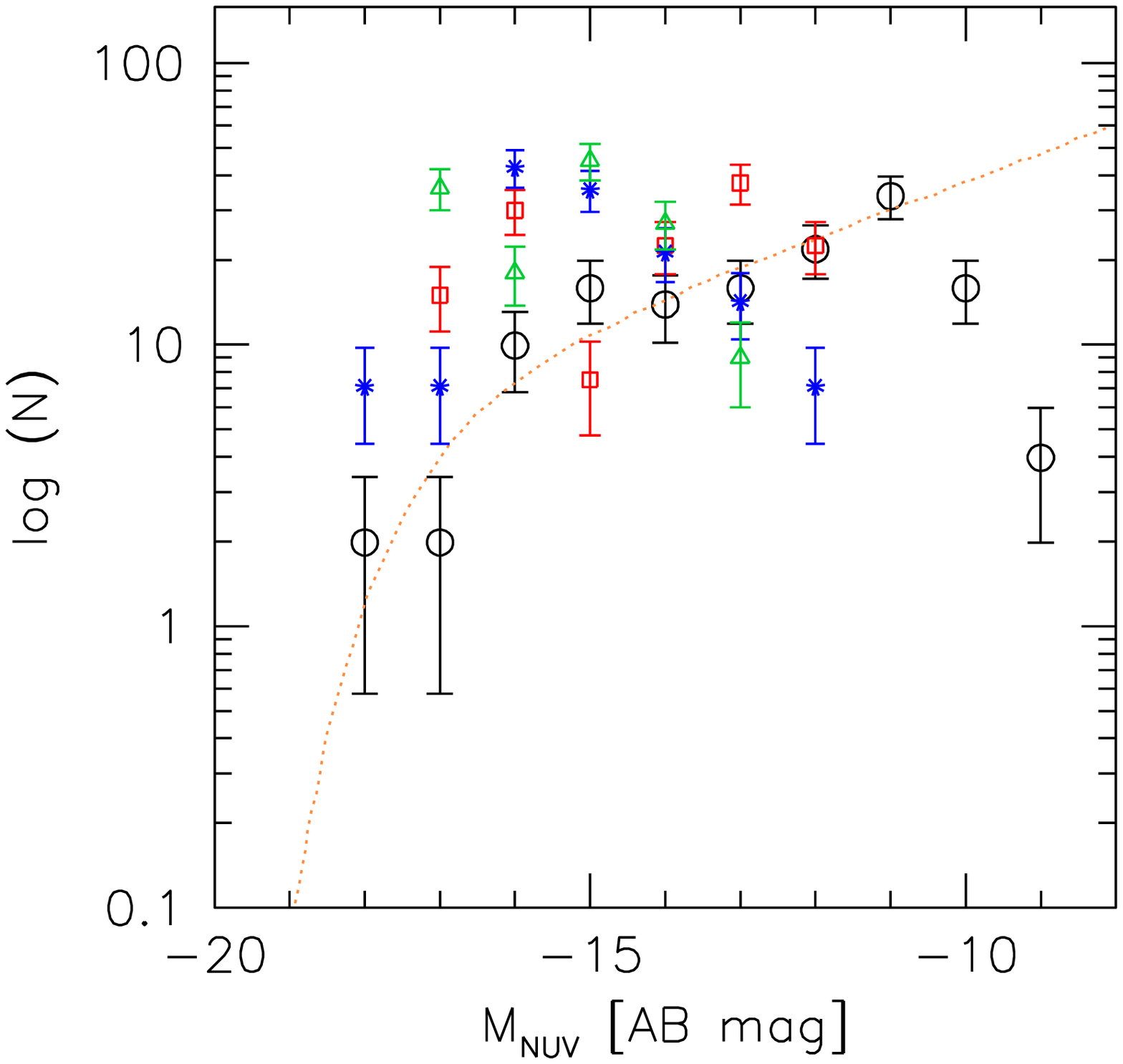}
 \caption{ The FUV and NUV LFs of NGC 5846 group (circles) compared to those of the Virgo cluster (dotted line \citet{Bos14}),  U376 (red squares), U268 (blue asterisks) and LGG225 (green triangles).  LFs have been normalized as in Fig. \ref{FNUV}.}
  \label{FNUV2}
\end{figure}

Figure \ref{FNUV2} shows that the shape of the FUV and NUV  LFs in less dense groups (see Figure \ref{pho}) is dominated by late-type galaxies. In the brightest magnitude bins, late type galaxies are more numerous in these groups than in denser groups like  NGC 5846 and Virgo. 

\section {Summary and conclusions}

This paper is the third of a series dedicated to the study of nearby groups with 
a different morphological mix of galaxy populations and sampling different dynamical phases. \\
\indent

We have obtained  FUV and NUV {\it GALEX}  and  SDSS - $u, g, r, i, z$  
AB  magnitudes of 90 spectroscopically confirmed members of  NGC~5846, the third most
massive nearby association after Virgo and Fornax nearby clusters.
The backbone of the group comes  from the catalog of \citet{Ramella02} that 
we enriched of members  applying kinematical and dynamical selection criteria  to galaxies with known optical redshift (see also Paper~I and ~II).\\
The group membership as well as the characteristics of the group have been
already investigated  with a different method by \citet{Mahdavi05} and  \citet{Eige10}. 
Our selection of 90 members includes all the spectroscopically confirmed members of \citet{Mahdavi05} and \citet{Eige10} plus two ETGs in HYPERLEDA meeting our criteria.

The kinematical and dynamical analysis of the group  indicates that it is in a evolved phase  according to \citet{Mahdavi05} analysis.
The main novelty of the present study is the  UV
analysis.
Our analysis of the  UV data shows that a large fraction of dEs (33 per cent, Section 4.3) does not reside  in  the red or in the blue sequence of the group CMD  but it lies in the  ``green valley", where ``rejuvenation" episodes of dEs  occur with higher frequency \citep{Maetal14}.  We  find that
only 5 per cent of the total ETG population of the group lies in the region of passively evolving ETGs (Section 4.3,  region c, Fig. \ref{CC1}) whereas dEs are found in the locus  of star forming galaxies  (region a, Fig. \ref{CC1}). Moreover,  by analysing the  cumulative NUV$-$i colour distributions of dEs and normal ETGs in the group,  the hypothesis that the two distributions are drawn from the same parent distribution can be rejected at a confidence level of $>$99 per cent.  We concluded that he UV-optical colours of normal ETGs in the group cannot be accounted  by dry mergers of the  dE population: gas dissipation, i.e. star formation, cannot be neglected  in the evolution of the group members.

 \citet{Bos14} found that Virgo galaxies are quenched by
ram pressure and  suggest that the quenching of the
star formation activity in dwarf systems and the formation of the faint end of the red sequence is a very recent phenomenon.
Although the UV LFs of both Virgo and NGC 5846 are very similar, only a small fraction of galaxies in NGC 5846 is passively evolving. Our UV-optical analysis
suggests that star formation events are still occurring in this
group, in particular in its dwarf ETG population, tracing
a picture of a still active phase notwithstanding its large
number of ETGs and its likely virialized configuration (Table 2). 

 \citet{Maetal14}, investigating the  evolution of the brightest ETGs in the U376 and LGG 225 groups, found that residual star formation, i.e. ``rejuvenation", is luminosity dependent so that  bursts of star formation can occur still today in dEs, as  found in this group.
 
In a forthcoming paper (Mazzei et al. 2016, in prep.), we will investigate further the evolution of the brightest ETG members  and  some dwarfs of this group using a smooth particle Hydrodynamic code with chemo-photometric implementation. We will also expand the study to other selected groups for which we have UV and optical images.

\section*{Acknowledgments} 
We acknowledge the usage of the {\tt HYPERLEDA}  
and NED databases.	    
Paola Mazzei and Roberto Rampazzo acknowledge support from INAF through grant 
PRIN-2014-14 `Star formation and evolution in galactic nuclei'.\\

 {\it Facilities:}  GALEX, Sloan\\
			   	    
\bibliographystyle{mn2e}   	    
 \bibliography{U677}

\begin{thebibliography}{}

\bibitem[\protect\citeauthoryear{{Adelman-McCarthy}, {Ag{\"u}eros}, {Allam},
  {Allende Prieto}, {Anderson}, {Anderson}, {Annis} \&
  {Bahcall}}{{Adelman-McCarthy} et~al.}{2008}]{Ade08}
{Adelman-McCarthy} J.~K.,  {Ag{\"u}eros} M.~A.,  {Allam} S.~S.,  {Allende
  Prieto} C.,  {Anderson} K.~S.~J.,  {Anderson} S.~F.,  {Annis} J.,
  {Bahcall} N.~A.,  2008, \apjs, 175, 297

\bibitem[\protect\citeauthoryear{{Balogh}, {Baldry}, {Nichol}, {Miller},
  {Bower} \& {Glazebrook}}{{Balogh} et~al.}{2004}]{Balogh04}
{Balogh} M.~L.,  {Baldry} I.~K.,  {Nichol} R.,  {Miller} C.,  {Bower} R.,
  {Glazebrook} K.,  2004, \apjl, 615, L101

\bibitem[\protect\citeauthoryear{{Barnes}}{{Barnes}}{2002}]{Barnes02}
{Barnes} J.~E.,  2002, \mnras, 333, 481

\bibitem[\protect\citeauthoryear{{Bianchi}}{{Bianchi}}{2009}]{Bianchi09}
{Bianchi} L.,  2009, \apss, 320, 11

\bibitem[\protect\citeauthoryear{{Bianchi}}{{Bianchi}}{2014}]{Bianchi14}
{Bianchi} L.,  2014, \apss, 354, 103

\bibitem[\protect\citeauthoryear{{Bianchi}, {Efremova}, {Herald}, {Girardi},
  {Zabot}, {Marigo} \& {Martin}}{{Bianchi} et~al.}{2011}]{Bianchi11}
{Bianchi} L.,  {Efremova} B.,  {Herald} J.,  {Girardi} L.,  {Zabot} A.,
  {Marigo} P.,    {Martin} C.,  2011, \mnras, 411, 2770

\bibitem[\protect\citeauthoryear{{Boselli} \& {Gavazzi}}{{Boselli} \&
  {Gavazzi}}{2014}]{Bos14}
{Boselli} A.,  {Gavazzi} G.,  2014, \aapr, 22, 74

\bibitem[\protect\citeauthoryear{{Burstein} \& {Heiles}}{{Burstein} \&
  {Heiles}}{1982}]{BH82}
{Burstein} D.,  {Heiles} C.,  1982, \aj, 87, 1165

\bibitem[\protect\citeauthoryear{{Cardelli}, {Clayton} \& {Mathis}}{{Cardelli}
  et~al.}{1989}]{Cardelli89}
{Cardelli} J.~A.,  {Clayton} G.~C.,    {Mathis} J.~S.,  1989, \apj, 345, 245

\bibitem[\protect\citeauthoryear{{Davis} \& {Geller}}{{Davis} \&
  {Geller}}{1976}]{Davis76}
{Davis} M.,  {Geller} M.~J.,  1976, \apj, 208, 13

\bibitem[\protect\citeauthoryear{{Di Matteo}, {Springel} \& {Hernquist}}{{Di
  Matteo} et~al.}{2005}]{Dimatteo05}
{Di Matteo} T.,  {Springel} V.,    {Hernquist} L.,  2005, in {A.~Merloni,
  S.~Nayakshin, \& R.~A.~Sunyaev} ed., Growing Black Holes: Accretion in a
  Cosmological Context {Black Holes in Galaxy Mergers}.
pp 340--345

\bibitem[\protect\citeauthoryear{{Dressler}}{{Dressler}}{1980}]{Dressler80}
{Dressler} A.,  1980, \apj, 236, 351

\bibitem[\protect\citeauthoryear{{Dressler} \& {Shectman}}{{Dressler} \&
  {Shectman}}{1988}]{Dressler88}
{Dressler} A.,  {Shectman} S.~A.,  1988, \aj, 95, 985

\bibitem[\protect\citeauthoryear{{Efron}}{{Efron}}{1982}]{Efron82}
{Efron} B.,  1982, {The Jackknife, the Bootstrap and other resampling plans}

\bibitem[\protect\citeauthoryear{{Eigenthaler} \& {Zeilinger}}{{Eigenthaler} \&
  {Zeilinger}}{2010}]{Eige10}
{Eigenthaler} P.,  {Zeilinger} W.~W.,  2010, \aap, 511, A12

\bibitem[\protect\citeauthoryear{{Eke}, {Baugh}, {Cole}, {Frenk}, {Norberg} \&
  {Peacock}}{{Eke} et~al.}{2004}]{Eke04}
{Eke} V.~R.,  {Baugh} C.~M.,  {Cole} S.,  {Frenk} C.~S.,  {Norberg} P.,
  {Peacock} J.~A.,  2004, \mnras, 348, 866

\bibitem[\protect\citeauthoryear{{Ferguson} \& {Sandage}}{{Ferguson} \&
  {Sandage}}{1990}]{Ferguson90}
{Ferguson} H.~C.,  {Sandage} A.,  1990, \aj, 100, 1

\bibitem[\protect\citeauthoryear{{Ferguson} \& {Sandage}}{{Ferguson} \&
  {Sandage}}{1991}]{Ferguson1991}
{Ferguson} H.~C.,  {Sandage} A.,  1991, \aj, 101, 765

\bibitem[\protect\citeauthoryear{{Firth}, {Evstigneeva}, {Jones}, {Drinkwater},
  {Phillipps} \& {Gregg}}{{Firth} et~al.}{2006}]{Firth06}
{Firth} P.,  {Evstigneeva} E.~A.,  {Jones} J.~B.,  {Drinkwater} M.~J.,
  {Phillipps} S.,    {Gregg} M.~D.,  2006, \mnras, 372, 1856

\bibitem[\protect\citeauthoryear{{Giuricin}, {Marinoni}, {Ceriani} \&
  {Pisani}}{{Giuricin} et~al.}{2000}]{Giuricin00}
{Giuricin} G.,  {Marinoni} C.,  {Ceriani} L.,    {Pisani} A.,  2000, \apj, 543,
  178

\bibitem[\protect\citeauthoryear{{G{\'o}mez}, {Nichol}, {Miller}, {Balogh},
  {Goto}, {Zabludoff}, {Romer}, {Bernardi}, {Sheth}, {Hopkins}, {Castander},
  {Connolly}, {Schneider}, {Brinkmann}, {Lamb}, {SubbaRao} \&
  {York}}{{G{\'o}mez} et~al.}{2003}]{Gomez03}
{G{\'o}mez} P.~L.,  {Nichol} R.~C.,  {Miller} C.~J.,  {Balogh} M.~L.,  {Goto}
  T.,  {Zabludoff} A.~I.,  {Romer} A.~K.,  {Bernardi} M.,  {Sheth} R.,
  {Hopkins} A.~M.,  {Castander} F.~J.,  {Connolly} A.~J.,  {Schneider} D.~P.,
  {Brinkmann} J.,  {Lamb} D.~Q.,  {SubbaRao} M.,    {York} D.~G.,  2003, \apj,
  584, 210

\bibitem[\protect\citeauthoryear{{Goto}, {Yamauchi}, {Fujita}, {Okamura},
  {Sekiguchi}, {Smail}, {Bernardi} \& {Gomez}}{{Goto} et~al.}{2003}]{Goto03}
{Goto} T.,  {Yamauchi} C.,  {Fujita} Y.,  {Okamura} S.,  {Sekiguchi} M.,
  {Smail} I.,  {Bernardi} M.,    {Gomez} P.~L.,  2003, \mnras, 346, 601

\bibitem[\protect\citeauthoryear{{Haynes} \& {Giovanelli}}{{Haynes} \&
  {Giovanelli}}{1991}]{Haynes1991}
{Haynes} M.~P.,  {Giovanelli} R.,  1991, \aj, 102, 841

\bibitem[\protect\citeauthoryear{{Hou}, {Parker}, {Wilman}, {McGee}, {Harris},
  {Connelly}, {Balogh}, {Mulchaey} \& {Bower}}{{Hou} et~al.}{2012}]{Hou12}
{Hou} A.,  {Parker} L.~C.,  {Wilman} D.~J.,  {McGee} S.~L.,  {Harris} W.~E.,
  {Connelly} J.~L.,  {Balogh} M.~L.,  {Mulchaey} J.~S.,    {Bower} R.~G.,
  2012, \mnras, 421, 3594

\bibitem[\protect\citeauthoryear{{Jedrzejewski}}{{Jedrzejewski}}{1987}]{Jedrze%
jewski87}
{Jedrzejewski} R.~I.,  1987, \mnras, 226, 747

\bibitem[\protect\citeauthoryear{{Kawata} \& {Mulchaey}}{{Kawata} \&
  {Mulchaey}}{2008}]{Kawata2008}
{Kawata} D.,  {Mulchaey} J.~S.,  2008, \apjl, 672, L103

\bibitem[\protect\citeauthoryear{{Lewis}, {Balogh}, {De Propris}, {Couch} \& et
  al.}{{Lewis} et~al.}{2002}]{Lewis02}
{Lewis} I.,  {Balogh} M.,  {De Propris} R.,  {Couch} W.,    et al. 2002,
  \mnras, 334, 673

\bibitem[\protect\citeauthoryear{{Machacek}, {Jerius}, {Kraft}, {Forman},
  {Jones}, {Randall}, {Giacintucci} \& {Sun}}{{Machacek}
  et~al.}{2011}]{Machacek2011}
{Machacek} M.~E.,  {Jerius} D.,  {Kraft} R.,  {Forman} W.~R.,  {Jones} C.,
  {Randall} S.,  {Giacintucci} S.,    {Sun} M.,  2011, \apj, 743, 15

\bibitem[\protect\citeauthoryear{{Mahdavi}, {Trentham} \& {Tully}}{{Mahdavi}
  et~al.}{2005}]{Mahdavi05}
{Mahdavi} A.,  {Trentham} N.,    {Tully} R.~B.,  2005, \aj, 130, 1502

\bibitem[\protect\citeauthoryear{{Makarov}, {Prugniel}, {Terekhova}, {Courtois}
  \& {Vauglin}}{{Makarov} et~al.}{2014}]{Makarov14}
{Makarov} D.,  {Prugniel} P.,  {Terekhova} N.,  {Courtois} H.,    {Vauglin} I.,
   2014, \aap, 570, A13

\bibitem[\protect\citeauthoryear{{Mamon}}{{Mamon}}{1992}]{Mamon92}
{Mamon} G.~A.,  1992, \apjl, 401, L3

\bibitem[\protect\citeauthoryear{{Marino}, {Bianchi}, {Mazzei}, {Rampazzo} \&
  {Galletta}}{{Marino} et~al.}{2014}]{Marino13a}
{Marino} A.,  {Bianchi} L.,  {Mazzei} P.,  {Rampazzo} R.,    {Galletta} G.,
  2014, Advances in Space Research, 53, 920

\bibitem[\protect\citeauthoryear{{Marino}, {Bianchi}, {Rampazzo}, {Buson} \&
  {Bettoni}}{{Marino} et~al.}{2010}]{Marino10}
{Marino} A.,  {Bianchi} L.,  {Rampazzo} R.,  {Buson} L.~M.,    {Bettoni} D.,
  2010, \aap, 511, A29

\bibitem[\protect\citeauthoryear{{Marino}, {Plana}, {Rampazzo}, {Bianchi},
  {Rosado}, {Bettoni}, {Galletta}, {Mazzei}, {Buson}, {Ambrocio-Cruz} \&
  {Gabbasov}}{{Marino} et~al.}{2013}]{Marino13}
{Marino} A.,  {Plana} H.,  {Rampazzo} R.,  {Bianchi} L.,  {Rosado} M.,
  {Bettoni} D.,  {Galletta} G.,  {Mazzei} P.,  {Buson} L.,  {Ambrocio-Cruz} P.,
     {Gabbasov} R.~F.,  2013, \mnras, 428, 476

\bibitem[\protect\citeauthoryear{{Martin}, {Fanson}, {Schiminovich} \& et
  al.}{{Martin} et~al.}{2005}]{Martin05}
{Martin} D.~C.,  {Fanson} J.,  {Schiminovich} D.,    et al. 2005, \apjl, 619,
  L1

\bibitem[\protect\citeauthoryear{{Mazzei}, {Marino} \& {Rampazzo}}{{Mazzei}
  et~al.}{2014}]{Maetal14}
{Mazzei} P.,  {Marino} A.,    {Rampazzo} R.,  2014, \apj, 782, 53

\bibitem[\protect\citeauthoryear{{Moore}, {Katz}, {Lake}, {Dressler} \&
  {Oemler}}{{Moore} et~al.}{1996}]{Moore1996}
{Moore} B.,  {Katz} N.,  {Lake} G.,  {Dressler} A.,    {Oemler} A.,  1996,
  \nat, 379, 613

\bibitem[\protect\citeauthoryear{{Morrissey}, {Conrow}, {Barlow}, {Small},
  {Seibert}, {Wyder} \& {Budav{\'a}ri}}{{Morrissey} et~al.}{2007}]{Morrissey07}
{Morrissey} P.,  {Conrow} T.,  {Barlow} T.~A.,  {Small} T.,  {Seibert} M.,
  {Wyder} T.~K.,    {Budav{\'a}ri} 2007, \apjs, 173, 682

\bibitem[\protect\citeauthoryear{{Mulchaey}, {Davis}, {Mushotzky} \&
  {Burstein}}{{Mulchaey} et~al.}{2003}]{Mulchaey03}
{Mulchaey} J.~S.,  {Davis} D.~S.,  {Mushotzky} R.~F.,    {Burstein} D.,  2003,
  \apjs, 145, 39

\bibitem[\protect\citeauthoryear{{Nolthenius}}{{Nolthenius}}{1993}]{Nolthenius%
93}
{Nolthenius} R.,  1993, \apjs, 85, 1

\bibitem[\protect\citeauthoryear{{Perea}, {del Olmo} \& {Moles}}{{Perea}
  et~al.}{1990}]{Perea90a}
{Perea} J.,  {del Olmo} A.,    {Moles} M.,  1990, \apss, 170, 347

\bibitem[\protect\citeauthoryear{{Poggianti}, {von der Linden}, {De Lucia} \&
  et al.}{{Poggianti} et~al.}{2006}]{Poggianti06}
{Poggianti} B.~M.,  {von der Linden} A.,  {De Lucia} G.,    et al. 2006, \apj,
  642, 188

\bibitem[\protect\citeauthoryear{{Ramella}, {Geller}, {Pisani} \& {da
  Costa}}{{Ramella} et~al.}{2002}]{Ramella02}
{Ramella} M.,  {Geller} M.~J.,  {Pisani} A.,    {da Costa} L.~N.,  2002, \aj,
  123, 2976

\bibitem[\protect\citeauthoryear{{Rampazzo}, {Panuzzo}, {Vega}, {Marino},
  {Bressan} \& {Clemens}}{{Rampazzo} et~al.}{2013}]{Rampazzo2013}
{Rampazzo} R.,  {Panuzzo} P.,  {Vega} O.,  {Marino} A.,  {Bressan} A.,
  {Clemens} M.~S.,  2013, \mnras, 432, 374

\bibitem[\protect\citeauthoryear{{Randall}, {Forman}, {Giacintucci}, {Nulsen},
  {Sun}, {Jones}, {Churazov}, {David}, {Kraft}, {Donahue}, {Blanton},
  {Simionescu} \& {Werner}}{{Randall} et~al.}{2011}]{Randall2011}
{Randall} S.~W.,  {Forman} W.~R.,  {Giacintucci} S.,  {Nulsen} P.~E.~J.,  {Sun}
  M.,  {Jones} C.,  {Churazov} E.,  {David} L.~P.,  {Kraft} R.,  {Donahue} M.,
  {Blanton} E.~L.,  {Simionescu} A.,    {Werner} N.,  2011, \apj, 726, 86

\bibitem[\protect\citeauthoryear{{Schawinski}, {Kaviraj}, {Khochfar}, {Yoon},
  {Yi}, {Deharveng}, {Boselli}, {Barlow} \& et al.}{{Schawinski}
  et~al.}{2007}]{Schawinski2007}
{Schawinski} K.,  {Kaviraj} S.,  {Khochfar} S.,  {Yoon} S.-J.,  {Yi} S.~K.,
  {Deharveng} J.-M.,  {Boselli} A.,  {Barlow} T.,    et al. 2007, \apjs, 173,
  512

\bibitem[\protect\citeauthoryear{{Silverman}}{{Silverman}}{1986}]{Silverman86}
{Silverman} B.~W.,  1986, {Density estimation for statistics and data analysis}

\bibitem[\protect\citeauthoryear{{Tago}, {Einasto}, {Saar}, {Tempel},
  {Einasto}, {Vennik} \& {M{\"u}ller}}{{Tago} et~al.}{2008}]{Tago08}
{Tago} E.,  {Einasto} J.,  {Saar} E.,  {Tempel} E.,  {Einasto} M.,  {Vennik}
  J.,    {M{\"u}ller} V.,  2008, \aap, 479, 927

\bibitem[\protect\citeauthoryear{{Tammann}}{{Tammann}}{1994}]{Tamm94}
{Tammann} G.~A.,  1994, in {Meylan} G.,  {Prugniel} P.,  eds, European Southern
  Observatory Conference and Workshop Proceedings Vol.~49 of European Southern
  Observatory Conference and Workshop Proceedings, {Dwarf Galaxies in the
  Past}.
p.~3

\bibitem[\protect\citeauthoryear{{Toomre} \& {Toomre}}{{Toomre} \&
  {Toomre}}{1972}]{Toomre72}
{Toomre} A.,  {Toomre} J.,  1972, \apj, 178, 623

\bibitem[\protect\citeauthoryear{{Trinchieri} \& {Goudfrooij}}{{Trinchieri} \&
  {Goudfrooij}}{2002}]{Trinchieri02}
{Trinchieri} G.,  {Goudfrooij} P.,  2002, \aap, 386, 472

\bibitem[\protect\citeauthoryear{{Tully}}{{Tully}}{1987}]{Tully87}
{Tully} R.~B.,  1987, \apj, 321, 280

\bibitem[\protect\citeauthoryear{{Tully}}{{Tully}}{1988}]{Tully88}
{Tully} R.~B.,  1988, {Nearby galaxies catalog}.
Cambridge and New York, Cambridge University Press, 1988, 221 p.

\bibitem[\protect\citeauthoryear{{Werner}, {Oonk}, {Sun}, {Nulsen}, {Allen},
  {Canning}, {Simionescu}, {Hoffer}, {Connor}, {Donahue}, {Edge}, {Fabian},
  {von der Linden}, {Reynolds} \& {Ruszkowski}}{{Werner}
  et~al.}{2014}]{Werner2014}
{Werner} N.,  {Oonk} J.~B.~R.,  {Sun} M.,  {Nulsen} P.~E.~J.,  {Allen} S.~W.,
  {Canning} R.~E.~A.,  {Simionescu} A.,  {Hoffer} A.,  {Connor} T.,  {Donahue}
  M.,  {Edge} A.~C.,  {Fabian} A.~C.,  {von der Linden} A.,  {Reynolds} C.~S.,
    {Ruszkowski} M.,  2014, \mnras, 439, 2291

\bibitem[\protect\citeauthoryear{{Werner}, {Zhuravleva}, {Churazov},
  {Simionescu}, {Allen}, {Forman}, {Jones} \& {Kaastra}}{{Werner}
  et~al.}{2009}]{Werner2009}
{Werner} N.,  {Zhuravleva} I.,  {Churazov} E.,  {Simionescu} A.,  {Allen}
  S.~W.,  {Forman} W.,  {Jones} C.,    {Kaastra} J.~S.,  2009, \mnras, 398, 23

\bibitem[\protect\citeauthoryear{{Yi}, {Lee}, {Sheen}, {Jeong}, {Suh} \&
  {Oh}}{{Yi} et~al.}{2011}]{Yi2011}
{Yi} S.~K.,  {Lee} J.,  {Sheen} Y.-K.,  {Jeong} H.,  {Suh} H.,    {Oh} K.,
  2011, \apjs, 195, 22

\end{thebibliography}
 
 \clearpage
\appendix

 \section{UV and optical images of galaxies in NGC 5846}

Figures \ref{A1}  show the UV and optical colour composite images of the members of
NGC 5846. The size of each image is 5 x 5  arcmin.  One arcmin (bar shown) corresponds to $\approx$7 kpc at the distance of the group.

 \begin{figure}
 \centering
 \vspace{-0.5cm}  
 \includegraphics[width=7cm]{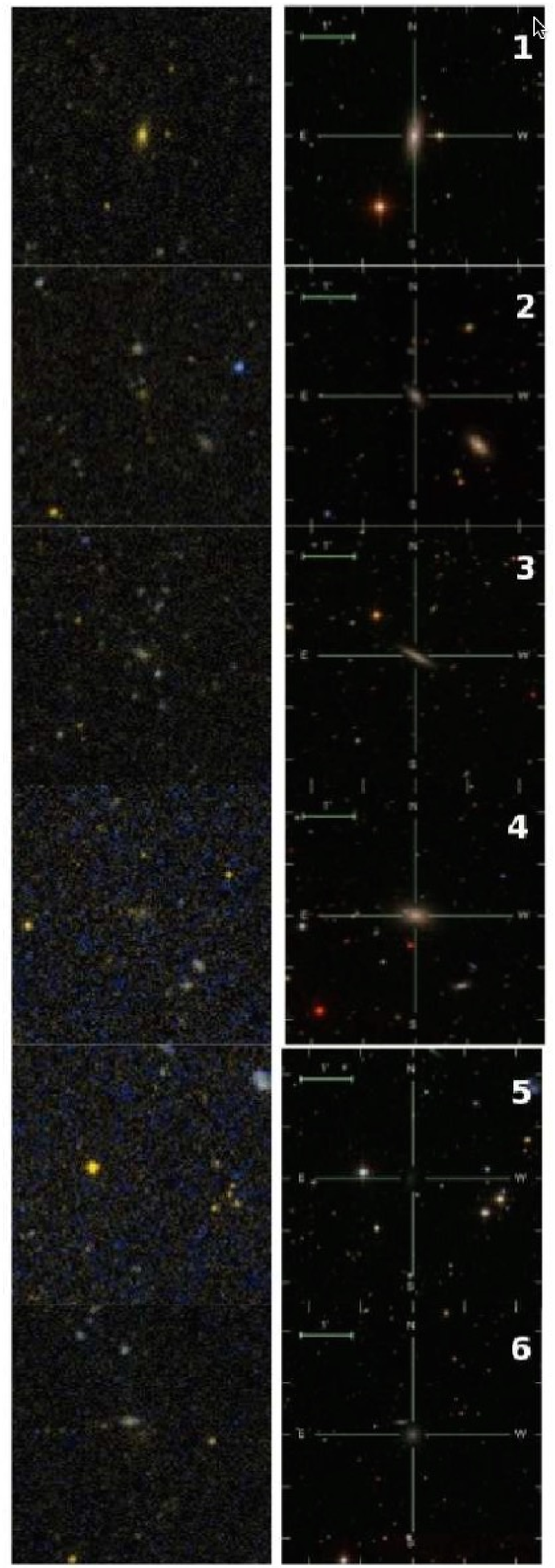}
   \caption{UV (FUV, blue, NUV, yellow, left) and optical (SDSS: g, r, i are blue, green and red, respectively, right) images of  members of NGC 5846.}
  \label{A1}
  \end{figure}

 \begin{figure}
 \centering
  \vspace{-0.5cm}  
 \includegraphics[width=7cm]{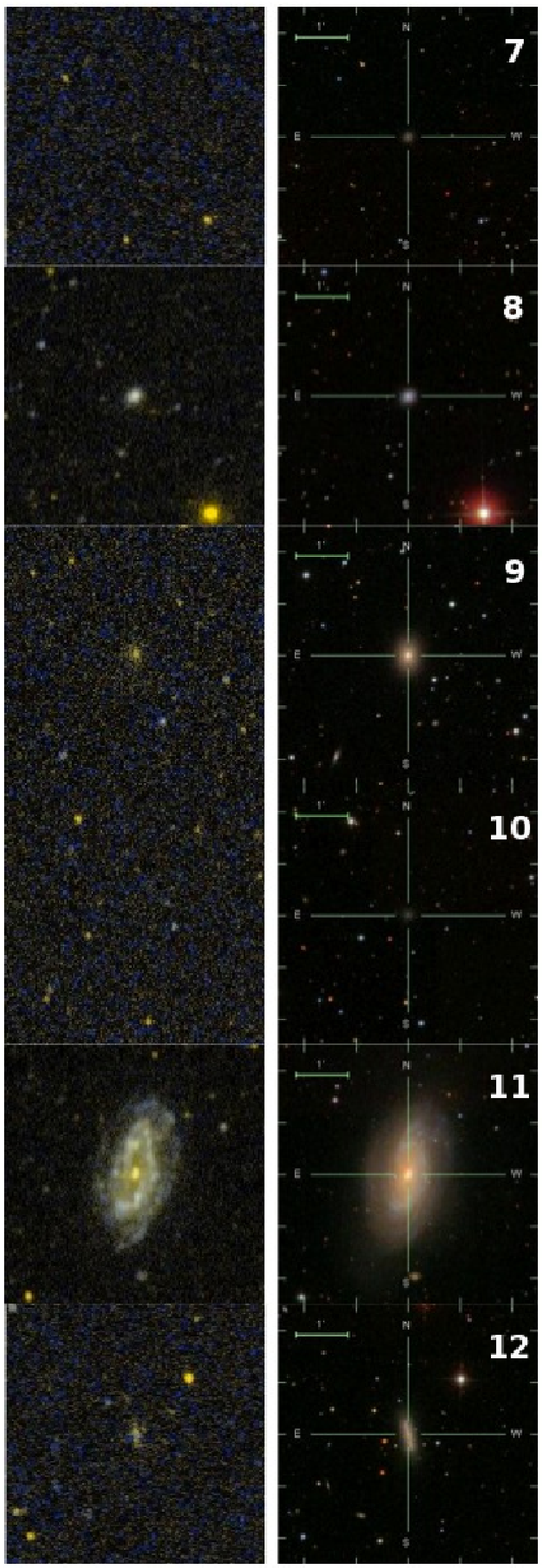}
 \addtocounter{figure}{-1}
 \caption{Continued.}
 \label{A1c}
 \end{figure}

 \begin{figure}
 \centering
  \vspace{-0.5cm}  
 \includegraphics[width=7cm]{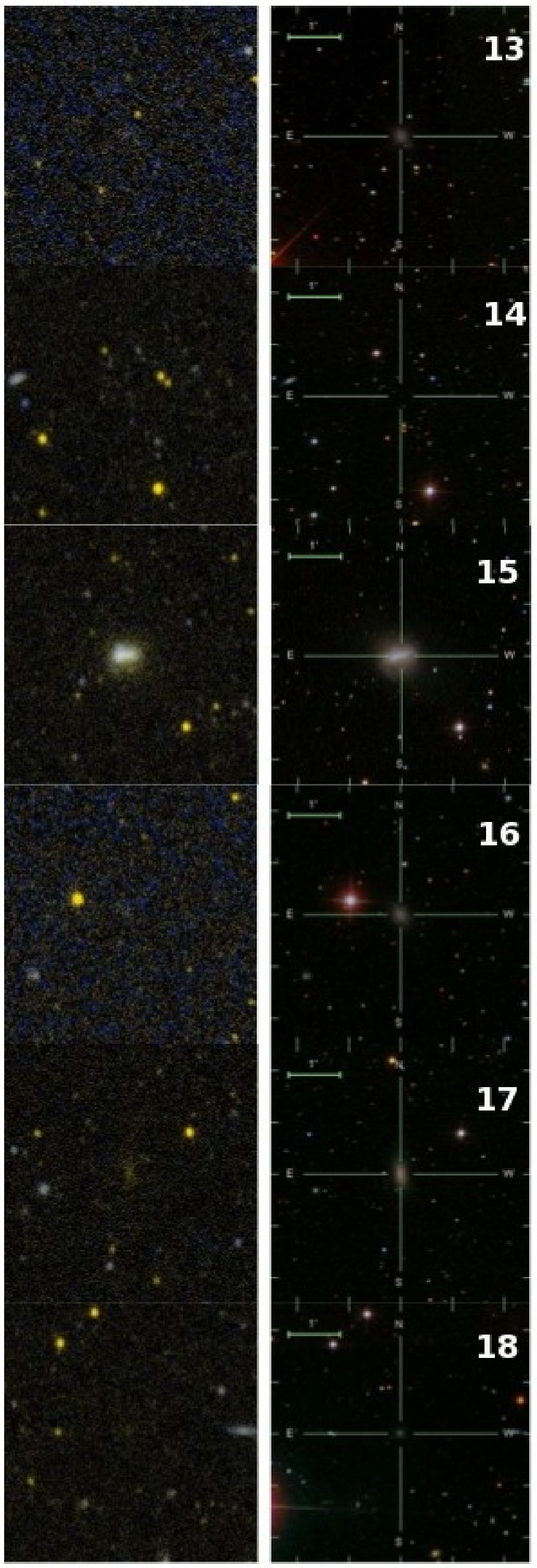}
 \addtocounter{figure}{-1}
   \caption{Continued.}
  \label{A1cc}
  \end{figure}

 \begin{figure}
  \centering
   \vspace{-0.5cm}  
 \includegraphics[width=7cm]{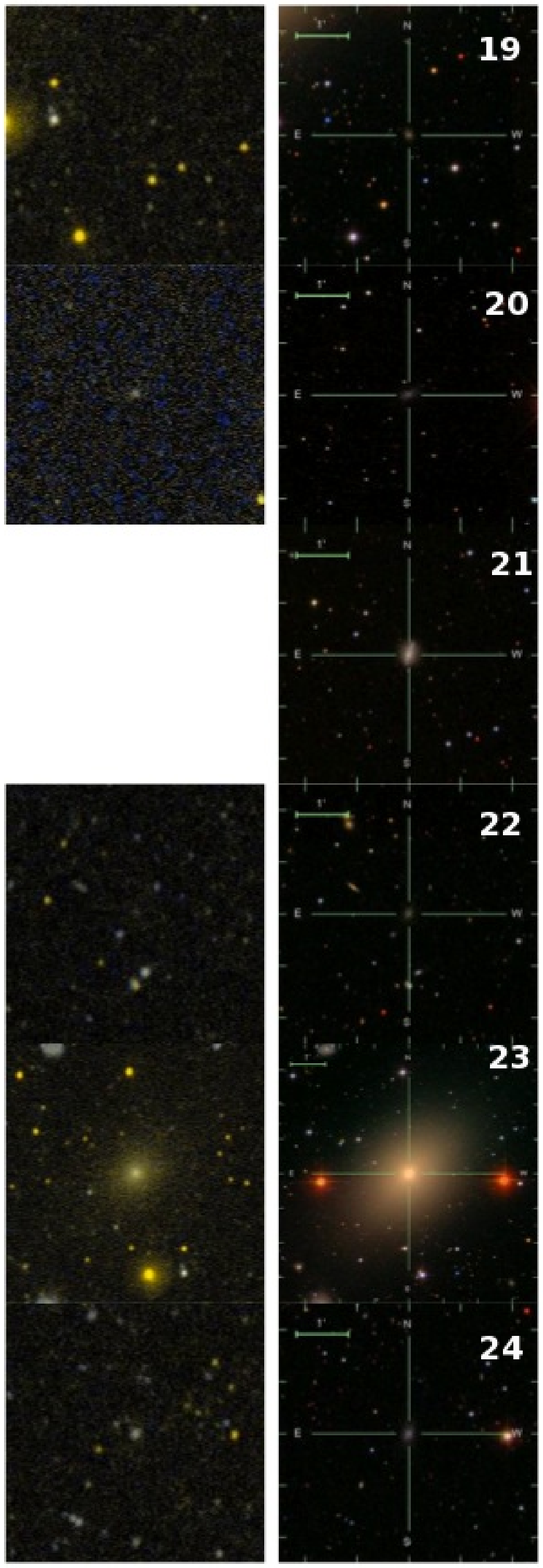}
 \addtocounter{figure}{-1}
   \caption{Continued.}
  \label{A1ccc}
  \end{figure}
	
 \begin{figure}
  \centering
   \vspace{-0.5cm}  
 \includegraphics[width=7cm]{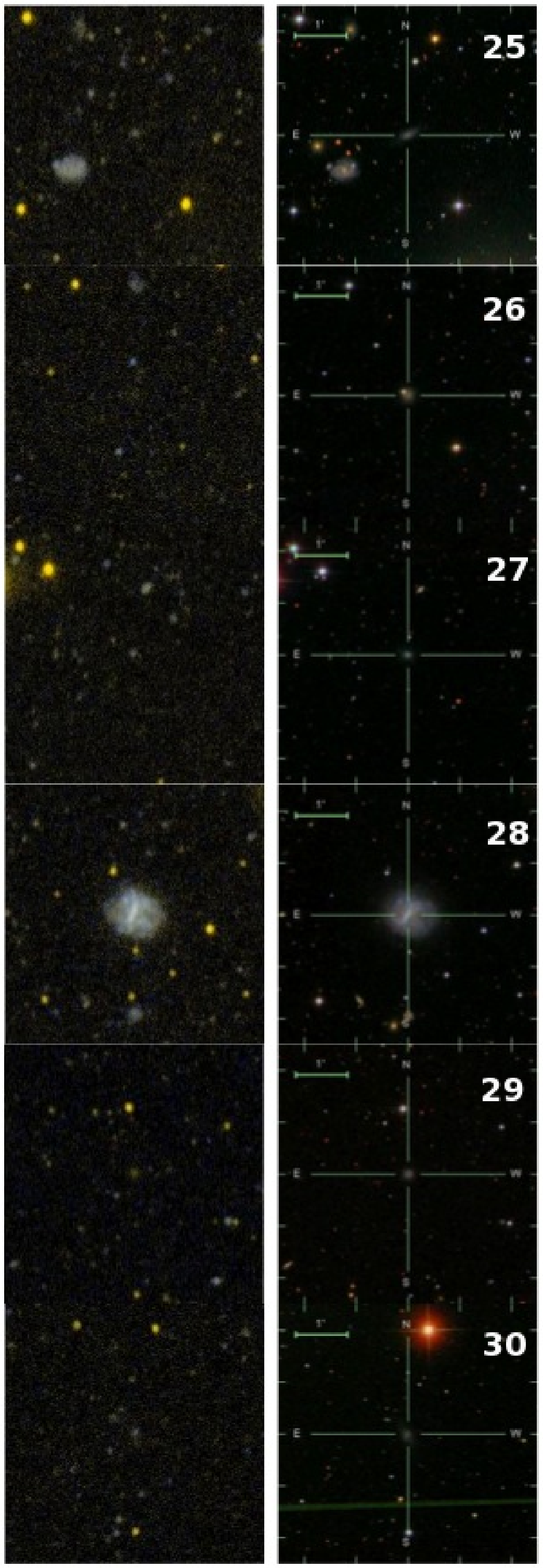}
 \addtocounter{figure}{-1}
   \caption{Continued.}
  \label{A1cccc}
  \end{figure}
		
 \begin{figure}
  \centering
   \vspace{-0.5cm}  
 \includegraphics[width=7cm]{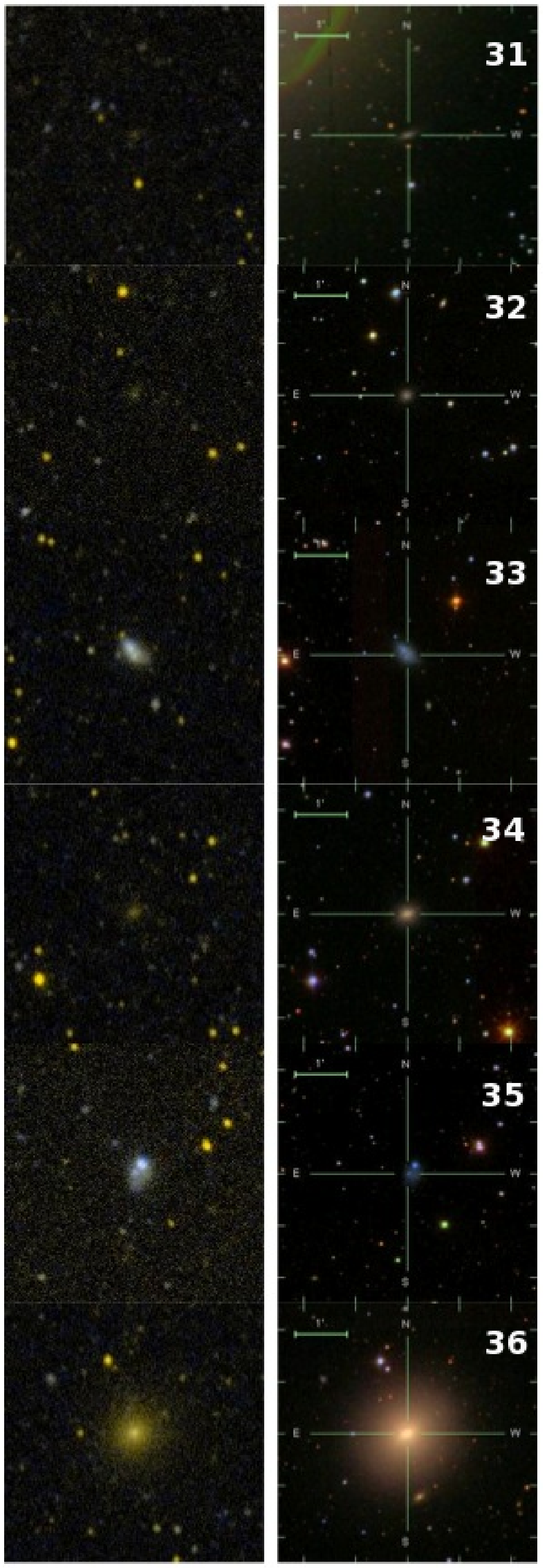}
 \addtocounter{figure}{-1}
   \caption{Continued.}
  \label{A1ccccc}
  \end{figure}
	
\begin{figure}
  \centering
   \vspace{-0.5cm}  
 \includegraphics[width=7cm]{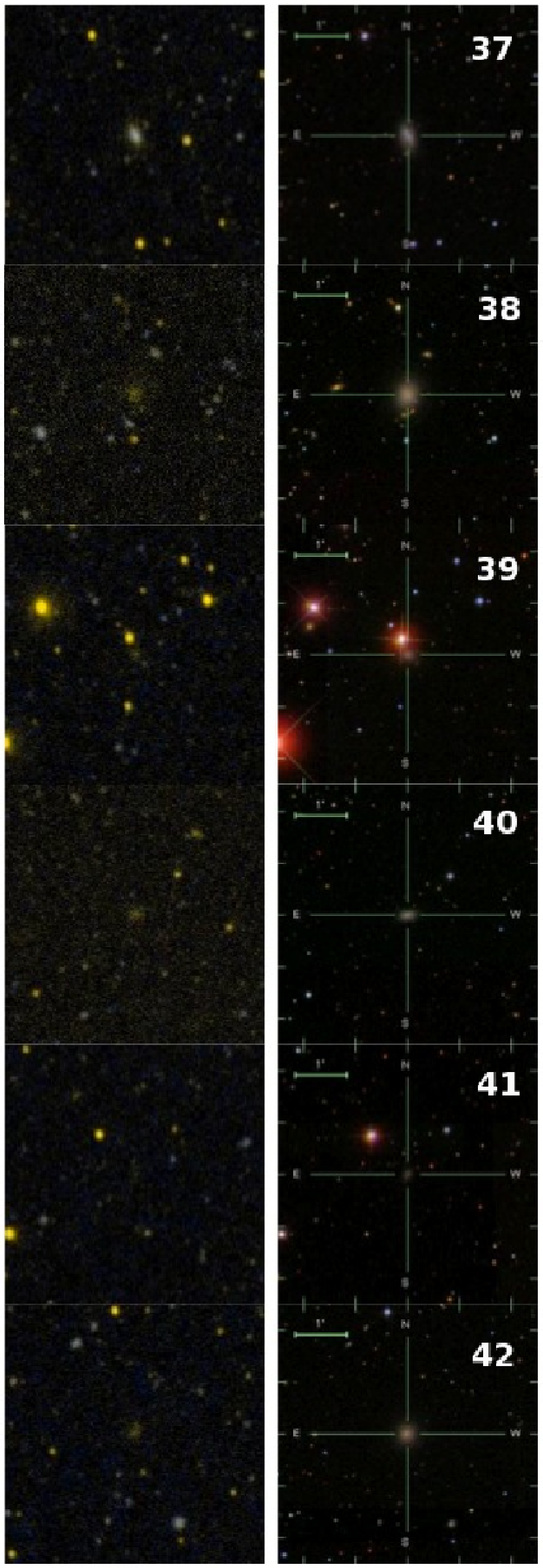}
 \addtocounter{figure}{-1}
   \caption{Continued.}
  \label{A1cccccc}
  \end{figure}

\begin{figure}
  \centering
   \vspace{-0.5cm}  
 \includegraphics[width=7cm]{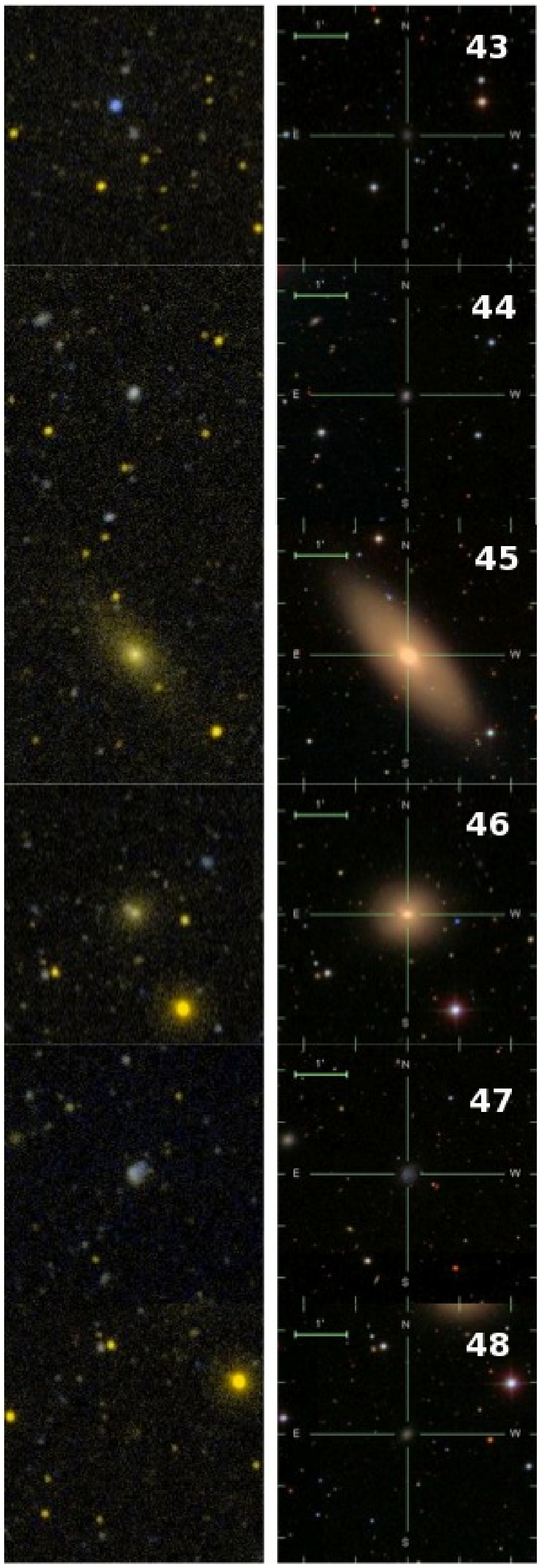}
 \addtocounter{figure}{-1}
   \caption{Continued.}
  \label{A1ccccccc}
  \end{figure}

\begin{figure}
  \centering
   \vspace{-0.5cm}  
 \includegraphics[width=7cm]{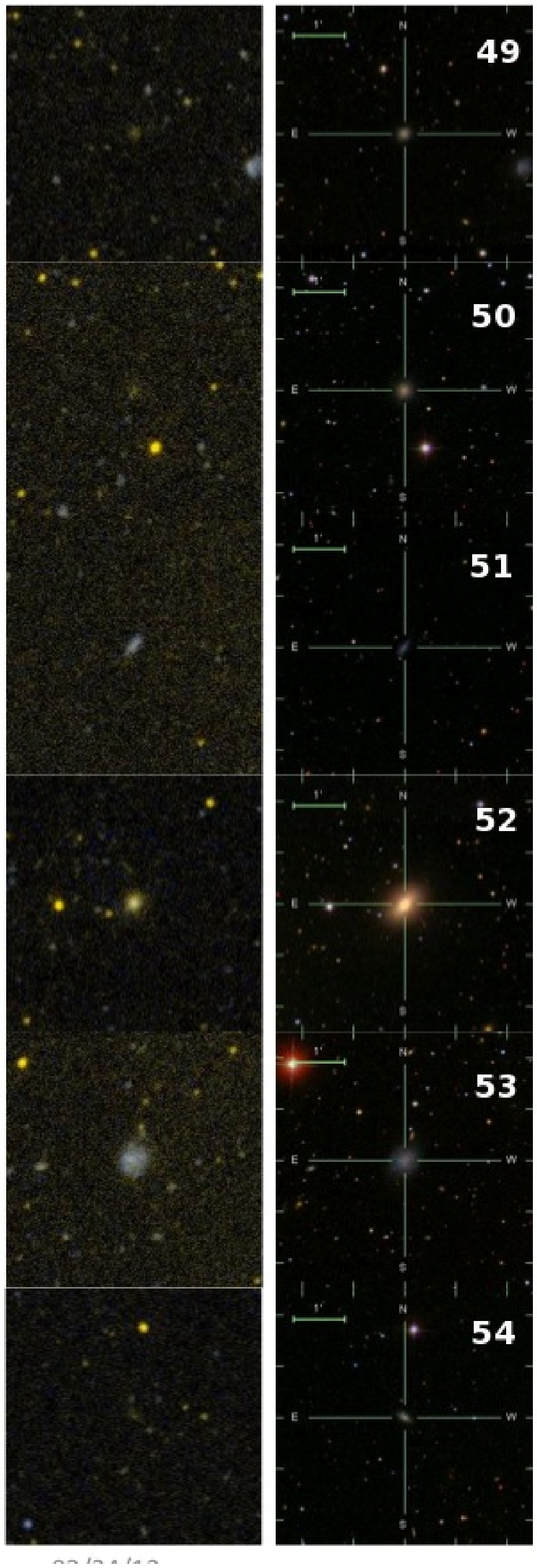}
 \addtocounter{figure}{-1}
   \caption{Continued.}
  \label{A1cccccccc}
  \end{figure}

\begin{figure}
  \centering
   \vspace{-0.5cm}  
 \includegraphics[width=7cm]{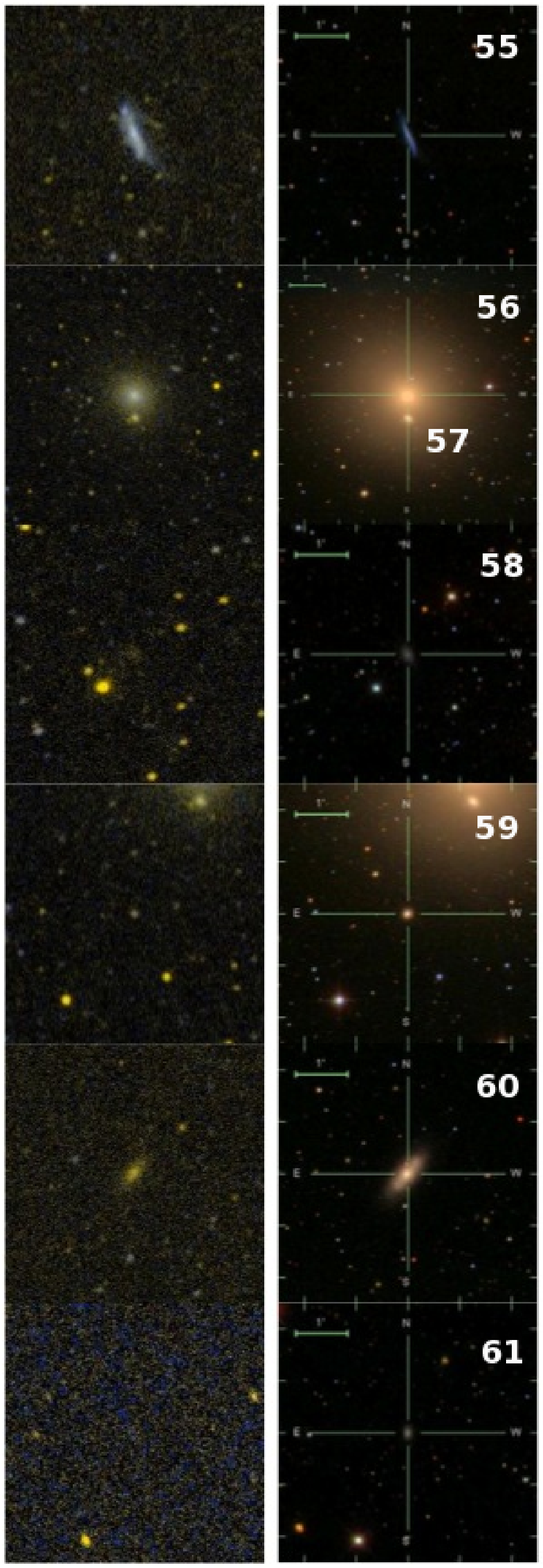}
 \addtocounter{figure}{-1}
   \caption{Continued.}
  \label{A1ccccccccc}
  \end{figure}

\begin{figure}
  \centering
   \vspace{-0.5cm}  
 \includegraphics[width=7cm]{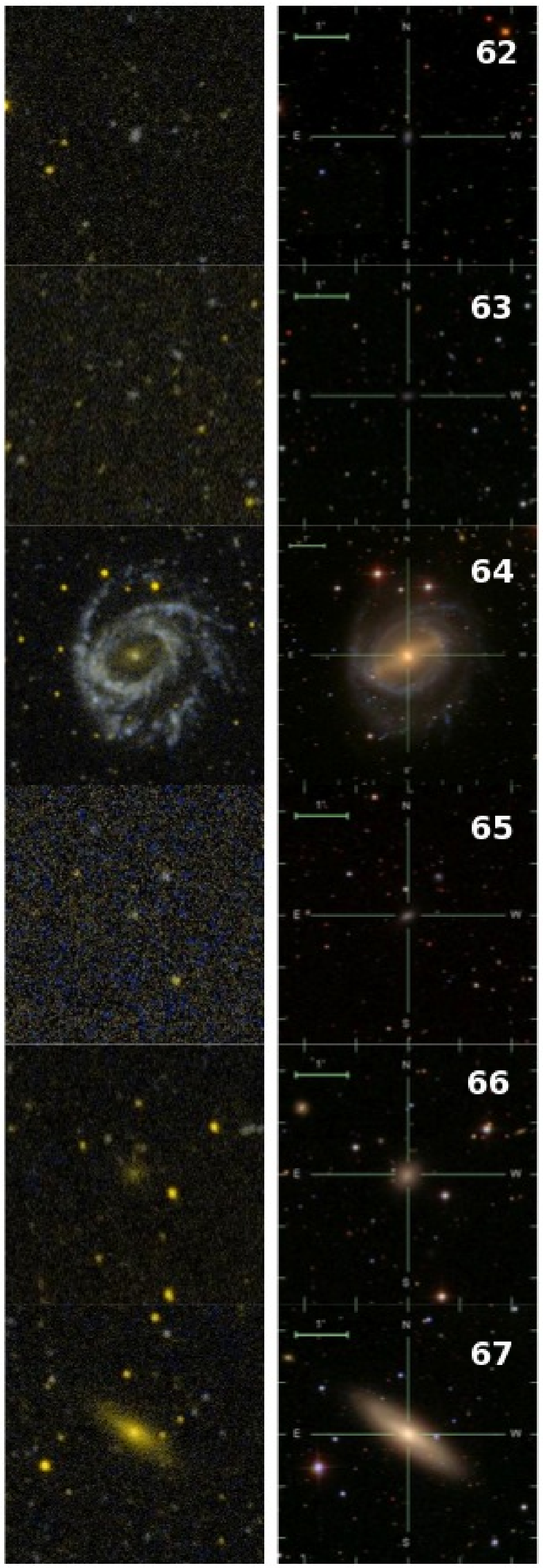}
 \addtocounter{figure}{-1}
   \caption{Continued.}
  \label{A1}
  \end{figure}

\begin{figure}
  \centering
   \vspace{-0.5cm}  
 \includegraphics[width=7cm]{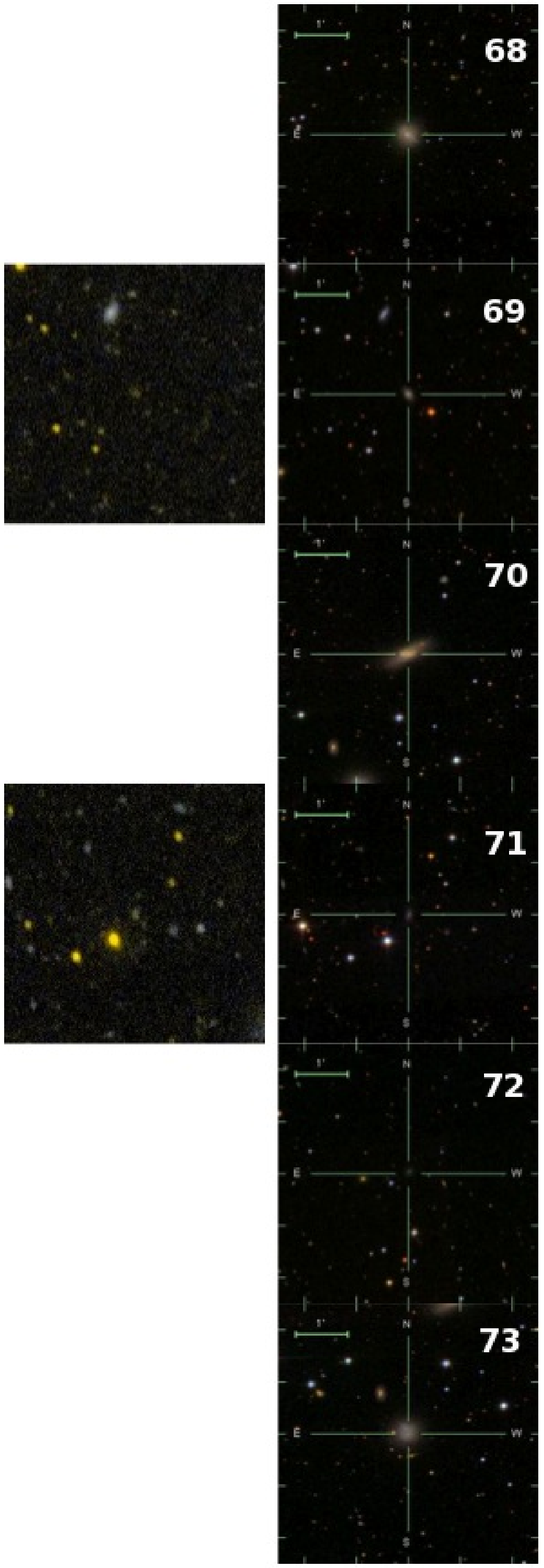}
 \addtocounter{figure}{-1}
   \caption{Continued.}
  \label{A1}
  \end{figure}

\begin{figure}
  \centering
   \vspace{-0.5cm}  
 \includegraphics[width=7cm]{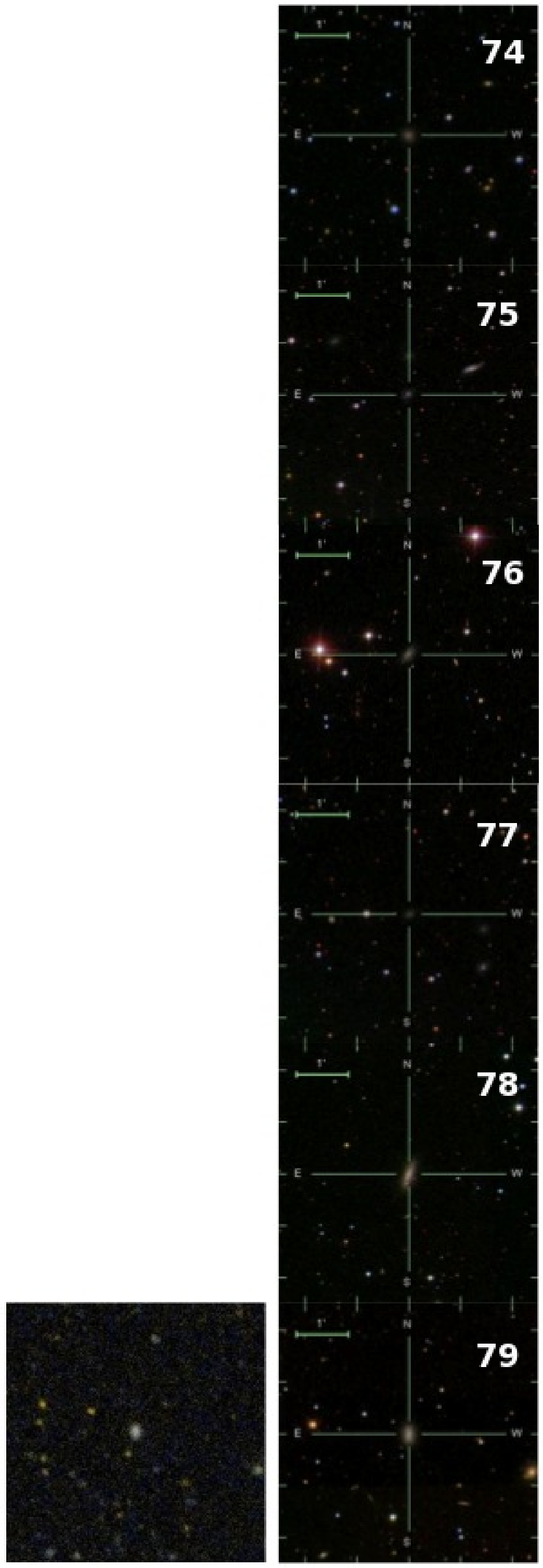}
 \addtocounter{figure}{-1}
   \caption{Continued.}
  \label{A1}
  \end{figure}

\begin{figure}
  \centering
   \vspace{-0.5cm}  
 \includegraphics[width=7cm]{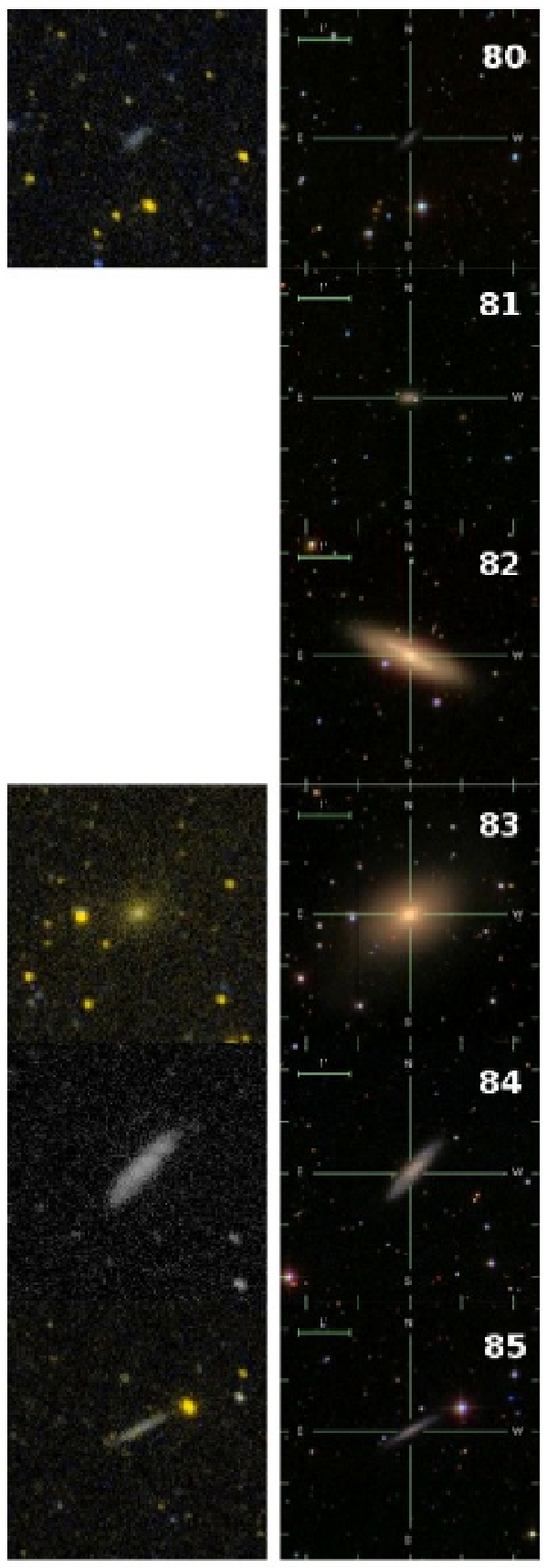}
 \addtocounter{figure}{-1}
   \caption{Continued.}
  \label{A1}
  \end{figure}

\begin{figure}
  \centering
   \vspace{-0.5cm}  
 \includegraphics[width=7cm]{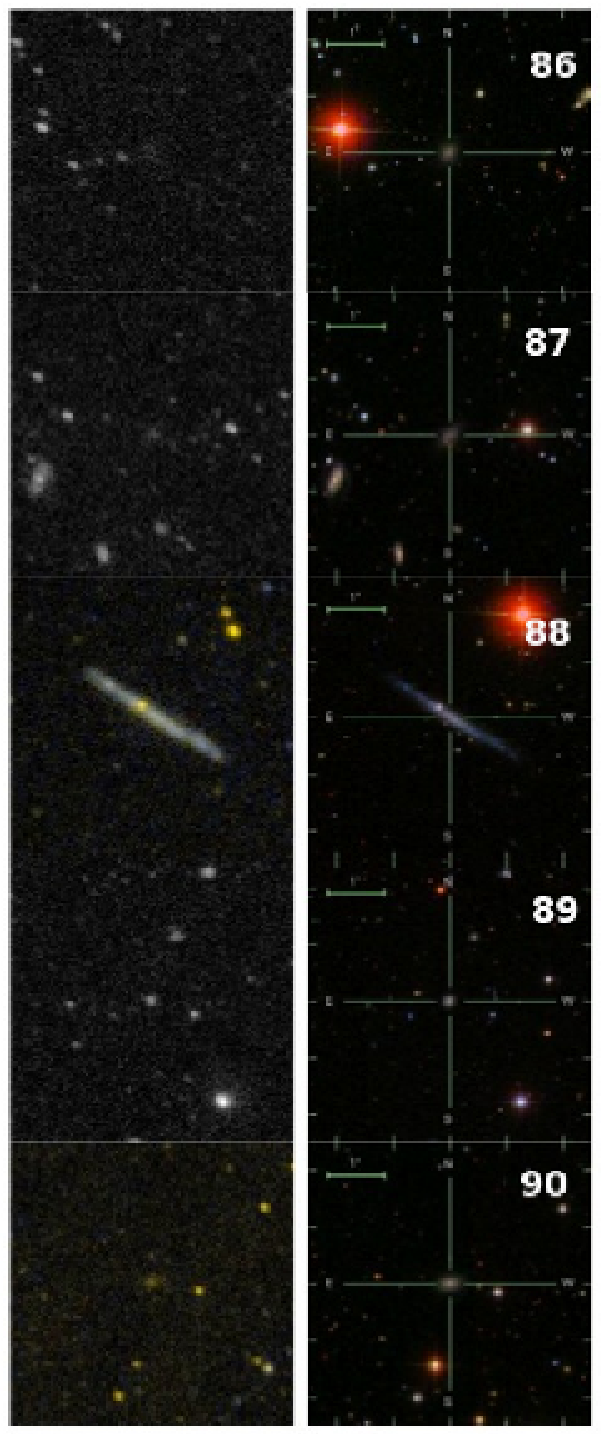}
 \addtocounter{figure}{-1}
   \caption{Continued.}
  \label{A1}
  \end{figure}

\clearpage

\section{The environment of NGC 5846}

\begin{table*}
	\caption{Galaxies in a box of 4 Mpc $\times$ 4 Mpc centered on NGC 5846 with heliocentric velocity between
	807 and 2600 km/s.}
	   \label{a2}	 
	 \scriptsize
	\begin{tabular}{llllllllllll}
	\hline\hline
	   Galaxies  & RA & Dec.&  Morph.& Mean Hel.& logD$_{25}$ & logr$_{25}$ & P.A. &B$_T$ \\ 
	                   & (J2000) & (J2000) & type &  Vel.& &  &   &  \\
		          & (h) & (deg)   & & [km/s]  &[arcmin] & &[deg]  & mag \\
				  
	\hline
PGC1150067 & 14.80685 & -0.17365 &    & 2350 $\pm$    60 & 0.34 & 0.07 & 68 & 18.83 $\pm$    0.33 \\ 
SDSSJ144834.42+052552.5 & 14.80956 & 5.43127 &    & 1698 $\pm$    1 &    &    & 153.9 & 18.18 $\pm$    0.5 \\ 
PGC1241857 & 14.8397 & 2.95819 &    & 1697 $\pm$    1 & 0.56 & 0.21 & 166.7 & 17.18 $\pm$    0.67 \\ 
SDSSJ145059.85+022016.4 & 14.84996 & 2.3379 &    & 1528 $\pm$    24 & 0.49 & 0.12 & 107.3 & 18.56 $\pm$    0.35 \\ 
SDSSJ145106.77+023127.0 & 14.85189 & 2.52415 &    & 2071 $\pm$    3 & 0.49 & 0.09 & 79.3 & 18.53 $\pm$    0.29 \\ 
SDSSJ145201.94+025841.8 & 14.86721 & 2.97824 &    & 1814 $\pm$    2 & 0.45 & 0.02 &    & 17.86 $\pm$    0.35 \\ 
NGC5768 & 14.86887 & -2.52976 & 5.3 & 1947 $\pm$    8 & 1.11 & 0.12 & 110.9 & 13.52 $\pm$    0.5 \\ 
SDSSJ145243.39+043616.7 & 14.8787 & 4.60467 &    & 1589 $\pm$    8 & 0.8 & 0.49 & 98.7 & 17.96 $\pm$    0.5 \\ 
IC1066 & 14.88413 & 3.29601 & 3.2 & 1567 $\pm$    4 & 1.08 & 0.26 & 69.5 & 14.27 $\pm$    0.28 \\ 
IC1067 & 14.88479 & 3.33175 & 3 & 1566 $\pm$    4 & 1.26 & 0.09 & 129 & 13.62 $\pm$    0.29 \\ 
NGC5770 & 14.88751 & 3.95977 & -2 & 1477 $\pm$    15 & 1.04 & 0.09 &    & 13.18 $\pm$    0.24 \\ 
NGC5774 & 14.89514 & 3.58253 & 6.9 & 1566 $\pm$    2 & 1.23 & 0.21 & 116.7 & 13.1 $\pm$    0.51 \\ 
IC1070 & 14.89759 & 3.48472 & 2.4 & 1677 $\pm$    15 & 0.89 & 0.4 & 121.7 & 15.94 $\pm$    0.66 \\ 
NGC5775 & 14.89933 & 3.54426 & 5.2 & 1676 $\pm$    2 & 1.57 & 0.64 & 148.9 & 12.23 $\pm$    0.13 \\ 
PGC1223887 & 14.90956 & 2.34392 &    & 2158 $\pm$    42 & 0.56 & 0.27 & 2.4 & 18.36 $\pm$    0.47 \\ 
PGC135871 & 14.91193 & 1.16187 & 10 & 1830 $\pm$    7 &    &    &    & 17.8 $\pm$    0.35 \\ 
PGC1197564 & 14.91551 & 1.52556 &    & 1759 $\pm$    8 & 0.49 & 0.11 & 70.6 & 18.77 $\pm$    0.87 \\ 
PGC1184577 & 14.91925 & 1.10064 &    & 1715 $\pm$    3 & 0.4 & 0.18 & 9.9 & 18.35 $\pm$    0.28 \\ 
PGC053365 & 14.92858 & -1.00899 & 9 & 1849 $\pm$    2 & 0.8 & 0.21 & 43.2 & 15.78 $\pm$    0.37 \\ 
UGC09601 & 14.93383 & -1.38787 & 5.9 & 1862 $\pm$    8 & 1.09 & 0.11 & 171.7 & 14.62 $\pm$    0.38 \\ 
PGC1083529 & 14.93898 & -2.76159 &    & 1888 $\pm$    9 & 0.41 & 0.06 & 85.4 & 18.66 $\pm$    0.3 \\ 
PGC184824 & 14.96245 & -2.98707 & 1.6 & 1800 $\pm$    64 & 0.61 & 0.13 & 130.7 & 17.14 $\pm$    0.37 \\ 
PGC184842 & 14.96883 & -1.31237 & 3.8 & 1947 $\pm$    7 & 0.75 & 0.33 & 149.5 & 16.45 $\pm$    0.31 \\ 
NGC5792 & 14.97296 & -1.09093 & 3 & 1924 $\pm$    2 & 1.55 & 0.41 & 88.5 & 12.12 $\pm$    0.12 \\ 
PGC053577 & 15.00036 & -1.09107 & 10 & 1886 $\pm$    2 & 0.66 & 0.03 &    & 15.81 $\pm$    0.36 \\ 
UGC09682 & 15.07505 & -0.85135 & 8.6 & 1810 $\pm$    4 & 1.21 & 0.6 & 175.2 & 15.43 $\pm$    0.36 \\ 
PGC2801020 & 15.07616 & -2.587 & 10 & 1624 $\pm$    2 & 0.89 & 0.14 & 147.8 & 16.51 $\pm$    0.32 \\ 
PGC1085904 & 15.11893 & -2.66278 &    & 2042 $\pm$    1 & 0.53 & 0.18 & 93.2 & 17.92 $\pm$    0.3 \\ 
PGC1128787 & 15.15933 & -1.02163 &    & 1858 $\pm$    1 & 0.5 & 0.19 & 133 & 17.56 $\pm$    0.35 \\ 
PGC054159 & 15.17978 & -0.34824 & 6 & 2159 $\pm$    2 & 0.91 & 0.6 & 86 & 16.25 $\pm$    0.57 \\ 
PGC1176138 & 15.20881 & 0.81256 & 10 & 1844 $\pm$    5 & 0.89 & 0.31 & 118.1 & 16.32 $\pm$    0.29 \\ 
PGC1200646 & 15.2115 & 1.62325 &    & 1873 $\pm$    4 & 0.62 & 0.31 & 56.3 & 17.25 $\pm$    0.3 \\ 
PGC258278 & 15.2125 & 6.16417 &    & 1487 $\pm$    8 &    &    &    &    $\pm$       \\ 
PGC1236445 & 15.25262 & 2.75183 &    & 1764 $\pm$    3 & 0.78 & 0.3 & 56.8 & 16.89 $\pm$    0.49 \\ 
PGC054452 & 15.25961 & 2.24823 & -1 & 1906 $\pm$    2 & 0.98 & 0.1 & 107.5 & 14.82 $\pm$    0.34 \\ 
UGC09787 & 15.2619 & 1.45576 & 9 & 1589 $\pm$    5 & 1 & 0.23 & 46.9 & 15.4 $\pm$    0.36 \\ 
PGC1234821 & 15.26832 & 2.69196 &    & 1463 $\pm$    2 & 0.72 & 0.62 & 169 & 17.23 $\pm$    0.45 \\ 
PGC3124577 & 15.29151 & 3.58548 &    & 1883 $\pm$    3 & 0.58 & 0.04 &    & 17.36 $\pm$    0.35 \\ 
PGC1168006 & 15.31585 & 0.51564 &    & 2083 $\pm$    14 & 0.49 & 0.25 & 141 & 18.16 $\pm$    0.33 \\ 
PGC1230249 & 15.32038 & 2.54503 & 4.2 & 1877 $\pm$    4 & 0.72 & 0.18 & 49.5 & 16.17 $\pm$    0.35 \\ 
PGC091432 & 15.32997 & 3.97806 & 7.9 & 1712 $\pm$    8 & 0.7 & 0.63 & 56 &    $\pm$       \\ 
NGC5913 & 15.34873 & -2.57796 & 1 & 2002 $\pm$    5 & 1.27 & 0.39 & 173.5 & 14.02 $\pm$    0.38 \\ 
NGC5921 & 15.36568 & 5.07041 & 4 & 1430 $\pm$    23 & 1.48 & 0.17 & 140 & 11.68 $\pm$    0.1 \\ 
PGC258471 & 15.37742 & 5.82917 &    & 1796 $\pm$    8 &    &    &    &    $\pm$       \\ 
UGC09830 & 15.38356 & 4.52917 & 5.9 & 1830 $\pm$    5 & 0.79 & 0.39 & 33 & 15.91 $\pm$    0.62 \\ 
PGC3123131 & 15.41392 & 3.08141 &    & 1754 $\pm$    1 & 0.5 & 0.15 & 83.5 & 17.58 $\pm$    0.35 \\ 
\hline
\end{tabular}
\end{table*}

\end{document}